\documentclass[11pt,graphicx,amsmath]{article}
\usepackage{amsmath}
\usepackage{graphicx}
\usepackage{CJK}
\usepackage{bm}
\usepackage{xcolor}
\usepackage[numbers,sort&compress]{natbib}

\def\gsim{\;\rlap{\lower 2.5pt  \hbox{$\sim$}}\raise 1.5pt\hbox{$>$}\;}
\def\lsim{\;\rlap{\lower 2.5pt  \hbox{$\sim$}}\raise 1.5pt\hbox{$<$}\;}
\def\edth{\;\raise1.0pt\hbox{$'$}\hskip-6pt\partial\;}
\def\baredth{\;\overline{\raise1.0pt\hbox{$'$}\hskip-6pt \partial}\;}
\def\be{\begin{equation}}
\def\ee{\end{equation}}
\def\ba{\begin{eqnarray}}
\def\ea{\end{eqnarray}}
\def\nn{\nonumber}

\def\bl#1\el{\begin{align}#1\end{align}}
\def\l{\left}
\def\r{\right}

\title{ Second-order cosmological perturbations. I. Produced by scalar-scalar coupling in synchronous gauge}

\author{\small   Bo Wang  \thanks{ymwangbo@mail.ustc.edu.cn}\ ,
          Yang  Zhang  \thanks{yzh@ustc.edu.cn}
           \\
 \small   Department of  Astronomy, Key Laboratory for Researches in Galaxies and Cosmology, \\
 \small    University of Science and Technology of China,   Hefei, Anhui, 230026,  China   }

 \date{}

\begin{document}

\maketitle

\large

\begin{center}
\text{\large\bf Abstract}
\end{center}

We present a systematic study  of
the  2nd-order scalar, vector   and tensor   metric perturbations
 in the Einstein-de Sitter Universe in synchronous coordinates.
For the scalar-scalar coupling between 1st-order perturbations,
 we  decompose  the 2nd-order  perturbed Einstein equation  into
the respective field equations of 2nd-order scalar, vector, and tensor perturbations,
and obtain their  solutions with  general initial conditions.
In particular,  the decaying modes of solution are included,
the 2nd-order vector  is   generated even  if the 1st-order vector is absent,
and  the solution of the 2nd-order tensor  corrects that in literature.
 We perform general synchronous-to-synchronous  gauge transformations
up to 2nd-order generated by
a 1st-order vector field $\xi^{(1)\mu}$ and a 2nd-order   $\xi^{(2)\mu}$.
All the residual gauge modes of
2nd-order metric perturbations  and density contrast are found,
and their number   is substantially reduced
when the transformed 3-velocity of dust is set to zero.
Moreover, we show that only  $\xi^{(2)\mu}$ is effective
in carrying out 2nd order transformations
 that we consider,
because $\xi^{(1)\mu}$  has been used in obtaining the 1st-order perturbations.
Holding the 1st-order perturbations fixed,
the   transformations  by $\xi^{(2)\mu}$ on the 2nd-order perturbations
have the same  structure as those  by $\xi^{(1)\mu}$ on the 1st-order perturbations.

\



\Large

\section{Introduction}

Studies of metric perturbations of   Robertson-Walker spacetimes
 within general relativity constitute the theoretical foundation  of cosmology.
In the past the   perturbations, both scalar and tensorial,
have been extensively explored   to linear order
\cite{Lifshitz1946,LifshitzKhalatnikov1963,
   PressVishniac1980,Bardeen1980,KodamSasaki1984,Grishchuk1994},
which have been used in calculations of large scale structure
    \cite{ Peebles1980},
cosmic microwave background radiation (CMB)
          \cite{BaskoPolnarev1984,Polnarev1985,
          MaBertschinger1995,ZaldarriagaHarari1995,Kosowsky1996,ZaldarriagaSeljak1997,
          Kamionkowski1997,KeatingTimbie1998,ZhaoZhang2006,Baskaran,Polnarevmiller2008}
and relic gravitational wave (RGW)
\cite{Grishchuk,FordParker1977GW,Starobinsky,Rubakov,Fabbri,AbbottWise1984,
    Allen1988,Giovannini,Tashiro,zhangyang05,Morais2014}.
In the era of precision cosmology,
it is necessary to study the 2nd-order perturbation beyond the linear perturbation,
to include  nonlinear effects of gravity
on CMB anisotropies and polarization \cite{PyneCarroll1996, MollerachHarariMatarrese2004},
the non-Gaussianality of primordial perturbation  \cite{Bartolo2010},
and on relic gravitational waves \cite{AnandaClarksonWands2007,Baumann2007}  etc.
Recently LIGO collaboration announced its direct detections
of gravitational waves  emitted  from  binary black holes   \cite{GW150914}.
RGW is not detected yet
and only constraints are given upon  the  background of GW, including RGW,
in a band $10-2000$ Hz \cite{aLIGO O1}.
The current observational constraint on RGW is given by CMB observations,
which is given in terms of the tensor-scalar ratio  $r < 0.1 $
over very low frequencies $10^{-18} \sim 10^{-16}$Hz
     \cite{WMAP9Bennett,Planck2014}.
Since both   scalar and tensor  metric perturbations are generated during inflation,
one might expect that they should be of the same order of amplitude.
 In some class of scalar inflation models,
 the ratio is predicted  to be $r=16\epsilon$,
where $\epsilon$ is the slow-roll parameter.
We like to see other possible mechanisms than the inflation,
which may change the tensor perturbation during the cosmic evolution.
We are mainly motivated by this issue,
and want to examine how the evolution of RGW  during  expansion
will be  affected by the  metric  perturbation itself.
This  is encoded in the  2nd-order perturbed Einstein equation,
in which
all three irreducible parts of metric perturbations will   appear:
scalar, vector, and tensor.
In literature  the 2nd-order perturbation has   been  studied
to certain extent.
Tomita started a study of
the 2nd-order perturbations in synchronous coordinates, in particular,
and gave the general equations of 2nd-order perturbations  \cite{Tomita1967},
analyzed the 2nd-order density contrast in some special cases  \cite{Tomita1971}.
In the context of large scale structure,
gravitational instability was studied in the 2nd-order
 perturbations  \cite{MatarresePantanoSa'ez1994,Russ1996,Salopek}.
 Ref.~\cite{MalikWands2004} studies the  gauge-invariant  definition
of  the  second  order curvature perturbation.
The 2nd-order perturbations  were studied in the  Arnowitt-Deser-Misner framework
\cite{NohHwang2004,HwangNoh2012},
and the equation of   density was given
with  the source of squared RGW \cite{HwangNoh2005}.
A gauge-invariant formulation of  2nd-order perturbations
was developed in Refs.~\cite{Nakamura2003,Domenech&Sasaki2017}.
Matarrese {\it et al} derived  the field equations of 2nd scalar and tensor perturbations
in Einstein-de Sitter model
in the case of scalar-scalar coupling  \cite{Bruni97, Matarrese98},
but the 2nd-order vector is not given,
and the  2nd-order tensor is not complete.
The 2nd-order vector due to scalar-scalar coupling is explored
 in Poisson gauge  \cite{MollerachHarariMatarrese2004,Lu2008}.
Ref.~\cite{Lu2008}  calculated only
the 2nd-order vector perturbation  in Poisson gauge,
and did not give the 2nd-order scalar and tensor perturbations.
In $\Lambda$CDM framework,
Ref.~\cite{Brilenkov&Eingorm2017} calculated 2nd-order
scalar and vector perturbations in the Poisson gauge.

In this paper we consider the  Einstein-de Sitter model   filled with irrotational matter,
the 1st-order vector perturbation can be dropped as a gauge mode.
However,
the 2nd-order vector perturbation will inevitably appear
along with the 2nd-order scalar and tensor perturbations.
Moreover,
these    metric perturbations  are coupled.
In the 2nd-order perturbed Einstein equation,
   there are three kinds of   coupling terms:
scalar-scalar, scalar-tensor, and tensor-tensor,
which are  products of 1st-order metric perturbations.
  The (00) component of the equation is the energy constraint
 and   the (0i) component is the momentum constraint;
neither   contain  second-order time derivatives.
The (ij) components are  evolution equations,
and involve second-order time derivatives of metric perturbations,
which  need to be decomposed  into equations of
  scalar, vector, and tensor, respectively.
This will involve lengthy calculations.
We shall write the  2nd-order perturbed Einstein equations
  into three  sets of equations,
each having,
respectively, the coupling of scalar-scalar,
scalar-tensor, and tensor-tensor.
For the   scalar-scalar  coupling,
we present a complete  decomposition,
derive the  equations of 2nd-order scalar, vector, and tensor perturbation, respectively,
 and obtain  their  solutions with general initial conditions.
Moreover, perturbations contain residual gauge  modes
in synchronous coordinates.
We shall also calculate 2nd-order residual gauge transformations from synchronous to synchronous,
and  identify  the  residual gauge modes
of the  2nd-order  scalar, vector, tensor metric perturbations.

In Sec. 2,
we briefly review  the  results of 1st-order  perturbations,
which will be  used  later.

In Sec. 3,  we  write  the 2nd-order Einstein equations into three sets,
according to scalar-scalar, scalar-tensor, and tensor-tensor couplings, respectively.

In Sec. 4,  for the scalar-scalar coupling,
     we derive  the  solutions of the 2nd-order scalar, vector, tensor perturbations.

In Sec. 5, we derive   the  residual gauge modes
       in the  2nd-order metric, and density  perturbations
       in synchronous coordinates.

 Section 6  is  the  conclusions and discussion.

In the Appendixes,
we attach  the technical  details of some formulas    used in the paper.
Appendix A lists  the 1st-order metric perturbations and the  density contrast.
Appendix B gives the 2nd-order perturbed Ricci  tensors.
Appendix C gives  the  synchronous-to-synchronous
              gauge transformations up to 2nd-order for a Robertson-Walker (RW) spacetime.
Appendix D  lists  the 2nd-order synchronous-to-Poisson gauge transformation.
We work with the synchronous coordinates,
  adopt  mostly the notation   in Ref.~\cite{Matarrese98} for comparison,
and take the speed of light $c=1$.

\section{   First-Order Perturbations}

In this section,
we  introduce notations and outline the  results of 1st-order  perturbations,
which will be used in later sections.
We consider the   Universe filled with the irrotational, pressureless dust
with the  energy-momentum tensor  $T^{\mu\nu}=\rho U^\mu U^\nu$,
where  $\rho $ is the mass density,
$U^\mu=(a^{-1},0,0,0)$
is 4-velocity such that     $U^\mu U_\mu =  -1$.
We take the perturbations of velocity  to be  $U^{(1)\mu} = U^{(2)\mu}= 0 $
[see relevant discussions in the paragraph around Eq.(\ref{U1Condition1}) in Appendix C],
but the density with perturbations  to be
\be
\rho= \rho^{(0)}  \l(1 + \delta^{(1)}+\frac12\delta^{(2)}  \r),
\ee
where $ \rho^{(0)} $ is the background density,
$\delta^{(1)}$, $\delta^{(2)}$ are the 1st, 2nd-order density contrasts.
The nonvanishing components are  $T_{00} = a^{2}\rho$ and  $T^{00} = a^{-2}\rho$.

The spatial flat RW metric in synchronous coordinates
\be \label{18q1}
ds^2=g_{\mu\nu}   dx^{\mu}dx^{\nu}=a^2(\tau)[-d\tau^2
+\gamma_{ij}   dx^idx^j] ,
\ee
where  $\tau$ is conformal time,
 $a(\tau)\propto\tau^2$ for the Einstein-de Sitter model,
$\gamma_{ij}$  is written as
\be\label{eq1}
\gamma_{ij} =\delta_{ij} + \gamma_{ij}^{(1)} + \frac{1}{2} \gamma_{ij}^{(2)}
\ee
where
$\gamma_{ij}^{(1)}$ and  $\gamma_{ij}^{(2)}$
are the 1st- and 2nd-order metric perturbations, respectively.
From (\ref{eq1}), one has $g^{ij}=a^{-2}\gamma^{ij}$ with
$\gamma^{ij}=\delta^{ij} -\gamma^{(1)ij}
-\frac{1}{2}\gamma^{(2)ij}+\gamma^{(1)ik}\gamma^{(1)j}_{k}$.
In this paper,
we use the same notations
 as in Ref. \cite {Matarrese98} for simple comparisons,
and use the indices  $\mu,\nu = 0, 1, 2, 3$ and  $i,j= 1, 2, 3$.
The Einstein equation is
\be\label{qwa1}
R_{\mu\nu}-\frac{1}{2} g_{\mu\nu}R=8\pi GT_{\mu\nu}.
\ee
The 0th order Einstein equation is
\be\label{Friedmann}
\left(\frac{a'}{a}\right)^2=\frac{8\pi G}{3}a^2\rho^{(0)}  \, ,
  ~~~ \,\,  2\frac{a''}{a}-\left(\frac{a'}{a}\right)^2 =0,
\ee
which also imply  the  continuity equation
$ \rho^{(0)}\, '+3\frac{a'}{a} \rho^{(0)}=0$,
where the prime denotes time derivative with respect to $\tau$.
The perturbed Einstein equation is
\be\label{pertEinstein}
G^{(A)}_{\mu\nu} =8\pi GT^{(A)}_{\mu\nu}, \,\, \, A= 1,2
\ee
where we shall study up to 2nd order.
For each  order of  (\ref{pertEinstein}),
the (00) component is the energy constraint,
 $(0i)$ components are the momentum constraints,
and  $(ij)$ components contain the evolution equations.
The set of equations (\ref{pertEinstein})   determines
the dynamics of gravitational systems,
and also implies conservation of energy and momentum of matter,
$T^{(A)\mu\nu}\,_{; \, \nu}=0$.

The 1st-order metric perturbation $\gamma^{(1)}_{ij}$
can be written as
\be  \label{gqamma1}
\gamma^{(1)}_{ij}=-2\phi^{(1)}\delta_{ij}  +\chi_{ij}^{(1)}
\ee
where  $\phi^{(1)}$ is the trace part of the scalar perturbation,
and $\chi_{ij}^{(1)}$ is  traceless and can be further decomposed into
a scalar and a tensor
\be \label{xqiij1}
\chi_{ij}^{(1)} =D_{ij}\chi^{\parallel(1)}
               +\chi^{\top(1)}_{ij},
\end{equation}
where $D_{ij} \equiv  \partial_i\partial_j-\frac{1}{3}\delta_{ij}\nabla^2 $
is a traceless operator $D^i\, _i =0$,
  $\chi^{\parallel(1)}$ is a scalar function,
and  $D_{ij}\chi^{\parallel(1)}$ is the traceless part of the scalar perturbation,
and $\chi^{\top(1)}_{ij}$  is the tensor part (relic gravitational wave),
satisfying the traceless and transverse conditions:
$\chi^{\top(1)i}\, _i=0$, $\partial^i\chi^{\top(1)}_{ij}=0$.
Thus the scalar perturbations have two modes:  $\phi^{(1)}$ and $D_{ij}\chi^{\parallel(1)}$.
In this paper,
we do not consider the 1st-order vector  perturbation
 which is a residual gauge mode and can be set $0$
since  the matter is an irrotational dust.
See the paragraph below (\ref{decompC}) in Appendix C  for an explanation.
However, as shall be seen later,
the 2nd-order vector  perturbation will inevitably appear
due to    couplings  of the 1st-order perturbations.
Thus,
 the 2nd-order   perturbation    is  written as
\begin{equation} \label{2c1rcerf}
\gamma^{(2)}_{ij}=-2\phi^{(2)}\delta_{ij}+
\chi^{(2)}_{ij}
\end{equation}
with the   traceless part
\be\label{chi2decompose}
\chi^{(2)}_{ij} = D_{ij} \chi^{\parallel(2)}
+\chi^{\perp(2) }_{ij}+\chi^{\top(2)}_{ij}.
\ee
where  the vector mode satisfies a condition
\be\label{chiperpDiver}
\partial^i\partial^j  \chi^{\perp(2) }_{ij}=0,
\ee
which can be written in terms of a curl  vector
\be\label{chiVec1}
\chi^{\perp(2) }_{ij}  = 2 A_{(i, \\j)}  \equiv \partial_i A_j+\partial_j A_i,
\,\,\,\,\,    \partial^i A_i =0.
\ee
 Since the 3-vector $A_i$ is divergenceless and has only two independent components,
  the vector metric perturbation $\chi^{\perp(2) }_{ij}$
has two independent polarization modes, correspondingly.

The 1st-order   perturbations  are well known (see  Appendix A).
We   list the solutions
that will appear  in  the 2nd-order equations as   the source.
From   Eq.(\ref{deltaphi}) in Appendix A,
the solution  of the 1st-order matter density contrast is
\be\label{continuity1}
\delta^{(1)}  =  \delta_0^{(1)} + 3   (\phi^{ (1)}-\phi^{ (1)}  _0),
\ee
where  $\delta_0^{(1)}$ is the  initial value of  the  1st-order density contrast
and $\phi^{ (1)}  _0$ is the   initial value of  scalar perturbation  $\phi^{ (1)}  $.
The 1st-order perturbed Einstein equation consists of
$  G^{(1)}_{00}    = 8\pi G  a^2 \rho^{(0)}\delta^{(1)}$,
$ G^{(1)}_{0i}    =0 $,
$ G^{(1)}_{ij} =0$,
where  $T^{(1)}_{i\mu }=0 $ for the dust.
The solutions of scalar are
\be \label{phi1sol}
\phi^{(1)}({\bf x}, \tau )
=   \frac{5}{3}  \varphi({\bf x})
         + \frac{\tau^2}{18}  \nabla^2 \varphi ({\bf x})
        +\frac{ X({\bf x })}{\tau^3},
\ee
\be\label{Dchi1sol}
D_{ij} \chi^{\parallel(1)}({\bf x}, \tau )
  =  -  \frac{\tau^2}{3} \l(\varphi({\bf x})_{,ij}
  - \frac{1}{3} \delta_{ij} \nabla^2 \varphi ({\bf x})\r)
  -\frac{6\nabla^{-2} D_{ij} X({\bf x })}{\tau^3}    ,
\ee
the 1st-order density contrast is
\be
\delta^{(1)} = \frac{1}{6} \tau^2   \nabla^2 \varphi     +\frac{3 X}{\tau^3} .
\ee
In the above,  $\varphi$ is the gravitational potential
at time $\tau_0$
\be\label{delta0}
\nabla^2 \varphi = \frac{6}{ \tau_0^2} \delta_{0g}^{(1)},
\ee
determined by    the initial density of the growing mode,
and  $ X$ represents  the  decaying mode  of perturbations such that
$\frac{3 X}{\tau_0^3}$ is the initial density contrast of decaying mode.
The reason to  keep the decaying mode   is the following:
in principle, for a full treatment,
the initial condition at $\tau_0$, i.e, at the radiation-matter equality ($z\sim 3500$),
should be determined by
the connection of physical quantities such as the energy density etc,
which  generally leads to continuous connection of perturbations
and their time derivatives.
For such a connection to be made consistently,
both growing and decaying modes in solutions of perturbations
should  be kept.
This is true for the 1st order, as well as the 2nd-order perturbations, respectively.

The  solution of tensor  is
\be  \label{Fourier}
\chi^{\top(1)}_{ij}  ( {\bf x},\tau)= \frac{1}{(2\pi)^{3/2}}
   \int d^3k   e^{i \,\bf{k}\cdot\bf{x}}
    \sum_{s={+,\times}} {\mathop \epsilon
    \limits^s}_{ij}(k) ~ {\mathop h\limits^s}_k(\tau)
       , \,\,\,\, {\bf k}=k\hat{k},
\ee
with two polarization tensors satisfying
\[
{\mathop \epsilon  \limits^s}_{ij}(k) \delta_{ij}=0,\,\,\,\,
{\mathop \epsilon  \limits^s}_{ij}(k)  k_i=0,\,\,\,
{\mathop \epsilon  \limits^s}_{ij}(k) {\mathop \epsilon  \limits^{s'}}_{ij}(k)
       =2\delta_{ss'}.
\]
During the matter dominant stage the mode is given by
\be\label{GWmode}
{\mathop h\limits^s}_k(\tau ) = \frac{1}{a(\tau)}\sqrt{\frac{\pi}{2}}
   \sqrt{\frac{\tau}{2}}
     \big[\, {\mathop d\limits^s}_1(k)  H^{(1)}_{\frac{3}{2} } (k\tau )
          +{\mathop d\limits^s}_2 (k) H^{(2)}_{\frac{3}{2} } (k\tau ) \big],
\ee
where the  coefficients ${\mathop d\limits^s}_1$, ${\mathop d\limits^s}_2$
are determined by the initial condition during inflation
and by subsequent evolutions
through the reheating, radiation dominant stages
\cite{Grishchuk,zhangyang05}.
Here cosmic processes, such as neutrino  free-streaming
        \cite{Weinberg2004,MiaoZhang2007},
QCD  transition,  and $e^+e^-$ annihilation \cite{WangZhang2008}
   only slightly modify  the amplitude of RGW
and will be  neglected  in  this study.
For relic gravitational waves (RGW) generated during inflation  \cite{Grishchuk,Rubakov,Fabbri,AbbottWise1984,
    Allen1988},
the two modes ${\mathop h\limits^s}_k(\tau)$ with  $s= {+,\times}$
are usually assumed to be statistically equivalent,
the superscript $s$ can be dropped.

Put  together,
the solution of  1st-order metric  perturbation is
\begin{equation}  \label{eq28}
  \gamma^{(1)}_{ij} =-\frac{10}{3}\varphi\delta_{ ij}
  -\frac{\tau^2}{3}\varphi_{,ij}
 -\frac{6 }{\tau^3} \nabla^{-2}X_{,ij}
  +\chi^{\top(1)}_{ij},
\end{equation}
which will be used later.
We remark that the evolution equations
[see  Eqs.(\ref {init}) (\ref {ggtw}) in Appendix A]
 of scalars $\phi^{(1) }$ and $ \chi^{\parallel(1)}$
are not wave equations,
in contrast to    Eq.(\ref {eq20}) for tensor $\chi^{\top(1)}_{ij} $.
Thus,   the scalar perturbations
do not propagate at speed of light,
but just  follow  where the density perturbation is distributed.

\section{ The Second-Order Perturbed Einstein Equations}

The $(00)$ component of  the 2nd-order perturbed  Einstein equation is
\be\label{G00}
G_{00}^{(2)} \equiv R_{00}^{(2)} -\frac{1}{2}g^{(0)}_{00}R^{(2)}
         = 4\pi G a^2\rho^{(0)} \delta^{(2)}  ,
\ee
where $ \delta^{(2)}$ is the 2nd-order density contrast.
Calculations give
\bl \label{G002eq}
&
\nabla^2\phi^{(2)}
-\frac{3a'}{a} \phi^{(2)'}
+\frac{1}{6}\nabla^2 \nabla^2 \chi^{||(2)}
\nn\\
=
&
4\pi G a^2\rho^{(0)} \delta^{(2)}
+12\frac{a'}a\phi^{(1)'}\phi^{(1)}
-3\phi^{(1) '}\phi^{(1) '}
-3\phi^{(1)}_{,\,k}\phi^{(1),\,k}
-8\phi^{(1)}\nabla^2\phi^{(1)}
\nn\\
    &
-2\phi^{(1)}D^{kl}\chi^{||(1)}_{,\,kl}
+\phi^{(1)}_{,\,kl}D^{kl}\chi^{||(1)}
+\frac{1}{8} D^{kl}\chi^{||(1)'}D_{kl}\chi^{||(1)'}
+\frac{a'}a D^{kl}\chi^{||(1)}D_{kl}\chi^{||(1)'}
            \nn\\
            &
+D_{ml}\chi^{||(1),\,l}_{,\,k}D^{km}\chi^{||(1)}
-\frac{1}{2}D^{km}\chi^{||(1)}\nabla^2D_{km}\chi^{||(1)}
+\frac{1}{2}D^{km}\chi^{||(1)}_{,\,k}D_{ml}\chi^{||(1),\,l}
\nn\\
&
-\frac{3}{8}D^{km}\chi^{||(1)}_{,\,l}D_{km}\chi^{||(1),\,l}
+\frac{1}{4}D^{km}\chi^{||(1)}_{,\,l}D^l_k\chi^{||(1)}_{,\,m}
\nn\\
&
+\chi^{\top(1)kl} \phi^{(1)}_{,\,kl}
+\frac{1}{4}\chi^{\top(1)'}_{kl}D^{kl}\chi^{||(1)'}
+\frac{a'}a \chi^{\top(1)kl}D_{kl}\chi^{||(1)'}
+\frac{a'}a \chi^{\top(1)'}_{kl}D^{kl}\chi^{||(1)}
    \nn\\
    &
+\chi^{\top(1)km}D_{ml}\chi^{||(1),\,l}_{,\,k}
-\frac{1}{2}\chi^{\top(1)km}\nabla^2D_{km}\chi^{||(1)}
-\frac{1}{2}D^{km}\chi^{||(1)}\nabla^2\chi^{\top(1)}_{km}
    \nn\\
    &
-\frac{3}{4}\chi^{\top(1),\,l}_{km}D^{km}\chi^{||(1)}_{,\,l}
+\frac{1}{2}\chi^{\top(1)l}_{k,\,m}D^{km}\chi^{||(1)}_{,\,l}
    \nn
    \\
    &
+\frac{1}{8}\chi^{\top(1)'kl}\chi^{\top(1)'}_{kl}
+\frac{a'}a \chi^{\top(1)kl}\chi^{\top(1)'}_{kl}
-\frac{1}{2}\chi^{\top(1)km}\nabla^2\chi^{\top(1)}_{km}
-\frac{3}{8}\chi^{\top(1)km}_{,\,l}\chi^{\top(1),\,l}_{km}
    \nn
    \\
    &
+\frac{1}{4}\chi^{\top(1)km}_{,\,l}\chi^{\top(1)l}_{k,\,m}
\el
where the coupling terms of 1st-order perturbations are moved to the rhs of the equation,
and they serve as an effective source of 2nd-order metric perturbations
besides   $4\pi G a^2 \rho^{0} \delta^{(2)}$.
By comparison,
the structure of Eq. (\ref {G002eq}) is similar to Eq.(\ref {G001eq}) of
the 1st-order  except these coupling terms.
Using  $\delta^{(2)}$ of  (\ref{delta2nd}) in Appendix A
and Eq.(\ref{phi1sol})  and Eq. (\ref{Dchi1sol})
to express  the 1st-order perturbations
in terms of the potential $\varphi$,
Eq.(\ref {G002eq}) is written as  the 2nd-order energy constraint \cite{Matarrese98}:
\be  \label{ieq3}
\frac{2}{\tau}\phi^{(2)'}
  +   \frac{6}{\tau^2}\phi^{(2)}
   -\frac{1}{3}\nabla^2\phi^{(2)}
   -\frac{1}{12}D^{ij}\chi^{\parallel(2)}_{,ij}
  =  E_S +E_{s(t)} +E_T,
\ee
where
\bl  \label{en}
E_S \equiv &
 (\frac{100}{27}
+\frac{20\tau^2_0}{9\tau^2})\varphi\nabla^2\varphi
+\frac{25}{9}\varphi_{,\,i}\varphi^{,\,i}
+(-\frac{5\tau^2}{54}
+\frac{\tau_0^4}{9\tau^2})\varphi_{,\,ij}\varphi^{,\,ij}
+\frac{5\tau^2}{27}\varphi_{,\,i}\nabla^2\varphi^{,\,i}     \nonumber   \\
 & +(\frac{4\tau^2}{27}+\frac{\tau^4_0}{18\tau^2})(\nabla^2\varphi)^2
 -\frac{\tau^4}{216}\nabla^2\varphi^{,\,i}\nabla^2\varphi_{,\,i}
 +\frac{\tau^4}{216}\varphi^{,\,ijk}\varphi_{,\,ijk}
 -\frac{2}{\tau^2}\delta_{S\,0}^{(2)}
 +\frac{6}{\tau^2}\phi^{(2)}_{S\,0}\nn \\
&
-\frac{9}{2\tau^8}X^2
+\frac{45}{2\tau^8}\nabla^{-2}X_{,\,kl}\nabla^{-2}X^{,\,kl}
-\frac{3}{2\tau^6}X^{,\,k}X_{,\,k}
+\frac{3}{2\tau^6}\nabla^{-2}X^{,\,kml}\nabla^{-2}X_{,\,kml}
\nn\\
&
+\frac{1}{3\tau^3}X\nabla^2\varphi
+\frac{10}{3\tau^3}X^{,\,k}\varphi_{,\,k}
+\frac{5}{3\tau^3}\varphi_{,kl}\nabla^{-2}X^{,\,kl}
+\frac{36}{\tau_0^6\tau^2}\nabla^{-2}X^{,ij}\nabla^{-2}X_{,ij}
\nn\\
&
+\frac{18}{\tau_0^6\tau^2}X^2
+\frac{40}{\tau_0^3\tau^2}\varphi X
+\frac{4}{\tau_0\tau^2}\varphi_{,ij}\nabla^{-2}X^{,ij}
+\frac{2}{\tau_0\tau^2}X\nabla^2\varphi
\nn\\
&
-\frac{1}{6\tau}X_{,\,k}\nabla^2\varphi^{,\,k}
+\frac{1}{6\tau}\varphi^{,kml}\nabla^{-2}X_{,\,kml} ,
\el
\bl \label{EST}
E_{s(t)} \equiv  & \frac{5\tau}{18}\chi^{\top(1)' ij}\varphi_{,\,ij}
+\frac{5}{9}\chi^{\top(1) ij}\varphi_{,\,ij}
-\frac{\tau^2}{18}\varphi_{,\,ij}\nabla^2\chi^{\top(1) ij}
-\frac{\tau^2}{36}\chi^{\top(1) ij,\,k}\varphi_{,\,ijk}
 \nn \\
& -\frac{2\tau_0^2}{3\tau^2}\varphi^{,\,ij}\chi^{\top(1)}_{0ij}
 -\frac{2}{\tau^2}\delta_{s(t)\,0}^{(2)}
 +\frac{6}{\tau^2}\phi^{(2)}_{s(t)\,0} \nn \\
& +\frac{5}{2\tau^4}\chi^{\top(1)'}_{kl}\nabla^{-2}X^{,kl}
-\frac{1}{\tau^3}\nabla^{-2}X^{,kl}\nabla^2\chi^{\top(1)}_{kl}
\nn\\
&
-\frac{1}{2\tau^3}\chi^{\top(1)}_{km,\,l}\nabla^{-2}X^{,klm}
-\frac{12}{\tau_0^3\tau^2}\chi^{\top(1)}_{0kl}\nabla^{-2}X^{,kl} ,
\el
\bl \label{ET}
E_T \equiv  &-\frac{1}{24}\chi^{\top(1) ' ij}\chi^{\top(1) '}_{ij}
-\frac{2}{3\tau}\chi^{\top(1) ' ij}\chi^{\top(1)}_{ij}
+\frac{1}{6}\chi^{\top(1) ij}\nabla^2\chi^{\top(1)}_{ij}
+\frac{1}{8}\chi^{\top(1) ij,k}\chi^{\top(1)}_{ij,k}
\nonumber  \\
 &-\frac{1}{12}\chi^{\top(1)ij,k}\chi^{\top(1)}_{kj,i}
 -\frac{1}{\tau^2}\chi^{\top(1) ij}\chi^{\top(1)}_{ij}
+\frac{1}{\tau^2}\chi^{\top(1) ij}_0 \chi^{\top(1)}_{0ij}
 -\frac{2}{\tau^2}\delta_{T\,0}^{(2)}
 +\frac{6}{\tau^2}\phi^{(2)}_{T\,0},
\el
are   the coupling terms  of the 1st-order perturbations.
All $E_S$, $E_{{s(t)} } $ and  $E_{{T} } $
contain   the  initial values $\delta^{(2)}_0$, $\phi^{(2)}_0$,
        $ \chi^{\top(1)}_{0ij}$ etc  at $\tau_0$.
The subscript $``S\,"$ denotes those contributed by
scalar-scalar couplings
$\varphi\varphi$ and $X\varphi$ and $XX$,
  $``s(t)"$  by scalar-tensor  $\varphi\chi^{\top(1)}_{ij}$ and $X\chi^{\top(1)}_{ij}$,
and   $``T\,"$   by  tensor-tensor $\chi^{\top(1)}_{ij}\chi^{\top(1)}_{kl}$.
We   notice that neither the  tensor    $\chi^{\top (2)}_{ij}$
nor  the vector    $\chi^{\perp (2)}_{ij}$
appears in the energy constraint (\ref{ieq3}).

The $(0i)$ component of the 2nd-order perturbed Einstein equation is
\be\label{Einstein0i}
 G_{0i}^{(2)} =  R_{0i}^{(2)}=0,
\ee
with $T^{(2)}_{0i}=0$.
Using $R^{(2) }_{0i}$ in Eq.(\ref{cnm13}) in Appendix B,
Eq.(\ref{Einstein0i}) leads to
the 2nd-order momentum constraint  \cite{Matarrese98}:
\be  \label{mconstr}
 2\phi^{(2)'}_{,\,j}
+ \frac{1}{2}   D_{ij} \chi^{||(2) \, ',\,i}
+\frac{1}{2}\chi^{\perp(2)',\,i}_{ij}
 = M_{Sj} + M_{s(t)j}+M_{Tj},
\ee
where
\bl    \label{MSj}
M_{Sj} \equiv&
-\frac{10\tau}{9}\varphi_{,\,j}\nabla^2\varphi
+\frac{\tau^3}{9}\varphi_{,\,kj}\nabla^2\varphi^{,\,k}
-\frac{10\tau}{9}\varphi^{,\,k}\varphi_{,\,kj}
-\frac{\tau^3}{9}\varphi^{,\,ik}
\varphi_{,\,ijk} \nn \\
&
+\frac{54}{\tau^7}\nabla^{-2}X^{,kl}\nabla^{-2}X_{,klj}
-\frac{54}{\tau^7}X^{,k}\nabla^{-2}X_{,jk}
+\frac{30}{\tau^4}\varphi_{,j}X
+\frac{30}{\tau^4}\varphi^{ ,\,k }\nabla^{-2}X_{,jk}
\nn\\
&
-\frac{3}{\tau^2}\nabla^{-2}X_{,jk}\nabla^2\varphi^{,k}
+\frac{2}{\tau^2}X^{,k}\varphi_{,jk}
+\frac{3}{\tau^2}\varphi_{,klj}\nabla^{-2}X^{,kl}
-\frac{2}{\tau^2}\varphi^{,kl}\nabla^{-2}X_{,klj}
\\
M_{s(t)j}  \equiv&  \frac{\tau^2}{3}\Big[\varphi_{,\,ik}(\chi^{\top(1)'ik}_{,\,j}
-\chi^{\top(1)'k,\,i}_{j})+\frac{1}{2}\chi^{\top(1)'ik}\varphi_{,\,ijk}
- \frac{1}{2}\chi^{\top(1)'}_{k j}\nabla^2\varphi^{,\,k}\Big]\nn\\
& +\frac{\tau}{3}\varphi_{,\,ik}\chi^{\top(1) ik}_{,\,j}
+\frac{5}{3}\varphi^{,\,k}\chi^{\top(1)'}_{k j} \nn \\
& -\frac{9}{\tau^4}\chi^{\top(1)}_{kl,\,j}\nabla^{-2}X^{,kl}
-\frac{6}{\tau^3}\chi^{\top(1)'}_{jk,\,l}\nabla^{-2}X^{,kl}
+\frac{6}{\tau^3}\chi^{\top(1)' }_{kl,  \, j}\nabla^{-2}X^{,kl}
   \nn\\
&
+\frac{3}{\tau^3}\chi^{\top(1)'}_{kl }\nabla^{-2}X^{,kl}_{,\,j}
 -\frac{3}{\tau^3}\chi^{\top (1)' }_{kj}X^{,k}
    \label{MsTj}
\\
M_{Tj} \equiv&\chi^{\top(1) ik}(\chi^{\top(1)'}_{kj,\,i}
-\chi^{\top(1)'}_{ki,\,j})-\frac{1}{2}\chi^{\top(1) ik}_{,\,j} \chi^{\top(1)'}_{ik} ,
   \label{MTj}
\el
are the couplings of the 1st-order perturbations,
formed from  various products
of  the  gravitational potential  $\varphi$  and tensor $ \chi^{\top(1)}_{ij} $.
Eq.(\ref{mconstr}) is   similar  to Eq.(\ref{momentconstr1}) for the 1st order,
except for the vector  on the lhs and  the coupling terms on the rhs.
Notice that the tensor   $\chi^{\top (2)}_{ij}$
does not appear in   (\ref{mconstr}).

The $(ij)$ component  of the 2nd-order perturbed Einstein equation is
\be\label{cnm33}
G_{ij}^{(2)} \equiv R^{(2)}_{ij}  -\frac{1}{2} \delta_{ij}a^2 R^{(2)}
-\frac{1}{4}a^2\gamma^{(2)}_{ij}  R^{(0)}-\frac{1}{2}a^2\gamma^{(1)}_{ij}  R^{(1)}
       =0.
\ee
To compare with  the  equation (4.30) in Ref.~\cite{Matarrese98},
one can  combine  Eq.(\ref {cnm33})   into the  the following
\be\label{EinToEvo2metric2}
2G_{ij}^{(2)}-\delta_{ij}G^{(2)}_{kl}\delta^{kl }  =0  .
\ee
Plugging   Eq.(\ref{EinsteinTensor2f}) and  Eq.(\ref{EinsteinTensorTrace})
and the 1st-order solutions  (\ref{phi1sol}) and (\ref{Dchi1sol})
into above,
and using Eq.(\ref{chiVec1})
gives   the   2nd-order  evolution equation  \cite{Matarrese98}:
\ba\label{eq34}
&& -(\phi^{(2)''}
+\frac{4}{\tau}\phi^{(2)'})\delta_{ij} +\phi^{(2)}_{,i j}
+\frac{1}{2} ( D_{ij}\chi^{||(2)''} +\frac{4}{\tau}D_{ij}\chi^{||(2)'})
+\frac{1}{2}(\chi^{\perp (2)''}_{ij} +\frac{4}{\tau}\chi^{\perp (2)'}_{ij})  \nn\\
&&
+\frac{1}{2}(\chi^{\top (2)''}_{ij}
+\frac{4}{\tau}\chi^{\top (2)'}_{ij}
-\nabla^2\chi^{\top(2)}_{i j}
)
 -\frac{1}{4}D_{kl}\chi^{||(2),\,kl}\delta_{ij}
+\frac{2}{3}\nabla^2\chi^{||(2)}_{,\,ij }
-\frac{1}{2}\nabla^2D_{ij}\chi^{||(2)}
\nonumber \\
&&
=S_{S\,ij}+S_{{s(t)}ij}+S_{T\,ij}
      ,
\ea
where
\bl \label{Ssij}
S_{Sij} \equiv  & -\frac{100}{9}\varphi\varphi_{,ij}
+\frac{25}{9}\varphi^{,k}\varphi_{,k}\delta_{ij}
-\frac{50}{3}\varphi_{,i}\varphi_{,j}
+\frac{24\tau^2}{9}\varphi^{,k}_{,i}\varphi_{,kj}
\nn \\
& +\frac{11\tau^2}{18}(\nabla^2\varphi)^2\delta_{ij}
 -\frac{22\tau^2}{9}\varphi_{,ij}\nabla^2\varphi
 -\frac{11\tau^2}{18}\varphi^{,kl}\varphi_{,kl}\delta_{ij}
 -\frac{5\tau^2}{9}\varphi_{,k}\varphi^{,k}_{,ij}
  \nn \\
& +\frac{\tau^4}{18}\varphi^{,k}_{,ij}\nabla^2\varphi_{,k}
-\frac{\tau^4}{72}\nabla^2\varphi^{,m}\nabla^2\varphi_{,m}\delta_{ij}
-\frac{\tau^4}{18}\varphi^{,kl}_{,j}\varphi_{,ikl}
+\frac{\tau^4}{72}\varphi^{,kl}_{,m}\varphi^{,m}_{,kl}\delta_{ij} \nn\\
& +
\frac{81}{2\tau^8} X^2\delta_{ij}
-\frac{81}{2\tau^8}\nabla^{-2}X_{,kl}\nabla^{-2}X^{,kl}\delta_{ij}
-\frac{162}{\tau^8}X\nabla^{-2}X_{,ij}
\nn\\
&
+\frac{324}{\tau^8}\nabla^{-2}X^{,k}_{,i}\nabla^{-2}X_{,kj}
-\frac{9}{2\tau^6}X^{,k}X_{,k}\delta_{ij}
+\frac{9}{2\tau^6}\nabla^{-2}X^{,kml}\nabla^{-2}X_{,kml}\delta_{ij}
\nn\\
&
+\frac{18}{\tau^6}X^{,k}\nabla^{-2}X_{,kij}
-\frac{18}{\tau^6}\nabla^{-2}X^{,kl}_{,i}\nabla^{-2}X_{,klj}
+\frac{7}{\tau^3}X\nabla^2\varphi\delta_{ij}
-\frac{7}{\tau^3}\varphi^{,kl}\nabla^{-2}X_{,kl}\delta_{ij}
\nn\\
&
-\frac{14}{\tau^3}\varphi_{,ij}X
-\frac{14}{\tau^3}\nabla^2\varphi\nabla^{-2}X_{,ij}
-\frac{10}{\tau^3}\varphi^{,k}\nabla^{-2}X_{,kij}
+\frac{8}{\tau^3}\varphi^{,k}_{,i}\nabla^{-2}X_{,kj}
\nn\\
&
+\frac{8}{\tau^3}\varphi_{,j}^{,k}\nabla^{-2}X_{,ki}
-\frac{1}{2\tau}X_{,k}\nabla^2\varphi^{,k}\delta_{ij}
+\frac{1}{2\tau}\varphi_{,klm}\nabla^{-2}X^{,klm}\delta_{ij}
\nn\\
&
+\frac{1}{\tau}X^{,k}\varphi_{,kij}
+\frac{1}{\tau}\nabla^{-2}X_{,kij}\nabla^2\varphi^{,k}
-\frac{1}{\tau}\varphi_{,klj}\nabla^{-2}X^{,kl}_{,i}
-\frac{1}{\tau}\varphi_{,kli}\nabla^{-2}X_{,j}^{,kl}
\el
\bl\label{Sstij}
S_{s(t)ij}
\equiv&
-\frac{\tau^2}{6} \varphi^{,kl}\chi^{\top(1)''}_{kl}\delta_{ij}
-\frac{2\tau}{3} \varphi^{,k}_{,i}\chi^{\top(1)'}_{kj}
-\frac{2\tau}{3} \varphi_{,j}^{,k}\chi^{\top(1)'}_{k i}
+\frac{\tau}{3} \chi^{\top(1)'}_{ij} \nabla^2 \varphi
-\frac{\tau}{2} \varphi^{,kl}\chi^{\top(1)'}_{kl}\delta_{ij}
\nn\\
&
+\frac{10}{3}\chi^{\top(1)}_{ij}  \nabla^2 \varphi
+\frac{5}{3}\varphi_{,kl}\chi^{\top(1)kl}\delta_{ij}
-\frac{10}{3}\varphi_{,j}^{,k}\chi^{\top(1)}_{ki}
-\frac{10}{3}\varphi_{,i}^{,k}\chi^{\top(1)}_{kj}
+\frac{10}{3}  \varphi\nabla^2\chi^{\top(1)}_{ij}
\nn\\
&
+\frac{\tau^2}{3} \varphi^{,kl}\chi^{\top(1)}_{ij,\,kl}
+\frac{\tau^2}{3} \varphi^{,kl}\chi^{\top(1)}_{kl,\,ij}
-\frac{\tau^2}{3} \varphi^{,kl}\chi^{\top(1)}_{li,\,jk}
-\frac{\tau^2}{3} \varphi^{,kl}\chi^{\top(1)}_{lj,\,ik}
+5  \varphi^{,\,k}\chi^{\top(1)}_{ij,\,k}
\nn
\\
&
-\frac{5}{3}  \varphi^{,\,k}\chi^{\top(1)}_{ki,j}
-\frac{5}{3}  \varphi^{,\,k}\chi^{\top(1)}_{kj,\,i}
+\frac{\tau^2}{6}  \chi^{\top(1)}_{ij,\,k}\nabla^2 \varphi^{,\,k}
-\frac{\tau^2}{6}\chi^{\top(1)}_{ki,\,j} \nabla^2\varphi^{,k}
- \frac{\tau^2}{6}  \chi^{\top(1)}_{kj,\,i}\nabla^2 \varphi^{,\,k}
\nn\\
&
+\frac{\tau^2}{6} \varphi^{,kl}_{,\,i}\chi^{\top(1)}_{kl,\,j}
+\frac{\tau^2}{6} \varphi^{,kl}_{,j}\chi^{\top(1)}_{kl,\,i}
-\frac{\tau^2}{12} \varphi^{,kml}\chi^{\top(1)}_{km,\,l}\delta_{ij} \nn\\
&
-\frac{33}{2\tau^4}\chi^{\top(1)'}_{kl}\nabla^{-2}X^{,kl}\delta_{ij}
-\frac{3}{2\tau^3}\chi^{\top(1)}_{kl,\,m}\nabla^{-2}X^{,klm}\delta_{ij}
-\frac{3}{\tau^3}\chi^{\top(1)''}_{kl}\nabla^{-2}X^{,kl}\delta_{ij}
\nn\\
&
-\frac{9}{\tau^4}X\chi^{\top(1)'}_{ij}
+\frac{18}{\tau^4}\chi^{\top(1)'}_{k i}\nabla^{-2}X^{,k}_{,j}
+\frac{18}{\tau^4}\chi^{\top(1)'}_{kj}\nabla^{-2}X^{,k}_{,i}
\nn\\
&
-\frac{3}{\tau^3}X^{,\,k}\chi^{\top(1)}_{k j,\,i}
-\frac{3}{\tau^3}X^{,\,k}\chi^{\top(1)}_{k i,\,j}
+\frac{3}{\tau^3}X^{,\,k}\chi^{\top(1)}_{ij,\,k}
\nn\\
&
-\frac{6}{\tau^3}\chi^{\top(1)}_{lj,\,ik}\nabla^{-2}X^{,kl}
-\frac{6}{\tau^3}\chi^{\top(1)}_{li,\,jk}\nabla^{-2}X^{,kl}
+\frac{6}{\tau^3}\chi^{\top(1)}_{ij,\,kl}\nabla^{-2}X^{,kl}
\nn\\
&
+\frac{6}{\tau^3}\chi^{\top(1)}_{kl,\,ij}\nabla^{-2}X^{,kl}
+\frac{3}{\tau^3}\chi^{\top(1)}_{kl,\,i}\nabla^{-2}X^{,kl}_{,j}
+\frac{3}{\tau^3}\chi^{\top(1)}_{kl,\,j}\nabla^{-2}X^{,kl}_{,\,i}
\el
\bl \label{Sstij1}
S_{Tij} \equiv
&\chi^{\top(1)'k}_{i}\chi^{\top(1)'}_{kj}
-\frac18\chi^{\top(1)'kl}\chi^{\top(1)'}_{kl}\delta_{ij}
+\chi^{\top(1)kl}\chi^{\top(1)}_{li,\,jk}
+\chi^{\top(1)kl}\chi^{\top(1)}_{lj,\,ik}
\nn\\
&
-\chi^{\top(1)kl}\chi^{\top(1)}_{kl,\,ij}
-\chi^{\top(1)kl}\chi^{\top(1)}_{ij,\,kl}
+\chi^{\top(1),\,k}_{l i}\chi^{\top(1),\,l}_{k j}
-\chi^{\top(1)k}_{i,\,l}\chi^{\top(1),\,l}_{j k}
\nn\\
&
-\frac12\chi^{\top(1)kl}_{,\,i}\chi^{\top(1)}_{kl,\,j}
+\frac38\chi^{\top(1)kl,\,m}\chi^{\top(1)}_{kl,\,m}\delta_{ij}
-\frac14\chi^{\top(1)}_{ml,\,k}\chi^{\top(1)m k,\,l}\delta_{ij}
\nn
\\
&
+\frac12\chi^{\top(1)kl}\chi^{\top(1)''}_{kl}\delta_{ij}
+\frac{2}{\tau}\chi^{\top(1)kl}\chi^{\top(1)'}_{kl}\delta_{ij},
\el
are the couplings
and play a role of source of evolution.
Note that  the  2nd-order perturbations,
scalar $\phi^{(2)}$, $\chi^{||(2) } $,
vector $\chi^{\perp (2)}_{ij} $,
and tensor $\chi^{\top(1)}_{ij}$,
all  appear in the evolution equation (\ref{eq34}).

There are three types of couplings:
  scalar-scalar, scalar-tensor,
  and  tensor-tensor
  in  (\ref{ieq3}), (\ref{mconstr}), and (\ref{eq34}).
In order to deal with these equations separately,
we  split the 2nd-order perturbations  into three parts
 according  to the type of couplings
\be\label{avd}
\phi^{(2)}\equiv \phi^{(2)}_S+\phi^{(2)}_{s(t)}+\phi^{(2)}_T,
\ee
\be\label{qw34achift}
D_{ij}\chi^{||(2) }\equiv D_{ij}\chi^{||(2) }_S
         +D_{ij}\chi^{||(2) }_{s(t)}+D_{ij}\chi^{||(2) }_{T},
\ee
\be\label{chiPerp}
\chi^{\perp(2)}_{ij}=\chi^{\perp(2)}_{S ij}
+\chi^{\perp(2)}_{s(t)ij}+\chi^{\perp(2)}_{T ij},
\ee
\be\label{achift}
\chi^{\top(2)}_{ij}=\chi^{\top(2)}_{S ij}
+\chi^{\top(2)}_{s(t)ij}+\chi^{\top(2)}_{T ij},
\ee
and   the equations (\ref{ieq3}), (\ref{mconstr}), and (\ref{eq34})
  into  three sets as follows.
The first set involves only  the   scalar-scalar coupling:
 \be \label{ens}
\frac{2}{\tau}\phi^{(2)'}_S-
\frac{1}{3}\nabla^2\phi^{(2)}_S+\frac{6}{\tau^2}\phi^{(2)}_S
    -\frac{1}{12}D^{ij}\chi^{\parallel(2) }_{S,ij}
          =E_S,
\ee
\be  \label{mos}
2\phi^{(2)'}_{S,j} +\frac{1}{2}   D_{ij}\chi^{\parallel(2)\, ',\,i}_{S }
+\frac{1}{2} \chi^{\perp(2)',\,i}_{S\,ij} =M_{Sj},
\ee
\ba\label{eq34s}
&& -(\phi^{(2)''}_S
+\frac{4}{\tau}\phi^{(2)'}_S)\delta_{ij} +\phi^{(2)}_{S,i j}
+\frac{1}{2} ( D_{ij}\chi^{||(2)''} _S
+\frac{4}{\tau}D_{ij}\chi^{||(2)'}_S) \nn\\
&&
+\frac{1}{2}(\chi^{\perp (2)''}_{S\,ij}
+\frac{4}{\tau}\chi^{\perp (2)'}_{S\,ij})+\frac{1}{2}(\chi^{\top (2)''}_{S\,ij}
+\frac{4}{\tau}\chi^{\top (2)'}_{S\,ij}
-\nabla^2\chi^{\top(2)}_{S\,i j}
)
\nonumber \\
&&
 -\frac{1}{4}D_{kl}\chi^{||(2),\,kl}_S\delta_{ij}
+\frac{2}{3}\nabla^2\chi^{||(2)}_{S,\,ij }
-\frac{1}{2}\nabla^2D_{ij}\chi^{||(2)}_S
=S_{S\,ij}
    \  .
\ea
Observe that $M_{Sj}$ on the rhs of  (\ref{mos}) has a  nonvanishing curl,
$\varepsilon^{ikj}\partial_k M_{Sj} \ne 0$
with $\varepsilon^{ikj}$ as the Levi-Civita symbol,
commanding the introduction of the 2nd-order vector
$ \chi^{\perp(2) }_{S\,ij} $
 on the lhs.

The second set involves the scalar-tensor coupling:
\be  \label{ieq3asdsd1q}
\frac{2}{\tau}\phi^{(2)'}_{s(t)}
-\frac{1}{3}\nabla^2\phi^{(2)}_{s(t)}
+\frac{6}{\tau^2}\phi^{(2)}_{s(t)}
  -\frac{1}{12} D^{ij}\chi^{\parallel(2) }_{{s(t)},ij}= E_{s(t)} \, ,
\ee
\be  \label{cfta1q}
2\phi^{(2)'}_{s(t),j}
     +\frac{1}{2} D_{ij}\chi^{\parallel(2)\, ',\,i}_{s(t)}
+\frac12\chi^{\perp (2)',\,i}_{{s(t)}i j}
               =M_{s(t)j}
\ee
\ba\label{afk1q}
&& -(\phi^{(2)''}_{s(t)}
+\frac{4}{\tau}\phi^{(2)'}_{s(t)})\delta_{ij} +\phi^{(2)}_{s(t),i j}
+\frac{1}{2} ( D_{ij}\chi^{||(2)''} _{s(t)}
+\frac{4}{\tau}D_{ij}\chi^{||(2)'}_{s(t)})
 \nn\\
&&
+\frac{1}{2}(\chi^{\perp (2)''}_{s(t)\,ij}
+\frac{4}{\tau}\chi^{\perp (2)'}_{s(t)\,ij})
+\frac{1}{2}(\chi^{\top (2)''}_{s(t)\,ij}
+\frac{4}{\tau}\chi^{\top (2)'}_{s(t)\,ij}
-\nabla^2\chi^{\top(2)}_{s(t)\,i j}
)
\nonumber \\
&&
 -\frac{1}{4}D_{kl}\chi^{||(2),\,kl}_{s(t)}\delta_{ij}
+\frac{2}{3}\nabla^2\chi^{||(2)}_{s(t),\,ij }
-\frac{1}{2}\nabla^2D_{ij}\chi^{||(2)}_{s(t)}
=S_{s(t)\,ij}
    \  .
\ea
The third set involves  the  tensor-tensor coupling:
\be  \label{ens26}
\frac{2}{\tau}\phi^{(2)'}_{T}
-\frac{1}{3}\nabla^2\phi^{(2)}_{T}
   +\frac{6}{\tau^2}\phi^{(2)}_{T}
    -\frac{1}{12}  D^{ij} \chi^{\parallel(2) }_{{T},ij}= E_{T},
\ee
\be  \label{mos26}
2\phi^{(2)'}_{T,j}
     +\frac{1}{2} D_{ij}\chi^{\parallel(2)\, ',\,i}_{\,T }
+\frac12\chi^{\perp (2)\,',\,i}_{{T}i j}
               =M_{Tj}
\ee
\ba\label{eq34s26}
&& -(\phi^{(2)''}_{T}
+\frac{4}{\tau}\phi^{(2)'}_{T})\delta_{ij} +\phi^{(2)}_{T,i j}
+\frac{1}{2} ( D_{ij}\chi^{||(2)''} _{T}
+\frac{4}{\tau}D_{ij}\chi^{||(2)'}_{T})
 \nn\\
&&
+\frac{1}{2}(\chi^{\perp (2)''}_{T\,ij}
+\frac{4}{\tau}\chi^{\perp (2)'}_{T\,ij})
+\frac{1}{2}(\chi^{\top (2)''}_{T\,ij}
+\frac{4}{\tau}\chi^{\top (2)'}_{T\,ij}
-\nabla^2\chi^{\top(2)}_{T\,i j}
)
\nonumber \\
&&
 -\frac{1}{4}D_{kl}\chi^{||(2),\,kl}_{T}\delta_{ij}
+\frac{2}{3}\nabla^2\chi^{||(2)}_{T,\,ij }
-\frac{1}{2}\nabla^2D_{ij}\chi^{||(2)}_{T}
=S_{T\,ij}
    \  .
\ea
The lhs of these equations involve  only the 2nd-order metric perturbations
with a similar structure  to those of the 1st-order equations,
but the difference is the   couplings on the rhs.

During the matter era, the 1st-order tensor has a lower amplitude
than the 1st-order scalar.
As  the 1st-order solutions (\ref{phi1sol}) (\ref{Dchi1sol})  (\ref{GWmode}) show,
the scalar  grows as $\phi ^{(1)},  \chi^{\parallel(1)} \propto a(\tau)$,
whereas  the  short wavelengths modes  of tensor
decrease as $h_k(\tau ) \propto 1/a(\tau)$,
and those of long wavelength remain
constant $h_k(\tau ) \propto const$ \cite{Grishchuk,zhangyang05}.
Given  the upper bound of tensor-scalar ratio $r<0.1$ \cite{WMAP9Bennett,Planck2014},
the amplitude of scalar is growing dominant over that of tensor.
This leads to an estimation that, among the three types of couplings,
the scalar-scalar is greater in magnitude
than the  scalar-tensor and tensor-tensor  for the matter stage.
In the following,
we shall focus on the set  of equations of scalar-scalar  coupling.
As for the scalar-tensor and tensor-tensor couplings,
they are more involved and  will be reported in a subsequent paper separately in the future.

\section{ 2nd-Order Perturbations  with the scalar-scalar coupling Source }

\subsection{Scalar perturbation $ \phi^{(2)}_S$}

We shall solve the set of equations (\ref {ens}),  (\ref {mos}), and (\ref {eq34s}),
which  have a structure similar to that of the 1st-order equations
[(\ref {momentconstr1}) (\ref {G001eq}) (\ref {cnm31}) in Appendix A]
except for the  coupling terms on the rhs.
Applying $ \partial^j \int_{\tau_0} ^\tau d\tau' $  on Eq.(\ref{mos})  gives
\be\label{mosInt1}
2\nabla^2\phi^{(2)}_{S}
 +\frac{1}{2}   D_{ij}\chi^{\parallel(2),\, ij }_{S}
 =\int_{\tau_0}^\tau d\tau' M_{Sj}^{,j}
+2\nabla^2\phi^{(2)}_{S\,0}
+\frac{1}{2}   D_{ij}\chi^{\parallel(2),\, ij }_{S\,0}
\ee
with   $ \chi^{\parallel(2)\, }_{S\,0}$ being  the value at $\tau_0$.
A combination
  [Eq.(\ref{ens}) + $\frac{1}{6}\partial^j \int_{\tau_0} ^\tau d\tau'$  Eq.(\ref{mos})],
yields the first-order differential equation of $ \phi^{(2)}_S$
as  follows:
\bl \label{phi'eq}
 \phi^{(2)\,'}_S +\frac{3}{\tau} \phi^{(2)}_S
         =&
 \frac{\tau}{2}  E_S
   - \frac{5\tau^3}{108}( \nabla^2\varphi_{} \nabla^2\varphi
   + 2\varphi^{,i}\nabla^2\varphi_{,i} + \varphi^{,ki}\varphi_{,ki} ) \nn \\
&  +\frac{\tau^5}{432} ( \nabla^2\varphi^{,k}\nabla^2\varphi_{,k}
           - \varphi^{,ijk} \varphi_{,ijk})
     -\frac{\tau}{12}  F \nn \\
&
+\frac{3}{4\tau^5}(X^{,k}X_{,k}
-\nabla^{-2}X^{,klm}\nabla^{-2}X_{,klm})
\nn\\
&
-\frac{5}{6\tau^2}(2\varphi_{,k}X^{,k}
+X\nabla^2\varphi
+\varphi^{ ,\,kl }\nabla^{-2}X_{,kl})
\nn\\
&
+\frac{1}{12}(X_{,k}\nabla^2\varphi^{,k}-\varphi_{,klm}\nabla^{-2}X^{,klm}),
\el
where
\bl\label{D}
F \equiv
&-2\nabla^2\phi^{(2)}_{S\,0}
-\frac{1}{3}\nabla^2\nabla^2\chi^{\parallel(2)\, }_{S\,0}
   \nn\\
&-\frac{5\tau_0^2}{9}
( \nabla^2\varphi_{} \nabla^2\varphi + 2\varphi^{,i}\nabla^2\varphi_{,i}
+ \varphi^{,ki}\varphi_{,ki} )
+\frac{\tau_0^4}{36}
( \nabla^2\varphi^{,k}\nabla^2\varphi_{,k} - \varphi^{,ijk} \varphi_{,ijk} )
  \nn \\
& +\frac{9}{\tau_0^6} (X^{,k}X_{,k} -\nabla^{-2}X^{,klm}\nabla^{-2}X_{,klm})
   -\frac{10}{\tau_0^3}
    (2\varphi_{,k}X^{,k}+X\nabla^2\varphi+\varphi^{ ,\,kl }\nabla^{-2}X_{,kl})
    \nn \\
&+\frac{1}{\tau_0} ( X_{,k}\nabla^2\varphi^{,k} -\varphi_{,klm}\nabla^{-2}X^{,klm}) \ ,
\el
depending on the initial values
 $\phi^{(2)}_{S0}$,   $\chi^{||(2)}_{S0}$  at $\tau_0$.
The solution of Eq.(\ref{phi'eq}) is
\bl \label{phi2S}
\phi^{(2)}_S
=& \frac{1}{7}\tau^4
\l(\frac{1}{36} \nabla^2\varphi_{} \nabla^2\varphi
- \frac{5}{54} \varphi_{,ki}\varphi^{,ki}\r)
+\tau^2 \l(
\frac{10}{27}\varphi\nabla^2\varphi
+\frac{5}{18}\varphi_{,i}\varphi^{,i} \r)
- \frac{\tau^2}{60} F \nn \\
&  -\frac{3}{4\tau^6} (5\nabla^{-2}X_{,\,kl}\nabla^{-2}X^{,\,kl}- X^2)
+\frac{Z({\bf x})}{\tau^3}-\frac{1}{3\tau}X\nabla^2\varphi
\nn\\
&
+\Big(
\phi^{(2)}_{S\,0}
-\frac{1}{3}\delta_{S\,0}^{(2)}
+\frac{6}{\tau_0^6}\nabla^{-2}X^{,kl}\nabla^{-2}X_{,kl}
+\frac{3}{\tau_0^6}X^2
+\frac{20}{3\tau_0^3}\varphi X
+\frac{2}{3\tau_0}\varphi_{,ij}\nabla^{-2}X^{,ij}
\nn\\
&
+\frac{1}{3\tau_0}X\nabla^2\varphi
+\frac{10\tau_0^2}{27}\varphi\nabla^2\varphi
+\frac{\tau_0^4}{54}\varphi_{,ij}\varphi^{,ij}
+\frac{\tau_0^4}{108}(\nabla^2\varphi)^2
\Big)   \ ,
\el
where
\bl\label{Zfixed}
Z
=&
 \frac{\tau_0^3}{3}\delta_{S\,0}^{(2)}
+ \frac{\tau_0^5}{60} F
-\tau_0^5
\l(\frac{20}{27}\varphi\nabla^2\varphi
+\frac{5}{18}\varphi_{,i}\varphi^{,i} \r)
-\frac{\tau_0^7}{7}
\l( \frac{5}{54} \nabla^2\varphi_{} \nabla^2\varphi
+\frac{1}{27} \varphi_{,ki}\varphi^{,ki} \r) \nn \\
&+\frac{3}{4\tau_0^3}(-3\nabla^{-2}X_{,\,kl}\nabla^{-2}X^{,\,kl}- 5X^2)
-\frac{20}{3 }\varphi X -\frac{2\tau_0^2}{3}\varphi_{,ij}\nabla^{-2}X^{,ij}
\ ,
\el
in which $\phi^{(2)}_{S0}$, $\chi^{||(2)}_{S0}$, $\delta^{(2)}_{S0}$
need to be fixed by joining  conditions with the Radiation Dominated stage.
In particular,
the 2nd-order density contrast $\delta^{(2)}_{S0}$ is generally nonvanishing
and has been inherited from the previous expansion stages.
In the above all the $X$ terms are contributed by
the 1st-order decaying modes.

 Note that,
the solution (\ref{phi2S}) can be also derived in another way,
using
the trace  part of the evolution equation  (\ref{eq34s})
and the energy constraint (\ref{ens}),
i.e,
the trace of Eq.(\ref{eq34s}) plus three times of Eq.(\ref{ens})
 is the Raychaudhuri equation of $\phi^{(2)}$,
and the  solution is the same as  (\ref{phi2S}).

The expression  (\ref{phi2S}) extends (4.31) of Ref.~\cite{Matarrese98}
to general initial conditions at $\tau_0$.
For a realistic cosmological model,
the initial metric perturbations
 at the radiation-matter equality
are important
and determine the spectra of  CMB abiotrophies and polarization \cite{ZhaoZhang2006}.

\subsection{Scalar perturbation $D_{ij}\chi^{||(2)}_S$}

Substituting  $M_{Sj}$ of   (\ref{MSj}) and  $\phi^{(2)}_S$ of  (\ref{phi2S})
into Eq.(\ref{mosInt1}), one obtains the scalar
\bl\label{chi||2S}
 \chi^{\parallel(2)}_{S}
=&  \frac{\tau^4}{84}\nabla^{-2}\Big[
\frac{20}{3}  \varphi^{,kl}\varphi_{,\,kl}
        -2 \nabla^2\varphi \nabla^2\varphi
+\nabla^{-2}\l(7 \nabla^2\varphi^{,k}\nabla^2\varphi_{,\,k}
-7  \varphi^{,\,klm}\varphi_{,\,klm}\r)
\Big] \nn \\
&  -\frac{5\tau^2}{18}\Big[
4\varphi \varphi
+4\nabla^{-2} \l(\varphi_{,\,k} \varphi^{,\,k}\r)
      +\nabla^{-2} \nabla^{-2} \l(6  \nabla^2 \varphi \nabla^2 \varphi
 -6 \varphi_{,\,kl}  \varphi^{,\,kl}\r)\Big]
\nonumber \\
&+ \nabla^{-2}\nabla^{-2} A
  + \frac{\tau^2 }{10} \nabla^{-2}F
   \nn\\
 & + \frac{9}{2\tau^6}
\Big[\nabla^{-2}\l(5\nabla^{-2}X_{,\,kl}\nabla^{-2}X^{,\,kl}- X^2\r)
\nn\\
&
+\nabla^{-2}\nabla^{-2}\l(6X^{,k}X_{,k}-6\nabla^{-2}X^{,klm}\nabla^{-2}X_{,klm}
      \r)\Big]
\nn\\
& +\frac{6}{\tau^3}
\Big[-\nabla^{-2}\l(Z+5\varphi X\r)+\nabla^{-2}\nabla^{-2}
\l(5\varphi\nabla^2X-5\varphi^{ ,\,kl }\nabla^{-2}X_{,kl}\r)\Big]
\nn\\
&
+\frac{1}{\tau}\Big[\nabla^{-2}\l(2 X\nabla^2\varphi\r)
+3\nabla^{-2}\nabla^{-2}\l(X_{,\,k}\nabla^2\varphi^{,k}
-\varphi_{,\,klm}\nabla^{-2}X^{,\,klm}\r)\Big] \ ,
\el
where
\bl\label{A}
A \equiv
&  \nabla^2\nabla^2\chi^{\parallel(2)\, }_{S\,0}
+2\nabla^2\delta_{S\,0}^{(2)}
+ \frac{5\tau_0^2}{3}\l[ \nabla^2\varphi_{} \nabla^2\varphi
- \varphi_{,\,kl}\varphi^{,kl}
+\nabla^2\l(\varphi_{,\,k}\varphi^{,k}
-\frac{4}{3}\varphi\nabla^2\varphi  \r) \r]
\nn\\
& -\frac{\tau_0^4}{12} \l[ \nabla^2\varphi_{,\,k}\nabla^2\varphi^{,\,k}
- \varphi_{,\,klm} \varphi^{,\,klm}
+\nabla^2\l(
\frac{4}{3}\varphi_{,\,kl}\varphi^{,kl}
+\frac{2}{3}\nabla^2\varphi\nabla^2\varphi  \r) \r] \nn \\
&  + \frac{9}{\tau_0^6} \Big[ 3\nabla^{-2}X^{,\,klm}\nabla^{-2}X_{,\,klm}
-3X^{,\,k}X_{,\,k} -\nabla^{2}
\l( 4\nabla^{-2}X^{,\,kl}\nabla^{-2}X_{,\,kl}
+2X^2 \r) \Big]
\nn\\
&
+\frac{10}{\tau_0^3} \Big[ 3\varphi^{ ,\,kl }\nabla^{-2}X_{,\,kl}
-3\varphi\nabla^2 X -\nabla^{2}\l(\varphi X\r) \Big]
\nn\\
&
+\frac{1}{\tau_0} \Big[ 3\varphi_{,\,klm}\nabla^{-2}X^{,\,klm}
-3X_{,\,k}\nabla^2\varphi^{,k} -2\nabla^{2}
\l( 2\varphi_{,ij}\nabla^{-2}X^{,ij}
+X\nabla^2\varphi \r)  \Big] \ ,
\el
which depends on the initial values at $\tau_0$.
Thus,  the scalar metric perturbation   $D_{ij}\chi^{||(2) }_S $
is obtained,
 \bl \label{ari3as1}
 D_{ij}\chi^{\parallel(2)}_{S}
=&
\frac{\tau^4}{84} D_{ij}\nabla^{-2}\Big[
\frac{20}{3}  \varphi^{,kl}\varphi_{,kl}
        -2 \nabla^2\varphi \nabla^2\varphi
+\nabla^{-2}\l(7 \nabla^2\varphi^{,k}\nabla^2\varphi_{,k}
-7  \varphi^{,klm}\varphi_{,klm}
\r)
\Big] \nn \\
&  -\frac{5\tau^2}{18} D_{ij}\Big[
  4\varphi \varphi
 +4\nabla^{-2} \l(\varphi_{,\,k} \varphi^{,\,k}\r)
     +\nabla^{-2} \nabla^{-2} \l(6  \nabla^2 \varphi \nabla^2 \varphi
  -6 \varphi_{,\,kl}  \varphi^{,\,kl}
  \r)  \Big]
     \nn \\
& +  D_{ij}\nabla^{-2}\nabla^{-2} A
   + \frac{\tau^2 }{10} D_{ij} \nabla^{-2}F
\nn\\
&  + \frac{9}{2\tau^6} D_{ij} \Big[ \nabla^{-2}\l(
5\nabla^{-2}X_{,\,kl}\nabla^{-2}X^{,\,kl} - X^2 \r)
\nn\\
&  +\nabla^{-2}\nabla^{-2} \l( 6X^{,k}X_{,k}
-6\nabla^{-2}X^{,klm}\nabla^{-2}X_{,klm} \r) \Big]
\nn\\
&
+\frac{6}{\tau^3} D_{ij} \Big[ -\nabla^{-2}\l(Z+5\varphi X\r)
+\nabla^{-2}\nabla^{-2} \l( 5\varphi\nabla^2X
-5\varphi^{ ,\,kl }\nabla^{-2}X_{,kl}  \r)  \Big]
\nn\\
&
 +\frac{1}{\tau}  D_{ij} \Big[  \nabla^{-2}\l(2 X\nabla^2\varphi\r)
+3\nabla^{-2}\nabla^{-2}  \l( X_{,\,k}\nabla^2\varphi^{,k}
-\varphi_{,\,klm}\nabla^{-2}X^{,\,klm} \r)  \Big] \, ,
 \el
 which holds for general initial conditions at $\tau_0$.
The   solution (\ref{ari3as1}) can be also obtained  by
the traceless part of the evolution equation (\ref{eq34s})  together with
the momentum constraint (\ref{mos}).

\subsection{Vector perturbation $\chi^{\perp(2)}_{S\, ij}$}

The time integration of the momentum constraint (\ref{mos})
from $\tau_0$ to $\tau$ yields
\be\label{mosint0}
2\phi^{(2)}_{S,j}
     +\frac{1}{3}\nabla^2\chi^{\parallel(2)}_{S,j}
+\frac12\chi^{\perp (2),\,i}_{{S}i j}
               =\int_{\tau_0}^{\tau}d\tau' M_{Sj}
               +2\phi^{(2)}_{S0,j}
     +\frac{1}{3}\nabla^2\chi^{\parallel(2)}_{S0,j}
+\frac12\chi^{\perp (2),\,i}_{{S}0i j}
\ ,
\ee
where $\chi^{\perp(2)}_{S 0ij}$ is the initial value of
$\chi^{\perp(2)}_{Sij}$ at $\tau_0$.
Plugging $M_{Sj}$ of  (\ref{MSj}),
  $\phi^{(2)}_{S}$ of  (\ref{phi2S}),
and $\chi^{\parallel(2)}_{S}$ of  (\ref{chi||2S})
into (\ref{mosint0}),
one   gets
\bl\label{chi2Sperp3}
 \chi^{\perp (2)\, ,j}_{S\,ij}
=&
\frac{10\tau^2}{9} \Big[
- \varphi_{,i} \nabla^2 \varphi
+\partial_i\nabla^{-2}\l(\varphi^{,k} \nabla^2\varphi_{,k}
   + \nabla^2 \varphi \nabla^2 \varphi\r)
\Big]
                        \nonumber \\
&
 + \frac{\tau^4}{18} \Big[
\varphi_{,ki} \nabla^2\varphi^{,k}
+\partial_i\nabla^{-2}\l(- \varphi^{,kl} \nabla^2 \varphi_{,kl}
-  \nabla^2 \varphi_{,k} \nabla^2\varphi^{,k}\r)
\Big]
+ G_i \nn \\
&  +\frac{18}{\tau^6} \Big[ X^{,k}\nabla^{-2}X_{,ik} +\partial_i\nabla^{-2}
\big( -X^{,kl}\nabla^{-2}X_{,kl} -X^{,k}X_{,k} \big) \Big]
\nn\\
&
+\frac{20}{\tau^3}
\Big[
-\varphi_{,i}X
-\varphi^{ ,\,k }\nabla^{-2}X_{,ik}
+\partial_i\nabla^{-2}
\big( X\nabla^2\varphi +2\varphi_{,k}X^{,k}
+\varphi^{ ,\,k l}\nabla^{-2}X_{,kl} \big)  \Big]
\nn\\
&
+\frac{2}{\tau} \Big[
3\nabla^2\varphi^{,k}\nabla^{-2}X_{,ik} -2 X^{,k}\varphi_{,ik}
+5\varphi^{,kl}\nabla^{-2}X_{,kli}
\nn\\
&
+\partial_i\nabla^{-2} \big( -3\nabla^2\varphi^{,kl}\nabla^{-2}X_{,kl}
-X_{,k}\nabla^2\varphi^{,k} -3\varphi^{,kl}X_{,kl}
-5\varphi^{,klm}\nabla^{-2}X_{,klm} \big) \Big] \,
,
\el
where
\bl\label{Gi}
G_i\equiv
& \chi^{\perp(2)\, ,j}_{S0\,ij}
+\frac{10\tau_0^2}{9} \Big[
\varphi_{,i} \nabla^2 \varphi
+\partial_i\nabla^{-2}\l(-\varphi^{,k} \nabla^2\varphi_{,k}
   - \nabla^2 \varphi \nabla^2 \varphi\r)
\Big]
                        \nonumber \\
&
 + \frac{\tau_0^4}{18} \Big[
-\varphi_{,ki} \nabla^2\varphi^{,k}
+\partial_i\nabla^{-2}\l( \varphi^{,kl} \nabla^2 \varphi_{,kl}
+ \nabla^2 \varphi_{,k} \nabla^2\varphi^{,k}\r)
\Big] \nn \\
& +\frac{18}{\tau_0^6}
\Big[ -X^{,k}\nabla^{-2}X_{,ik} +\partial_i\nabla^{-2}
\big( X^{,kl}\nabla^{-2}X_{,kl} +X^{,k}X_{,k} \big)  \Big]
\nn\\
&
+\frac{20}{\tau_0^3}
\Big[ \varphi_{,i}X +\varphi^{ ,\,k }\nabla^{-2}X_{,ik}
+\partial_i\nabla^{-2} \big( -X\nabla^2\varphi
-2\varphi_{,k}X^{,k} -\varphi^{ ,\,k l}\nabla^{-2}X_{,kl} \big) \Big]
\nn\\
&
+\frac{2}{\tau_0}  \Big[ -3\nabla^2\varphi^{,k}\nabla^{-2}X_{,ik}
+2 X^{,k}\varphi_{,ik} -5\varphi^{,kl}\nabla^{-2}X_{,kli}
\nn\\
&  +\partial_i\nabla^{-2} \big( 3\nabla^2\varphi^{,kl}\nabla^{-2}X_{,kl}
+X_{,k}\nabla^2\varphi^{,k} +3\varphi^{,kl}X_{,kl}
+5\varphi^{,klm}\nabla^{-2}X_{,klm}  \big)  \Big] \ ,
\el
depending  on the initial values at $\tau_0$.
To get $ \chi^{\perp (2) }_{S\,ij}$ from Eq.(\ref{chi2Sperp3}),
one need  to remove $\partial^j$.
By writing  $\chi^{\perp(2)}_{Sij} = A_{Si,j} + A_{Sj,i}$
in terms of a  3-vector $ A_{Si }$ as Eq.(\ref{chiVec1}),
Eq.(\ref{chi2Sperp3}) becomes an equation of $A_{Si }$,
whose solution is
\bl\label{ASi}
 A_{Si}
=&
 \frac{10\tau^2}{9}\nabla^{-2} \Big[
- \varphi_{,i} \nabla^2 \varphi
+\partial_i\nabla^{-2}\l(\varphi^{,k} \nabla^2\varphi_{,k}
   + \nabla^2 \varphi \nabla^2 \varphi\r)
\Big]
          \nonumber \\
&
 + \frac{\tau^4}{18}\nabla^{-2} \Big[
\varphi_{,ki} \nabla^2\varphi^{,k}
+\partial_i\nabla^{-2}\l(- \varphi^{,kl} \nabla^2 \varphi_{,kl}
-  \nabla^2 \varphi_{,k} \nabla^2\varphi^{,k}\r)
\Big]
+ \nabla^{-2}G_i \nn \\
& +\frac{18}{\tau^6}
\nabla^{-2}\Big[
X^{,k}\nabla^{-2}X_{,ik}
+\partial_i\nabla^{-2}
\big(
-X^{,kl}\nabla^{-2}X_{,kl}
-X^{,k}X_{,k}
\big)
\Big]
\nn\\
&
+\frac{20}{\tau^3}
\nabla^{-2}\Big[
-\varphi_{,i}X
-\varphi^{ ,\,k }\nabla^{-2}X_{,ik}
+\partial_i\nabla^{-2}
\big(
X\nabla^2\varphi
+2\varphi_{,k}X^{,k}
+\varphi^{ ,\,k l}\nabla^{-2}X_{,kl}
\big)
\Big]
\nn\\
&
+\frac{2}{\tau}
\nabla^{-2}\Big[
3\nabla^2\varphi^{,k}\nabla^{-2}X_{,ik}
-2 X^{,k}\varphi_{,ik}
+5\varphi^{,kl}\nabla^{-2}X_{,kli}
\nn\\
&
+\partial_i\nabla^{-2}
\big(
-3\nabla^2\varphi^{,kl}\nabla^{-2}X_{,kl}
-X_{,k}\nabla^2\varphi^{,k}
-3\varphi^{,kl}X_{,kl}
-5\varphi^{,klm}\nabla^{-2}X_{,klm}
 \big) \Big] \ .
\el
Thus, the vector perturbation  is obtained
\bl\label{chi2Sperp4}
\chi^{\perp(2)}_{Sij}
=&
 \frac{10\tau^2}{9}\nabla^{-2} \Big[
- \partial_i\big(\varphi_{,j} \nabla^2 \varphi\big)
- \partial_j\big(\varphi_{,i} \nabla^2 \varphi\big)
\nn\\
&
+\partial_i\partial_j\nabla^{-2}\l(2\varphi^{,k} \nabla^2\varphi_{,k}
   + 2\nabla^2 \varphi \nabla^2 \varphi\r)
\Big]
+ \frac{\tau^4}{18}\nabla^{-2} \Big[
\partial_i\big(\varphi_{,kj} \nabla^2\varphi^{,k}\big)
                        \nonumber \\
&
+\partial_j\big(\varphi_{,ki} \nabla^2\varphi^{,k}\big)
+\partial_i\partial_j\nabla^{-2}\l(- 2\varphi^{,kl} \nabla^2 \varphi_{,kl}
-2\nabla^2 \varphi_{,k} \nabla^2\varphi^{,k}\r)
\Big]  + \nabla^{-2}(G_{i,j}+G_{j,i}) \nn \\
&
+\frac{18}{\tau^6} \nabla^{-2} \Big[
\partial_i\big(X^{,k}\nabla^{-2}X_{,jk}\big)
+\partial_j\big(X^{,k}\nabla^{-2}X_{,ik}\big)  \nn\\
&
+\partial_i\partial_j\nabla^{-2}
\big(  -2X^{,kl}\nabla^{-2}X_{,kl} -2X^{,k}X_{,k} \big)   \Big]
\nn\\
&
+\frac{20}{\tau^3}
\nabla^{-2}\Big[
\partial_i
\big(
-\varphi_{,j}X
-\varphi^{ ,\,k }\nabla^{-2}X_{,jk}
\big)
+\partial_j
\big(
-\varphi_{,i}X
-\varphi^{ ,\,k }\nabla^{-2}X_{,ik}
\big)
\nn\\
&
+\partial_i\partial_j\nabla^{-2}
\big(
2X\nabla^2\varphi
+4\varphi_{,k}X^{,k}
+2\varphi^{ ,\,k l}\nabla^{-2}X_{,kl}
\big)
\Big]
\nn\\
&
+\frac{2}{\tau}
\nabla^{-2}\Big[
\partial_i
\big(
3\nabla^2\varphi^{,k}\nabla^{-2}X_{,jk}
-2 X^{,k}\varphi_{,jk}
+5\varphi^{,kl}\nabla^{-2}X_{,klj}
\big)
\nn\\
&
+\partial_j \big( 3\nabla^2\varphi^{,k}\nabla^{-2}X_{,ik}
-2 X^{,k}\varphi_{,ik}  +5\varphi^{,kl}\nabla^{-2}X_{,kli}  \big)
\nn\\
&
+\partial_i\partial_j\nabla^{-2}
\big(  -6\nabla^2\varphi^{,kl}\nabla^{-2}X_{,kl}
-2X_{,k}\nabla^2\varphi^{,k} -6\varphi^{,kl}X_{,kl}
-10\varphi^{,klm}\nabla^{-2}X_{,klm} \big) \Big] \ .
\el
 As mentioned earlier,
$\chi^{\perp(2)}_{Sij}$ has two polarizations,
their solutions as above contain
two unknown functions in $\chi^{\perp(2)\, ,j}_{S0\,ij}(\bf x )$
with $\partial^i \chi^{\perp(2)\, ,j}_{S0\,ij}=0$
through $G_{i}$ as the initial values.
This is consistent since their initial first order time derivatives
are fixed via the momentum constraint.
The  solution   (\ref{chi2Sperp4})
can be also derived from  the curl portion of
the momentum constraint (\ref{mos}) itself
without explicitly using the solutions
$\phi^{(2)}_{S}$ of (\ref{phi2S}),
and $\chi^{\parallel(2)}_{S}$ of  (\ref{chi||2S}).
The result (\ref{chi2Sperp4}) tells us that
the 2nd-order vector perturbation is
 generated by the coupling of 1st-order scalar perturbations,
even though the 1st-order vector perturbation is absent.

\subsection{ Tensor perturbation  $\chi^{\top(2)}_{S\, ij}$}

Finally  consider the traceless part of the evolution equation (\ref{eq34s})
\bl\label{RGWS1}
\chi^{\top (2)''}_{S\,ij}
    +\frac{4}{\tau} \chi^{\top (2)'}_{S\,ij}
    -\nabla^2\chi^{\top(2)}_{S\,i j}
   =&2\bar S_{S\,ij}
  -\l(2D_{ij}\phi^{(2)}_{S} +\frac{1}{3}\nabla^2D_{ij}\chi^{||(2)}_{S\,}   \r)
\nonumber \\
&
-\l( D_{ij}\chi^{||(2)''}_{S\,}
 +\frac{4}{\tau}  D_{ij}\chi^{||(2)'}_{S\,}\r)
-\l(\chi^{\perp (2)''}_{S\,ij}
    +\frac{4}{\tau}  \chi^{\perp (2)'}_{S\,ij}
\r)
,
\el
where $\bar{S}_{S\,ij} \equiv S_{S\,ij}-\frac{1}{3}\delta_{ij}S^k_{S\,k} $,
explicitly given by
\bl\label{barSSij}
\bar S_{Sij} =
&  \tau^4\Big( \frac{1}{18}\varphi^{,k}_{,ij}\nabla^2\varphi_{,k}
-\frac{1}{18}\varphi_{,i kl} \varphi^{,kl}_{,j}\Big)
+ \tau^2\Big( \frac{8}{3}\varphi^{,k}_{,i}\varphi_{,kj}
-\frac{22}{9}  \varphi_{,i j} \nabla^2\varphi
-\frac{5}{9}\varphi_{,k}\varphi^{,k}_{,ij}
\Big)
 \nn \\
&
-\frac{100}{9}\varphi\varphi_{,ij}
-\frac{50}{3}\varphi_{,i}\varphi_{,j}
      \nn \\
& +\delta_{ij}\Big[
\tau^4\Big(\frac{1}{54} \varphi_{,klm}\varphi^{,klm}
-\frac{1}{54}\nabla^2\varphi_{,k}\nabla^2\varphi^{,k} \Big) \nn \\
& +\tau^2\Big(-\frac{8}{9} \varphi^{,k l}\varphi_{,k l}
+\frac{22}{27}   \nabla^2\varphi \nabla^2\varphi
+\frac{5}{27}  \varphi_{,k} \nabla^2 \varphi^{,k} \Big)
+\frac{100}{27}\varphi \nabla^2 \varphi
+\frac{50}{9} \varphi_{,k}\varphi^{,k}
  \nn \\
& + \frac{1}{\tau^8}
\Big(
54 X^2
-108\nabla^{-2}X_{,kl}\nabla^{-2}X^{,kl}
\Big)
+\frac{1}{\tau^6}
\Big(
6\nabla^{-2}X^{,klm}\nabla^{-2}X_{,klm}
-6X^{,k}X_{,k}
\Big)
\nn\\
&
+\frac{1}{\tau^3}
\Big(
\frac{28}{3}X\nabla^2\varphi
-\frac{16}{3}\varphi^{,kl}\nabla^{-2}X_{,kl}
+\frac{10}{3}\varphi^{,k}X_{,k}
\Big)
+ \frac{2}{3\tau}\Big(
\varphi_{,klm}\nabla^{-2}X^{,klm}
    - X_{,k}\nabla^2\varphi^{,k}  \Big)
\Big]
\nn\\
&
+\frac{1}{\tau^8}
\Big(
324\nabla^{-2}X^{,k}_{,i}\nabla^{-2}X_{,kj}
-162X\nabla^{-2}X_{,ij}
\Big)
+\frac{1}{\tau^6}
\Big(
18 X^{,k}\nabla^{-2}X_{,kij}
\nn\\
& -18\nabla^{-2}X^{,kl}_{,i}\nabla^{-2}X_{,klj} \Big)
+\frac{1}{\tau^3}
\Big(
-14\varphi_{,ij}X
-14\nabla^2\varphi\nabla^{-2}X_{,ij}
-10\varphi^{,k}\nabla^{-2}X_{,kij}
\nn
\\
&
+8\varphi^{,k}_{,i}\nabla^{-2}X_{,kj}
+8\varphi_{,j}^{,k}\nabla^{-2}X_{,ki}\Big)
+\frac{1}{\tau}
\Big(
X^{,k}\varphi_{,kij}
+\nabla^{-2}X_{,kij}\nabla^2\varphi^{,k}
\nn\\
&
-\varphi_{,klj}\nabla^{-2}X^{,kl}_{,i}
-\varphi_{,kli}\nabla^{-2}X_{,j}^{,kl}
\Big)  \ .
\el
One can substitute the known  $\phi^{(2)}_{S}$,
$ D_{ij}\chi^{||(2) }_{S} $,
$ \chi^{\perp (2)}_{S\,ij}$ into  Eq.(\ref{RGWS1}),
and solve for $\chi^{\top (2)}_{S\,ij}$.
But the following calculation is   simpler and will yield the same result.
Applying consecutively
$\partial^i\partial^j$,
  $    \nabla^{-2}   \nabla^{-2} $,   and  $D_{ij}$ on  (\ref {RGWS1}) leads to
\be\label{RGWS2}
-\l(2D_{ij}\phi^{(2)}_{S\,}
+\frac{1}{3}\nabla^2D_{ij}\chi^{||(2)}_{S\,}
\r)
- \l(D_{ij}\chi^{||(2)''}_{S\,}
 +\frac{4}{\tau}D_{ij}\chi^{||(2)'}_{S\,}\r)
=-3D_{ij}   \nabla^{-2}\nabla^{-2}   \bar S_{S\,kl}^{,\,kl}
\,,
\ee
Substituting  Eq.(\ref{RGWS2}) into  the rhs of Eq.(\ref{RGWS1})
gives
\be\label{RGWS3}
\chi^{\top (2)''}_{S\,ij}
    +\frac{4}{\tau}\chi^{\top (2)'}_{S\,ij}
    -\nabla^2\chi^{\top(2)}_{S\,i j}
=2\bar S_{S\,ij}
-3D_{ij}\nabla^{-2}\nabla^{-2}\bar S_{S\,kl}^{,\,kl}
-\l(\chi^{\perp (2)''}_{S\,ij}
    +\frac{4}{\tau}\chi^{\perp (2)'}_{S\,ij}
\r)
.
\ee
Applying $\partial^j$ to (\ref {RGWS3}) and
together with Eq.(\ref{chiVec1})
leads  to an equation of  $A_{S\,i}$ as the following
\be\label{AperpS2}
0=2\bar S_{S\,ij}^{,j}
-2\nabla^{-2}\bar S_{S\,kl,\,i}^{,\,kl}
-\nabla^2\l(A^{''}_{S\,i}
    +\frac{4}{\tau}A^{'}_{S\,i}
\r)
.
\ee
With the help of Eq.(\ref{AperpS2}),
one has
\bl\label{RGWS4}
-\l(\chi^{\perp (2)''}_{S\,ij}
    +\frac{4}{\tau}\chi^{\perp (2)'}_{S\,ij}
\r)
=&-\partial_j\l(A^{''}_{S\,i}
    +\frac{4}{\tau}A^{'}_{S\,i}\r)
-\partial_i\l(A^{''}_{Sj}
    +\frac{4}{\tau}A^{'}_{Sj}\r)
\nn\\
=&-2\nabla^{-2}\bar S_{S\,ki,j}^{,k}
-2\nabla^{-2}\bar S_{S\,kj,\,i}^{,k}
+4\nabla^{-2}\nabla^{-2}\bar S_{S\,kl,\,ij}^{,\,kl}
\ .
\el
Substituting   (\ref{RGWS4}) into the rhs of  Eq.(\ref{RGWS3})
yields   the  equation of 2nd-order  tensor perturbation
\ba \label{gwS11}
& & \chi^{\top (2)''}_{S\,ij}   +\frac{4}{\tau}\chi^{\top (2)'}_{S\,ij}
      -\nabla^2\chi^{\top(2)}_{S\,i j}   \nn \\
 & = & 2\bar S_{S\,ij}
+\nabla^{-2}\nabla^{-2}\bar S_{S\,kl,\,ij}^{,\,kl}
+\delta_{ij}\nabla^{-2}\bar S_{S\,kl}^{,\,kl}
-2\nabla^{-2}\bar S_{S\,ki,j}^{,k}
-2\nabla^{-2}\bar S_{S\,kj,\,i}^{,k} \ .
\ea
This is a second-order, hyperbolic differential wave equation
with the source  constructed from  $\bar S_{S\,ij}$.
Substituting (\ref{barSSij}) into the rhs of (\ref {gwS11})
and regrouping by  powers of $\tau$,
 one obtains the  equation
\be\label{eqRGW2}
\chi^{\top (2)''}_{S\,ij}
+\frac{4}{\tau}\chi^{\top (2)'}_{S\,ij}
-\nabla^2\chi^{\top(2)}_{S\,i j}
=B_{1\,ij}
+ \tau^2 B_{2\,ij}
+ \tau^4 B_{3\,ij}
+
\frac{B_{4\, ij}}{\tau^8}
+\frac{B_{5\, ij}}{\tau^6}
+\frac{B_{6\, ij}}{\tau^3}
+\frac{B_{7\, ij}}{\tau} \,  ,
\ee
where
\bl\label{Bij1}
B_{1\,ij}=&
\delta_{ij}\nabla^{-2}\Big(
\frac{50}{9}   \varphi_{,\,kl} \varphi^{,\,kl}
-  \frac{50}{9}  \nabla^2 \varphi \nabla^2 \varphi\Big)
+\nabla^{-2}\Big(
\frac{200}{9}\varphi_{,ij} \nabla^2 \varphi
-\frac{200}{9}\varphi_{,ki}\varphi_{,j}^{,k}
\Big)
\nn\\
&
+\nabla^{-2}\nabla^{-2}\partial_i\partial_j\Big(
\frac{50}{9}\varphi^{,kl}\varphi_{,kl}
-\frac{50}{9}\nabla^2 \varphi \nabla^2 \varphi
\Big)
\el
\bl\label{Bij2}
B_{2\,ij}=& \delta_{ij}\Big[
\frac{11}{9}   \nabla^2\varphi \nabla^2\varphi
-\frac{11}{9} \varphi^{,k l}\varphi_{,k l}
+ \nabla^{-2}\Big(
\frac{7}{9}  \nabla^2 \varphi_{,\,k}\nabla^2 \varphi^{,\,k}
- \frac{7}{9}\varphi_{,\,klm}\varphi^{,\,klm}
\Big)
\Big]
\nn\\
&
+\frac{44}{9}\varphi^{,k}_{,i}\varphi_{,kj}
-\frac{44}{9}  \varphi_{,i j} \nabla^2\varphi
+\nabla^{-2}\Big(
\frac{28}{9} \varphi_{,\,kli} \varphi^{,\,kl}_{,j}
-\frac{28}{9}\varphi_{,kij} \nabla^2\varphi^{,k}
\Big)
\nonumber \\
&
+\nabla^{-2}\nabla^{-2}\partial_i\partial_j\Big(
\frac{7}{9}\nabla^2 \varphi_{,k} \nabla^2\varphi^{,k}
-\frac{7}{9}\varphi^{,klm}\varphi_{,klm}
\Big)
\nn\\
&
+\nabla^{-2}\partial_i\partial_j\Big(
\frac{11}{9} \nabla^2 \varphi\nabla^2\varphi
-\frac{11}{9} \varphi_{,\,kl} \varphi^{,\,kl}
\Big)
,
\el
\bl\label{Bij3}
B_{3\,ij}=&\delta_{ij}\Big(\frac{1}{36} \varphi_{,klm}\varphi^{,klm}
-\frac{1}{36}\nabla^2\varphi_{,k}\nabla^2\varphi^{,k}
\Big)
+\frac{1}{9}\varphi^{,k}_{,ij}\nabla^2\varphi_{,k}
-\frac{1}{9}\varphi_{,i kl} \varphi^{,kl}_{,j}
\nn\\
&
+ \nabla^{-2}\partial_i\partial_j\Big(
\frac{1}{36} \varphi_{,\,klm}\varphi^{,\,klm}
-\frac{1}{36}\nabla^2\varphi_{,\,k} \nabla^2 \varphi^{,\,k}
\Big).
\el
\bl\label{Bij4n}
B_{4\, ij}=&  \delta_{ij} \Big[ 81 X^2 -162\nabla^{-2}X_{,kl}\nabla^{-2}X^{,kl}
+\nabla^{-2} \Big( 162X^{,kl}\nabla^{-2}X_{,kl} +162X^{,k}X_{,k} \Big) \Big]
\nn\\
&
-324X\nabla^{-2}X_{,ij} +648\nabla^{-2}X^{,k}_{,i}\nabla^{-2}X_{,kj}
\nn\\
&
-\nabla^{-2}\partial_i \,\Big(
324X^{,k}\nabla^{-2}X_{,kj} \Big)
-\nabla^{-2}\partial_j \,\Big(
324X^{,k}\nabla^{-2}X_{,ki}
\Big)
\nn\\
&
+\nabla^{-2}\partial_i\partial_j
\,\Big(
81X^2
-162\nabla^{-2}X^{,kl}\nabla^{-2}X_{,kl}
\Big)
\nn\\
&
+\nabla^{-2}\nabla^{-2}\partial_i\partial_j
\,\Big(
162X^{,kl}\nabla^{-2}X_{,kl}
+162X^{,k}X_{,k}
\Big)
 \ .
\el
\bl\label{Bij5n}
B_{5\, ij}=&
\delta_{ij}
\Big[
-9X^{,k}X_{,k}
+9\nabla^{-2}X^{,klm}\nabla^{-2}X_{,klm}
\Big]
+36X^{,k}\nabla^{-2}X_{,kij}
-36\nabla^{-2}X^{,kl}_{,i}\nabla^{-2}X_{,klj}
\nn\\
&
+\nabla^{-2}\partial_i\partial_j
\,\Big(
-9X^{,k}X_{,k}
+9\nabla^{-2}X^{,klm}\nabla^{-2}X_{,klm}
\Big)
 \ .
\el
\bl\label{Bij6n}
B_{6\, ij}=&
\delta_{ij}
\Big[
14X\nabla^2\varphi
-14\varphi^{,kl}\nabla^{-2}X_{,kl}
+\nabla^{-2}
\Big(
2\varphi^{,klm}\nabla^{-2}X_{,klm}
-2X_{,k}\nabla^2\varphi^{,k}
\Big)
\Big]
\nn\\
&
-28\varphi_{,ij}X
-28\nabla^2\varphi\nabla^{-2}X_{,ij}
-20\varphi^{,k}\nabla^{-2}X_{,kij}
+16\varphi^{,k}_{,i}\nabla^{-2}X_{,kj}
+16\varphi_{,j}^{,k}\nabla^{-2}X_{,ki}
\nn\\
&
+\nabla^{-2}\partial_i
\,\Big(
-8\varphi_{,j}^{,k}X_{,k}
+12\nabla^2\varphi^{,k}\nabla^{-2}X_{,kj}
+20\varphi^{,kl}\nabla^{-2}X_{,klj}
\Big)
\nn\\
&
+\nabla^{-2}\partial_j
\,\Big(
-8\varphi_{,i}^{,k}X_{,k}
+12\nabla^2\varphi^{,k}\nabla^{-2}X_{,ki}
+20\varphi^{,kl}\nabla^{-2}X_{,kli}
\Big)
\nn\\
&
+\nabla^{-2}\partial_i\partial_j
\,\Big(
14X\nabla^2\varphi
-14\varphi^{,kl}\nabla^{-2}X_{,kl}
+20\varphi^{,k}X_{,k}
\Big)
\nn\\
&
+\nabla^{-2}\nabla^{-2}\partial_i\partial_j
\,\Big(
2\varphi^{,klm}\nabla^{-2}X_{,klm}
-2X_{,k}\nabla^2\varphi^{,k}
\Big)
 \ .
\el
\bl  \label{Bij7n}
B_{7\, ij}=&
\delta_{ij}
\Big[
-X_{,k}\nabla^2\varphi^{,k}
+\varphi_{,klm}\nabla^{-2}X^{,klm}
\Big]
\nn\\
&
+2X^{,k}\varphi_{,kij}
+2\nabla^{-2}X_{,kij}\nabla^2\varphi^{,k}
-2\varphi_{,klj}\nabla^{-2}X^{,kl}_{,i}
-2\varphi_{,kli}\nabla^{-2}X_{,j}^{,kl}
\nn\\
&
+\nabla^{-2}\partial_i\partial_j
\,\Big(
-X^{,k}\nabla^2\varphi_{,k}
+\varphi_{,klm}\nabla^{-2}X^{,klm}
\Big)
 \ .
\el
The solution of Eq.(\ref{eqRGW2}) is
\be \label{chi2S1}
\chi^{\top(2)}_{Sij}
=  \frac{1}{(2\pi)^{3/2}}
   \int d^3k   e^{i \,\bf{k}\cdot\bf{x}} \,
     \left[  \bar y_{ij} (\mathbf k, \tau)
    + \sum_{s={+,\times}} {\mathop \epsilon
    \limits^s}_{ij}(k) ~ {\mathop h\limits^s}_k(\tau) \right] ,
\ee
where
\bl \label{yijbar}
\bar y_{ij} =&
\frac{\bar B_{1\,ij}}{k^2}
-\frac{10\bar B_{2\,ij}}{k^4}
 + \frac{280\bar B_{3\,ij}}{k^6}
+\tau^2\frac{\bar B_{2\,ij}}{k^2}
       -\tau^2\frac{28\bar B_{3\,ij}}{k^4}
+\tau^4\frac{\bar B_{3\,ij}}{k^2} \nn \\
& +  \Big(
\frac{k\cos(k\tau)\bar B_{5\,ij}}{8\tau^3}
-\frac{k^3\cos(k\tau)\bar B_{4\,ij}}{144\tau^3}
+\frac{k^2\sin(k\tau)\bar B_{5\,ij}}{8\tau^2}
\nn\\
&
-\frac{k^4\sin(k\tau)\bar B_{4\,ij}}{144\tau^2}
\Big)\l(\int_{0}^{k\tau}\frac{\sin s}{s}ds\r)
\nn\\
&
+\Big(
\frac{k\sin(k\tau)\bar B_{5\,ij}}{8\tau^3}
-\frac{k^3\sin(k\tau)\bar B_{4\,ij}}{144\tau^3}
-\frac{k^2\cos(k\tau)\bar B_{5\,ij}}{8\tau^2}
\nn\\
&
+\frac{k^4\cos(k\tau)\bar B_{4\,ij}}{144\tau^2}
\Big)\l(\int_{k\tau}^{+\infty}\frac{\cos s}{s}ds\r)
\nn\\
&
+\frac{\bar B_{4\,ij}}{18\tau^6}
+\frac{\bar B_{5\,ij}}{4\tau^4}
-\frac{k^2\bar B_{4\,ij}}{72\tau^4}
+\frac{2\bar B_{7\,ij}}{k^4\tau^3}
+\frac{\bar B_{6\,ij}}{k^2\tau^3}
+\frac{\bar B_{7\,ij}}{k^2\tau} \,  ,
\el
$\bar B_{1\,ij}$,  $\bar B_{2\,ij}$,  $\bar B_{3\,ij}$,
  $\bar B_{4\,ij}$,  $\bar B_{5\,ij}$,  $\bar B_{6\,ij}$,  $\bar B_{7\,ij}$
are the Fourier components of $B_{1\,ij}$,   $B_{2\,ij}$,  $B_{3\,ij}$,
  $ B_{4\,ij}$,  $ B_{5\,ij}$,  $ B_{6\,ij}$,  $ B_{7\,ij}$
   respectively,
and the term $\sum_{s} {\mathop \epsilon
    \limits^s}_{ij} ~ {\mathop h\limits^s}_k$
in (\ref{chi2S1}) is a 2nd-order homogeneous solution similar to
 (\ref{Fourier}) and (\ref{GWmode}), whose coefficients
  ${\mathop d\limits^s}_1$, ${\mathop d\limits^s}_2$
 are to be determined by the initial condition of 2nd-order RGW.
Using the  relation
\be\label{B2ijF2}
\frac{1}{(2\pi)^{3/2}}
    \int d^3k e^{i\mathbf k\cdot\mathbf x}\frac{1}{k^2}\bar B_{1ij}
=-\nabla^{-2}B_{1 ij}
\ee
and the like,  the solution  (\ref{chi2S1})  is written as
\bl \label{chi2S21}
\chi^{\top(2)}_{Sij}
=&
-\nabla^{-2}B_{1\,ij}
 -10\nabla^{-2}\nabla^{-2} B_{2\,ij}
-280 \nabla^{-2}\nabla^{-2}\nabla^{-2}B_{3\,ij}
\nn\\
&
-\tau^2\nabla^{-2}B_{2\,ij}
       -28\tau^2\nabla^{-2}\nabla^{-2} B_{3\,ij}
-\tau^4\nabla^{-2}B_{3\,ij}
\nn\\
&
+ \frac{ B_{4\,ij}}{18\tau^6}
+\frac{ B_{5\,ij}}{4\tau^4}
+\frac{\nabla^2 B_{4\,ij}}{72\tau^4}
+\frac{2\nabla^{-2}\nabla^{-2} B_{7\,ij}}{\tau^3}
-\frac{\nabla^{-2} B_{6\,ij}}{\tau^3}
-\frac{\nabla^{-2} B_{7\,ij}}{\tau}
\nn\\
&
+\frac{1}{(2\pi)^{3/2}}
   \int d^3k   e^{i \,\bf{k}\cdot\bf{x}} \,
     \Bigg[
      \Big(
\frac{k\cos(k\tau)\bar B_{5\,ij}}{8\tau^3}
-\frac{k^3\cos(k\tau)\bar B_{4\,ij}}{144\tau^3}
\nn\\
&
+\frac{k^2\sin(k\tau)\bar B_{5\,ij}}{8\tau^2}
-\frac{k^4\sin(k\tau)\bar B_{4\,ij}}{144\tau^2}
\Big)\l(\int_{0}^{k\tau}\frac{\sin s}{s}ds\r)
\nn\\
&
+\Big(
\frac{k\sin(k\tau)\bar B_{5\,ij}}{8\tau^3}
-\frac{k^3\sin(k\tau)\bar B_{4\,ij}}{144\tau^3}
-\frac{k^2\cos(k\tau)\bar B_{5\,ij}}{8\tau^2}
\nn\\
&
+\frac{k^4\cos(k\tau)\bar B_{4\,ij}}{144\tau^2}
\Big)\l(\int_{k\tau}^{+\infty}\frac{\cos s}{s}ds\r)
+ \sum_{s={+,\times}} {\mathop \epsilon
    \limits^s}_{ij}(k) ~ {\mathop h\limits^s}_k(\tau)
    \Bigg]       \,  .
\el
Plugging   (\ref{Bij4n})-(\ref{Bij7n}) into Eq.(\ref{chi2S21}),
we obtain the solution of 2nd-order tensor perturbation:
\bl \label{chi2S2}
&
\chi^{\top(2)}_{Sij}
=
+\Big[
\frac{20}{3} \delta_{ij}\nabla^{-2}\nabla^{-2}(
\varphi_{,\,kl} \varphi^{,\,kl}
- \nabla^2 \varphi \nabla^2 \varphi)
+\frac{80}{3}\nabla^{-2}\nabla^{-2}  (\varphi_{,i j} \nabla^2\varphi
-\varphi^{,k}_{,i}\varphi_{,kj})
\nn\\
&
+\frac{20}{3} \nabla^{-2}\nabla^{-2}\nabla^{-2}\partial_i\partial_j(
\varphi_{,\,kl} \varphi^{,\,kl}
- \nabla^2 \varphi\nabla^2\varphi
)
\Big]
\nn\\
&
+\tau^2
\Big[
\frac{11}{9}\delta_{ij}\nabla^{-2}(
\varphi^{,k l}\varphi_{,k l}
- \nabla^2\varphi \nabla^2\varphi
)
+\frac{44}{9}\nabla^{-2}(\varphi_{,i j} \nabla^2\varphi
-\varphi^{,k}_{,i}\varphi_{,kj}
)
\nn\\
&
+\frac{11}{9}\nabla^{-2}\nabla^{-2}\partial_i\partial_j(
\varphi_{,\,kl} \varphi^{,\,kl}
-\nabla^2 \varphi\nabla^2\varphi
)
\Big]
\nn\\
&
+\tau^4
\Big[
\delta_{ij}\nabla^{-2}(-\frac{1}{36} \varphi_{,klm}\varphi^{,klm}
+\frac{1}{36}\nabla^2\varphi_{,k}\nabla^2\varphi^{,k}
)
+\frac{1}{9}\nabla^{-2}(\varphi_{,i kl} \varphi^{,kl}_{,j}
-\varphi^{,k}_{,ij}\nabla^2\varphi_{,k})
\nn\\
&
+ \frac{1}{36}\nabla^{-2}\nabla^{-2}\partial_i\partial_j(
\nabla^2\varphi_{,\,k} \nabla^2 \varphi^{,\,k}
-\varphi_{,\,klm}\varphi^{,\,klm}    )\Big]
            \nn\\
& +\frac{9}{2\tau^6}
\Big[
\delta_{ij}
\Big(
 X^2
-2\nabla^{-2}X_{,kl}\nabla^{-2}X^{,kl}
+\nabla^{-2}
(
2X^{,kl}\nabla^{-2}X_{,kl}
+2X^{,k}X_{,k}
)
\Big)
\nn\\
&
-4X\nabla^{-2}X_{,ij}
+8\nabla^{-2}X^{,k}_{,i}\nabla^{-2}X_{,kj}
-\nabla^{-2}\partial_i
\,\Big(
4X^{,k}\nabla^{-2}X_{,kj}
\Big)
-\nabla^{-2}\partial_j
\,\Big(
4X^{,k}\nabla^{-2}X_{,ki}
\Big)
\nn\\
&
+\nabla^{-2}\partial_i\partial_j
\,\Big(
X^2
-2\nabla^{-2}X^{,kl}\nabla^{-2}X_{,kl}
\Big)
+\nabla^{-2}\nabla^{-2}\partial_i\partial_j
\,\Big(
2X^{,kl}\nabla^{-2}X_{,kl}
+2X^{,k}X_{,k}
\Big)
\Big]
\nn\\
&
+\frac{9}{8\tau^4}
\Big[
\delta_{ij}
\nabla^2\Big(
X^2
-\nabla^{-2}X^{,kl}\nabla^{-2}X_{,kl}
\Big)
+\partial_i\partial_j
\,\Big(
 X^2
-\nabla^{-2}X^{,kl}\nabla^{-2}X_{,kl}
\Big)
\nn\\
&
+\nabla^2 \Big(
4\nabla^{-2}X^{,k}_{,i}\nabla^{-2}X_{,kj}
-4 X\nabla^{-2}X_{,ij}
\Big)
\Big]
\nn\\
&
+\frac{7}{\tau^3}
\Big[
\delta_{ij}
\nabla^{-2}\Big(
2\varphi^{,kl}\nabla^{-2}X_{,kl}
-2X\nabla^2\varphi
\Big)
+\nabla^{-2}\nabla^{-2}\partial_i\partial_j
\,\Big(
-2X\nabla^2\varphi
+2\varphi^{,kl}\nabla^{-2}X_{,kl}
\Big)
\nn\\
&
+\nabla^{-2}
\Big(
4\varphi_{,ij}X
+4\nabla^2\varphi\nabla^{-2}X_{,ij}
-4\varphi^{,k}_{,i}\nabla^{-2}X_{,kj}
-4\varphi_{,j}^{,k}\nabla^{-2}X_{,ki}
\Big)
\Big]
\nn\\
&
+\frac{1}{\tau}
\Big[
\delta_{ij}\nabla^{-2}
\Big(
X_{,k}\nabla^2\varphi^{,k}
-\varphi_{,klm}\nabla^{-2}X^{,klm}
\Big)
+\nabla^{-2}\nabla^{-2}\partial_i\partial_j
\,\Big(
X^{,k}\nabla^2\varphi_{,k}
-\varphi_{,klm}\nabla^{-2}X^{,klm}
\Big)
\nn\\
&
+\nabla^{-2}\Big(
-2X^{,k}\varphi_{,kij}
-2\nabla^{-2}X_{,kij}\nabla^2\varphi^{,k}
+2\varphi_{,klj}\nabla^{-2}X^{,kl}_{,i}
+2\varphi_{,kli}\nabla^{-2}X_{,j}^{,kl}
\Big)
\Big]
\nn\\
&
+\frac{1}{(2\pi)^{3/2}}
   \int d^3k   e^{i \,\bf{k}\cdot\bf{x}} \,
     \Bigg[
      \Big(
\frac{k\cos(k\tau)\bar B_{5\,ij}}{8\tau^3}
-\frac{k^3\cos(k\tau)\bar B_{4\,ij}}{144\tau^3}
+\frac{k^2\sin(k\tau)\bar B_{5\,ij}}{8\tau^2}
\nn\\
&
-\frac{k^4\sin(k\tau)\bar B_{4\,ij}}{144\tau^2}
\Big)\l(\int_{0}^{k\tau}\frac{\sin s}{s}ds\r)
+\Big(
\frac{k\sin(k\tau)\bar B_{5\,ij}}{8\tau^3}
-\frac{k^3\sin(k\tau)\bar B_{4\,ij}}{144\tau^3}
\nn\\
&
-\frac{k^2\cos(k\tau)\bar B_{5\,ij}}{8\tau^2}
+\frac{k^4\cos(k\tau)\bar B_{4\,ij}}{144\tau^2}
\Big)\l(\int_{k\tau}^{+\infty}\frac{\cos s}{s}ds\r)
+ \sum_{s={+,\times}} {\mathop \epsilon
    \limits^s}_{ij}(k) ~ {\mathop h\limits^s}_k(\tau)
    \Bigg]
\,.
\el

Let us examine  the  traceless metric perturbation
given by Eq.(4.33)  in Ref.~\cite{Matarrese98}
as the following
\bl\label{GeneralSol}
\chi^{(2)}_{Sij} = \pi_{Sij} & +
 \frac{5\tau^2}{9}(-6 \varphi_{,i}\varphi_{,j}-4\varphi\varphi_{,ij}
+2\delta_{ij}\varphi_{,k}\varphi^{,k}+\frac{4}{3} \delta_{ij} \varphi\nabla^2\varphi)
\nn\\
&
 +\frac{\tau^4}{126}(19 \varphi^{,k}_{,i}\varphi_{,jk}
-12 \varphi_{,ij} \nabla^2\varphi+4(\nabla^2\varphi)^2\delta_{ij}
-\frac{19}{3} \delta_{ij} \varphi^{,kl}\varphi_{,kl})
\el
due to  the  scalar-scalar  coupling,
where $\pi_{Sij}$ is the  tensor   given by (4.37) in Ref.~\cite{Matarrese98}.
(\ref{GeneralSol}) contains the growing modes only,
which correspond to the case $X=0$.
We notice that the last  two terms on the rhs of  (\ref{GeneralSol})
still contain a tensor portion beside  $\pi_{S ij}$.
Now we can decompose   Eq.(\ref{GeneralSol}) into the scalar, vector, and tensor:
$\chi^{(2)}_{Sij}= D_{ij}    \chi^{||(2) }_S+  \chi^{\perp(2)}_{Sij}
+ \chi^{\top(2)}_{Sij}$.
By calculation,
we find that the scalar $D_{ij} \chi^{||(2)}_S$ is just the expression of Eq.(\ref{ari3as1})
(with $X=0$,  $\delta_{S\,0}^{(2)}, \phi^{(2)}_{S\,0}, \chi^{\parallel(2)}_{S\,0}=0$ at $\tau_0=0$),
 the vector $ \chi^{\perp(2) }_{S\, ij}$  is just the expression of  Eq.(\ref{chi2Sperp4})
(with $X=0$,   $\chi^{\perp(2)}_{S0\,ij}=0$ at  $\tau_0=0$),
and  the tensor is
\bl\label{tensor22}
 \chi^{\top(2)}_{S\,ij}
=& \pi_{Sij}
+ \frac{5\tau^2}{9}\Big[\delta_{ij}\nabla^{-2}(
\varphi_{,kl}\varphi^{,kl}-
 \nabla^2\varphi\nabla^2\varphi
)
-2 \varphi_{,i}\varphi_{,j}
+\partial_i\nabla^{-2}(2\varphi_{,j}\nabla^2\varphi)
\nn\\
&
+\partial_j\nabla^{-2}(2\varphi_{,i}\nabla^2\varphi)
-\partial_i\partial_j\nabla^{-2}\nabla^{-2}(
 \nabla^2\varphi\nabla^2\varphi
-\varphi_{,kl}\varphi^{,kl}
)
\Big]
\nn\\
&
 +\frac{\tau^4}{252}\Big[\delta_{ij}\Big(
6\nabla^2\varphi\nabla^2\varphi
-6\varphi^{,kl}\varphi_{,kl}
+\nabla^{-2}(
7\nabla^2\varphi_{,k} \nabla^2\varphi^{,k}
-7\varphi_{,klm} \varphi^{,klm})
\Big)
\nn\\
&
 +38 \varphi^{,k}_{,i}\varphi_{,jk}
-24 \varphi_{,ij} \nabla^2\varphi
-\partial_i\nabla^{-2}(14\varphi_{,jk} \nabla^2\varphi^{,k})
-\partial_j\nabla^{-2}(14\varphi_{,ik} \nabla^2\varphi^{,k})
\nn\\
&
+\partial_i\partial_j\nabla^{-2}(6\nabla^2\varphi\nabla^2\varphi
-6\varphi^{,kl}\varphi_{,kl}
)
 \nn \\
&+\partial_i\partial_j\nabla^{-2}\nabla^{-2}(
-7\varphi_{,klm}\varphi^{,klm}
+7\nabla^2\varphi_{,k} \nabla^2\varphi^{,k}
)
\Big].
\el
The expression   $\pi_{Sij}$ in Ref.~\cite{Matarrese98}
  can be written as the following
\bl\label{piij3}
\pi_{Sij}
=&
\frac{20}{3}\delta_{ij}\nabla^{-2}\nabla^{-2}(
\varphi_{,kl}\varphi^{,kl}
-\nabla^2\varphi\nabla^2\varphi)
+\frac{20}{3}\partial_i\partial_j \nabla^{-2}\nabla^{-2}\nabla^{-2}
(
\varphi_{,kl}\varphi^{,kl}
-\nabla^2\varphi\nabla^2\varphi)
\nn\\
&
+\frac{80}{3} \nabla^{-2}\nabla^{-2}(\varphi_{,ij}\nabla^2\varphi
-\varphi_{,ik}\varphi^{,k}_{,j})
+\tau^2 \Big[
\frac{2}{3} \delta_{ij}\nabla^{-2}(
\varphi_{,kl}\varphi^{,kl}
-\nabla^2\varphi\nabla^2\varphi
)
\nn\\
&
+\frac{2}{3} \partial_i\partial_j\nabla^{-2}\nabla^{-2}
(\varphi_{,kl}\varphi^{,kl}
-\nabla^2\varphi\nabla^2\varphi
)
+\frac{8}{3}\nabla^{-2} (\varphi_{,ij}\nabla^2\varphi
-\varphi_{,ik}\varphi^{,k}_{,j})
\Big]
\nn\\
&
 + \tau^4 \Big[
\frac{1}{42}\delta_{ij}(
\varphi_{,kl}\varphi^{,kl}
-\nabla^2\varphi\nabla^2\varphi
)
+\frac{1}{42}\partial_i\partial_j\nabla^{-2}
(\varphi_{,kl}\varphi^{,kl}
-\nabla^2\varphi\nabla^2\varphi
)
\nn\\
&
+\frac{2}{21}(\varphi_{,ij}\nabla^2\varphi
-\varphi_{,ik}\varphi^{,k}_{,j})
\Big]
+\frac{1}{(2\pi)^{3/2}}
   \int d^3k   e^{i \,\bf{k}\cdot\bf{x}} \,
     \left[  \sum_{s={+,\times}} {\mathop \epsilon
    \limits^s}_{ij}(k) ~ {\mathop h\limits^s}_k(\tau) \right]
.
\el
(Notice that the term containing  $j_1(k\tau)/k\tau$
in (4.38) of Ref.~\cite{Matarrese98}
  can be absorbed in the homogenous solution.)
Substituting  (\ref{piij3}) into
(\ref{tensor22})  recovers  our  solution (\ref{chi2S2})
of  the case of $X=0$.
Thus,   we have proven that,
the solution  (\ref{chi2S2}) is the full  expression of tensor,
whereas    $\pi_{Sij}$  is only a portion of tensor.

We can also derive the 2nd-order density contrast in terms of gravitational potential.
Substituting the 1st order of (\ref{delta0}) and (\ref{eq28})
 and the 2nd orders of (\ref{2c1rcerf}) and (\ref{phi2S})
into Eq.(\ref{delta2nd}) in Appendix A yields
\bl\label{delta2phi}
\delta^{(2)}_S
 =&
 \frac{\tau^4}{126} \l(5\nabla^2\varphi_{} \nabla^2\varphi
     +2\varphi_{,ki}\varphi^{,ki} \r)
+ \frac{\tau^2}{18}\l(40\varphi\nabla^2\varphi
+15\varphi_{,\,k}\varphi^{,\,k}\r)
- \frac{\tau^2}{20} F
   \nn \\
&
+ \frac{9}{4\tau^6}(
3\nabla^{-2}X_{,kl}\nabla^{-2}X^{,kl}
+5 X^2
)
+\frac{3}{\tau^3}Z
+\frac{20}{\tau^3}\varphi X
+\frac{2}{\tau}\varphi_{,kl}\nabla^{-2}X^{,kl} \, .
\el
This extends  (4.39) of Ref.~\cite{Matarrese98}
to general initial conditions
  $\delta_{S\,0}^{(2)}$,   $\phi _{S\,0}^{(2)}$,  $\chi _{S\,0}^{||(2)}$
through  $Z$ and $F$.

So far,  the  solutions of
 the 2nd-order scalar, vector, and tensor metric perturbations,
 as well as the 2nd-order density contrast,
 have been obtained.
However, in synchronous coordinates,
there are still residual gauge transformations,
and, correspondingly, the  solutions
will  contain 2nd-order residual gauge modes.
In  regard to  applications,
one must  find out these 2nd-order gauge modes.
We shall address this   issue in the next section.

\section{The 2nd-order residual gauge transformations in synchronous coordinates }

The general 2nd-order gauge  transformations of metric perturbations
are generated by  a 1st-order  vector field $\xi^{(1)\mu}$ and
a  2nd-order vector field $\xi^{(2)\mu}$,
which are specified by  Eqs.(\ref{xmutransf})-(\ref{xi_2})
in Appendix C.

Consider the special case
from synchronous-to-synchronous for the dust model
with  $a(\tau) \propto \tau^2$ in this section.
The 1st-order  vector field  $\xi^{(1)\mu } $
is listed in  (\ref {xi_r}), (\ref{alpha1}), (\ref{beta1}), and (\ref{d1}),
and the 1st-order gauge transformation of metric perturbations
is listed in  (\ref {gaugetrchiT})   (\ref {phigauge}) (\ref{phigauge2}) in Appendix C.
We shall give  the corresponding 2nd-order ones for the case of scalar-scalar coupling.
From the general formulas (\ref{alpha2_3}),  (\ref{beta2_1}), and (\ref{d2_2})
for the case  $a \propto \tau^2$,
we get the  2nd-order vector field  $\xi^{(2)\mu }$
in the presence of $\xi^{(1)\mu }$
 as the following
\bl\label{alpha2_2}
\alpha^{(2)}
=\frac{ A^{(2)}(\mathbf x)}{\tau^2} \, ,
\el
\bl\label{beta2_2}
\beta^{(2)}
=&
\nabla^{-2}\Big[
-\frac{20}{3\tau}\varphi^{,k}A^{(1)}_{,k}
-\frac{20}{3\tau}\varphi\nabla^2A^{(1)}
+\frac{2\tau}{3}A^{(1),k}\nabla^2\varphi_{,k}
+\frac{2\tau}{3}\varphi_{,kl}A^{(1),kl}
\nn\\
&
-\frac{2}{\tau}A^{(1),k}\nabla^2C^{||(1)}_{,k}
-\frac{2}{\tau}C^{||(1)}_{,kl}A^{(1),kl}
-\frac{3}{\tau^4}A^{(1)}_{,k}X^{,k}
-\frac{3}{\tau^4}A^{(1),\,kl}\nabla^{-2}X_{,kl}
\Big]
\nn \\
&
-\frac{1}{2\tau^4}A^{(1)}A^{(1)}
+\frac{1}{2\tau^2}A^{(1)}_{,k}A^{(1),k}
-\frac1\tau A^{(2)}
+C^{||(2)}    \,,
\el

\bl\label{d2_1}
d^{(2)}_i
=&
\partial_{i}\nabla^{-2}\Big[
\frac{3}{\tau^4}A^{(1)}_{,k}X^{,k}
+\frac{ 3}{\tau^4}A^{(1),\,kl}\nabla^{-2}X_{,kl}
+\frac{20}{3\tau}\varphi^{,k}A^{(1)}_{,k}
+\frac{20}{3\tau}\varphi\nabla^2A^{(1)}
\nn\\
&
-\frac{2\tau}{3}A^{(1),k}\nabla^2\varphi_{,k}
-\frac{2\tau}{3}\varphi_{,kl}A^{(1),kl}
+\frac{2}{\tau}A^{(1),k}\nabla^2C^{||(1)}_{,k}
+\frac{2}{\tau}C^{||(1)}_{,kl}A^{(1),kl}
\Big]
\nn\\
&
-\frac{3}{\tau^4}A^{(1),k}\nabla^{-2}X_{,ki}
-\frac{2}{\tau} \l[\frac{10 }{3 }\varphi A^{(1)}_{,\,i}
 + C^{||(1)}_{,\,ik} A^{(1),\,k} \r]
+\frac{2\tau}{3}\varphi_{,\,ik} A^{(1),\,k}
+C^{\perp(2)}_{\,i}
.
\el
where   the solutions (\ref{phi1sol}) and (\ref{Dchi1sol}) of scalar perturbations
have  been used,
and the  $\chi^{\top(1)}_{ij } $-dependent parts
have been dropped
as they belong to the scalar-tensor coupling.
In the above, $A^{(1)} ({\bf x}) $ is an arbitrary function,
and $C^{(1)}_i ({\bf x}) $ is  an arbitrary  3-vector
which can be written into two parts:
$C ^{(1)}_{i}  = C\, ^{ \parallel(1)}_{, i}
+  C\, ^{ \bot(1) }_{i}$,
with  $ C^{\parallel (1)}_{,i }  $ being the longitudinal part
and    $C^{\perp(1)}_{i} $   the  transverse part.
See  Eq.(\ref{gi0}) and Eq.(\ref{decompC}) in Appendix C.
Similar for  $A^{(2)} ({\bf x}) $ and $C^{(2)}_i ({\bf x}) $.
The expressions of components of $\xi^{(2)\mu}$  in (\ref {beta2_2}) (\ref {d2_1})
are rather lengthy as they involve the complicated functions
of the 1st-order metric perturbations and components of $\xi^{(1)\mu}$.
This  complication of $\xi^{(2)\mu}$ is caused by
the requirement $\bar g^{(2)}_{0i}=0$  in the presence of $\xi^{(1)\mu}$.
The general formulas
(\ref{phi2transform0}),  (\ref{chi||2transF2}), (\ref{chiPerp2TransF}), and (\ref {chiT2transF2}),
follow the  required residual gauge transformations of 2nd-order metric perturbations
in the Einstein-de Sitter model
\bl   \label{phi2tr}
\bar \phi^{(2)}_S=&\phi^{(2)}_S
-\frac{2}{\tau^6}
\Big[
XA^{(1)}
+A^{(1)}A^{(1)}
\Big]
+\frac{1}{\tau^4}
\Big[
A^{(1)}_{,k}X^{,k}
+3A^{(1),\,kl}\nabla^{-2}X_{,kl}
\nn\\
&
+2A^{(1)}\nabla^2A^{(1)}
+\frac{5}{3}A^{(1)}_{,\,k}A^{(1),k}
\Big]
+\frac{1}{\tau^3}
\Big[
-2X_{,\,k}C^{||(1),k}
-4C^{||(1),\,kl}\nabla^{-2}X_{,kl}
\nn\\
&
-\frac{40}{3}\varphi A^{(1)}
-\frac{8}{3}A^{(1)}\nabla^2C^{||(1)}
-2A^{(1)}_{,\,k}C^{||(1),k}
\Big]
-\frac{1}{3\tau^2}A^{(1)}_{,kl}A^{(1),kl}
\nn\\
&
+\frac{1}{\tau}\Big[
-\frac{2}{3} A^{(1)}\nabla^2\varphi
+\frac{10}{9}\varphi_{,\,k}A^{(1),k}
+\frac{1}{3}C^{||(1),k}\nabla^2A^{(1)}_{,k}
\nn\\
&
-\frac{1}{3} A^{(1),k}\nabla^2C^{||(1)}_{,k}
+\frac{2}{3}A^{(1)}_{,kl}C^{||(1),kl}
\Big]
\nn\\
&
+\Big[
-\frac{20}9\varphi\nabla^2C^{||(1)}
-\frac{10}3\varphi_{,\,k}C^{||(1),k}
-\frac13 C^{||(1),k}\nabla^2C^{||(1)}_{,k}
-\frac23C^{||(1)}_{,kl}C^{||(1),kl}
\Big]\nn\\
&
+\tau\Big[
\frac{1}{3}A^{(1),k}\nabla^2\varphi_{,\,k}
+\frac{4}{9}\varphi_{,kl} A^{(1),kl} \Big]
+\tau^2\Big[
-\frac{1}{9}\nabla^2\varphi_{,\,k}C^{||(1),k}
-\frac{2}{9}\varphi_{,kl} C^{||(1),kl} \Big]
\nn \\
&
+   \frac{2A^{(2)}}{\tau^3}
-\frac{1}{3\tau} \nabla^2A^{(2)}
+\frac13\nabla^2C^{||(2)}
,
\el

\bl\label{chi||2trans2}
{D_{ij} \bar \chi}^{||(2)}_S
&=
D_{ij}\chi^{||(2)}_S
+\frac{1}{\tau^6}D_{ij}\Big[
12\nabla^{-2}\big(A^{(1)}X\big)
+18\nabla^{-2}\nabla^{-2}
\big(
A^{(1),kl}\nabla^{-2}X_{,kl}
-X\nabla^2 A^{(1)}
\big)
\Big]
\nn \\
&
+ \frac{1}{\tau^4}D_{ij}\Big[
-6\nabla^{-2}
 \big(
A^{(1)}_{,k}X^{,k}
+3A^{(1),\,kl}\nabla^{-2}X_{,kl}
\big)
\nn \\
& +18\nabla^{-2}\nabla^{-2} \big(A^{(1),klm}\nabla^{-2}X_{,klm}
-X^{,k}\nabla^2 A^{(1)}_{,k}\big)
+A^{(1)}A^{(1)}
-14\nabla^{-2}\big(A^{(1)}\nabla^2A^{||(1)}\big)  \nn \\
& +21\nabla^{-2}\nabla^{-2}\big(-A^{(1),kl}A^{||(1)}_{,kl}
+\nabla^2A^{(1)}\nabla^2A^{||(1)}
\big)
\Big]
\nn\\
&
+\frac{1}{\tau^3}
D_{ij}\Big[
12C^{||(1),\,k}\nabla^{-2}X_{,k}
-12\nabla^{-2}
\big(
\nabla^{2}C^{||(1),\,k}\nabla^{-2}X_{,k}
\big)
\nn\\
&
+18\nabla^{-2}\nabla^{-2}\big(
X^{,k}\nabla^2C^{||(1)}_{,k}
-C^{||(1),klm}\nabla^{-2}X_{,klm}
\big)
\nn \\
&
+16\nabla^{-2}\big(A^{(1)}\nabla^2C^{||(1)}\big)
+24\nabla^{-2}\nabla^{-2}\big(A^{(1),kl}C^{||(1)}_{,kl}
-\nabla^2A^{(1)}\nabla^2C^{||(1)}\big)
\Big]
\nn  \\
&
+\frac{1}{\tau^2}D_{ij}\Big[ A^{(1)}_{,\,k}A^{(1),k}
-2\nabla^{-2}\big(A^{(1),\,k}\nabla^2A^{(1)}_{,\,k}\big)
\nn\\
&
+3\nabla^{-2}\nabla^{-2}\big(
\nabla^2A^{(1),\,k}\nabla^2A^{(1)}_{,\,k}
-A^{(1),klm}A^{(1)}_{,\,klm}
\big)
\Big]
+\frac{1}{\tau}D_{ij}\Big[
\frac{20 }{3}\varphi A^{(1)}
\nn\\
&
-2A^{(1),k}C^{||(1)}_{,k}
+\frac{4}{3}\nabla^{-2}\big(
-5\varphi\nabla^2A^{(1)}
-4A^{(1)}\nabla^2\varphi
+3A^{(1),\,k}\nabla^2C^{||(1)}_{,\,k}\big)
\nn\\
&
+6\nabla^{-2}\nabla^{-2}\big(
-3\varphi^{,kl}A^{(1)}_{,kl}
+3\nabla^2\varphi\nabla^2A^{(1)}
+A^{(1),klm}C^{||(1)}_{,\,klm}
-\nabla^2A^{(1),\,k}\nabla^2C^{||(1)}_{,\,k}
\big)
\Big]
\nn\\
&
+D_{ij}\Big[
2 C^{||(1),k}C^{||(1)}_{,k}
-\nabla^{-2}\big(-\frac{40}3\varphi\nabla^2C^{||(1)}
+2 C^{||(1)}_{,k}\nabla^2C^{||(1),k}\big)
\nn\\
&
+\nabla^{-2}\nabla^{-2}\big(
-20\nabla^2\varphi\nabla^2C^{||(1)}
+20\varphi^{,kl}C^{||(1)}_{,kl}
+3\nabla^2C^{||(1)}_{,k}\nabla^2C^{||(1),k}
\nn\\
&
-3C^{||(1),klm}C^{||(1)}_{,klm}
\big)
\Big]
+\tau D_{ij}\Big[
-\frac{4}{3} A^{(1),k}\varphi_{,k}
+\nabla^{-2}\big(
\frac43\varphi_{,k}\nabla^2A^{(1),k}
-\frac23A^{(1)}_{,k}\nabla^2\varphi^{,k}
\big)
\nn\\
&
+\nabla^{-2}\nabla^{-2}\big(
-\nabla^2A^{(1)}_{,k}\nabla^2\varphi^{,k}
+A^{(1),klm}\varphi_{,klm}\big)
\Big]
+\tau^2D_{ij}\Big[
\frac{2}{3}C^{||(1),k}\varphi_{,\,k}
\nn\\
&
-\nabla^{-2}\big(\frac23\varphi^{,k}\nabla^2C^{||(1)}_{,k}\big)
+\nabla^{-2}\nabla^{-2}\big(
-\varphi^{,klm}C^{||(1)}_{,klm}
+\nabla^2\varphi^{,k}\nabla^2C^{||(1)}_{,k}
\big)
\Big] \nn \\
&
+ \frac{2}{\tau} D_{ij}A^{(2)}
-2D_{ij}C^{||(2)}
\el
\bl\label{chiPerp2Trans}
&\bar\chi^{\perp(2)}_{S\,ij}
=
\chi^{\perp(2)}_{S\,ij}
+\frac{12}{\tau^6}
\Big[
\partial_i\nabla^{-2}
\Big(
A^{(1),k}\nabla^{-2}X_{,kj}
-A^{(1)}_{,j}X
\Big)
+\partial_i\partial_j\nabla^{-2}\nabla^{-2}
\Big(
X \nabla^2A^{(1)}
\nn\\
&
-A^{(1),kl}\nabla^{-2}X_{,kl}
\Big)
\Big]
+\frac{1}{\tau^4}
\Big[
-\partial_i\Big(
9A^{(1),\,k}\nabla^{-2}X_{,kj}
\Big)
+\partial_i\nabla^{-2}
\Big(
12A^{(1),\,k}X_{,kj}
    \nn\\
    &
-12A^{(1),kl}_{,j}\nabla^{-2}X_{,kl}
\Big)
+\partial_i\partial_j\nabla^{-2}
\Big(
9X_{,k}A^{(1),\,k}
-3 A^{(1),\,kl}\nabla^{-2}X_{,kl}
\Big)
\nn \\
&
+\partial_i\partial_j\nabla^{-2}\nabla^{-2}
\Big(
-12A^{(1),k}\nabla^2X_{,k}
+24\nabla^{-2}X_{,kl}\nabla^2A^{(1),\,kl}
+36A^{(1),klm}\nabla^{-2}X_{,klm}
\Big)
\nn\\
&
+14\partial_i\nabla^{-2}\big(
A^{(1)}_{,j}\nabla^2A^{(1)}\big)
-7\partial_i\partial_j\nabla^{-2}\big(A^{(1),k}A^{(1)}_{,k}\big)
+14\partial_i\partial_j\nabla^{-2}\nabla^{-2}\big(A^{(1),kl}A^{(1)}_{,kl}
\nn\\
&
-\nabla^2A^{(1)}\nabla^2A^{(1)}\big)
\Big]
+\frac{1}{\tau^3}
\Big[
\partial_i\nabla^{-2}
\Big(
12C^{||(1),kl}\nabla^{-2}X_{,klj}
+12\nabla^{-2}X_{,kj}\nabla^2C^{||(1),\,k}
\Big)
\nn\\
&
+\partial_i\partial_j\nabla^{-2}
\Big(
12C^{||(1),kl}\nabla^{-2}X_{,kl}
-12X_{,k}C^{||(1),\,k}
\Big)
+\partial_i\partial_j\nabla^{-2}\nabla^{-2}
\Big(
12C^{||(1),k}\nabla^2X_{,k}
\nn\\
&
-24\nabla^{-2}X_{,kl}\nabla^2C^{||(1),\,kl}
-36C^{||(1),klm}\nabla^{-2}X_{,klm}
\Big)
+16\partial_i\nabla^{-2}\big(A^{(1),k}C^{||(1)}_{,jk}
-A^{(1)}_{,j}\nabla^2C^{||(1)}\big)
\nn\\
&
-16\partial_i\partial_j\nabla^{-2}\nabla^{-2}\big(A^{(1),kl}C^{||(1)}_{,kl}
-\nabla^2A^{(1)}\nabla^2C^{||(1)}\big)
\Big]
 \nn \\
&
+\frac{1}{\tau^2}\Big[
2\partial_i\nabla^{-2}\big(
A^{(1)}_{,jk}\nabla^2A^{(1),k}\big)
-\partial_i\partial_j\nabla^{-2}\big(A^{(1),kl}A^{(1)}_{,kl}\big)
+2\partial_i\partial_j\nabla^{-2}\nabla^{-2}\big(
A^{(1),klm}A^{(1)}_{,klm}
\nn\\
&
-\nabla^2A^{(1)}_{,k}\nabla^2A^{(1),k}\big)
\Big]
+\frac{1}{\tau}\Big[
\frac{4}{3}\partial_i\nabla^{-2}\big(
-5\varphi^{,k}A^{(1)}_{,jk}
-4A^{(1),k}\varphi_{,jk}
+5\varphi_{,j}\nabla^2A^{(1)}
+4A^{(1)}_{,j}\nabla^2\varphi
\nn\\
&
+3A^{(1),kl}C^{||(1)}_{,jkl}
-3A^{(1)}_{,jk}\nabla^2C^{||(1),k}\big)
\nn \\
&
+4\partial_i\partial_j\nabla^{-2}\nabla^{-2}\big(
-3\nabla^2\varphi\nabla^2A^{(1)}
+3A^{(1),kl}\varphi_{,kl}
+\nabla^2A^{(1)}_{,k}\nabla^2C^{||(1),k}
-A^{(1),klm}C^{||(1)}_{,klm}\big)
\Big]
\nn\\
&
+\Big[
2\partial_j\nabla^{-2}\big(
\frac{20}3\varphi^{,k}C^{||(1)}_{,ik}
-\frac{20}3\varphi_{,i}\nabla^2C^{||(1)}
-C^{||(1),kl}C^{||(1)}_{,ikl}
+C^{||(1)}_{,ik}\nabla^2C^{||(1),k}
\big)
\nn\\
&
+2\partial_i\partial_j\nabla^{-2}\nabla^{-2}\big(
-\frac{20}3\varphi^{,kl}C^{||(1)}_{,kl}
+\frac{20}3\nabla^2\varphi\nabla^2C^{||(1)}
+C^{||(1),klm}C^{||(1)}_{,klm}
-\nabla^2C^{||(1),k}\nabla^2C^{||(1)}_{,k}\big)
\Big]
\nn   \\
& +\tau\Big[
\partial_j\nabla^{-2}\big(
\frac{4}{3}A^{(1),kl}\varphi_{,ikl}
-\frac{2}{3}A^{(1),ikl}\varphi_{,kl}
+\frac{2}{3}A^{(1),k}_{,i}\nabla^2\varphi_{,k}
-\frac{4}{3}\varphi^{,k}_{,i}\nabla^2A^{(1)}_{,k}
\big)
\nn\\
&
+\partial_i\partial_j\nabla^{-2}\nabla^{-2}\big(
-\frac{2}{3}A^{(1),klm}\varphi_{,klm}
+\frac{2}{3}\nabla^2A^{(1),k}\nabla^2\varphi_{,k}
\big) \Big]
\nn\\
&
+\tau^2\Big[
\partial_j\nabla^{-2}\big(
-\frac{2}{3}\varphi^{,kl}C^{||(1)}_{,ikl}
+\frac{2}{3}\varphi_{,ik}\nabla^2C^{||(1),k}
\big)
\nn\\
&
+\frac{2}{3}\partial_i\partial_j\nabla^{-2}\nabla^{-2}\big(
\varphi^{,klm}C^{||(1)}_{,klm}
-\nabla^2\varphi^{,k}\nabla^2C^{||(1)}_{,k}\big)
\Big] -C^{\perp(2)}_{i,j}
\nn\\
&
+( i \leftrightarrow   j )
\el
{\large
\bl\label{chiT2trans2}
&
\bar\chi^{\top(2)}_{S\,ij}
=\chi^{\top(2)}_{S\,ij}
+
\frac{1}{\tau^6}
\Big[
12A^{(1)}\nabla^{-2}X_{,ij}
+\delta_{ij}\nabla^{-2}\big(
6A^{(1),kl}\nabla^{-2}X_{,kl}
-6X \nabla^2A^{(1)}
\big)
+12\partial_i\nabla^{-2}
\big(
A^{(1)}_{,j}X
\nn\\
&
-A^{(1),k}\nabla^{-2}X_{,kj}
\big)
+12\partial_j\nabla^{-2}
\big(
A^{(1)}_{,i}X
-A^{(1),k}\nabla^{-2}X_{,ki}
\big)
-\partial_i\partial_j\nabla^{-2}
\big(
12A^{(1)}X
\big)
\nn\\
&
+6\partial_i\partial_j\nabla^{-2}\nabla^{-2}
\big(
A^{(1),kl}\nabla^{-2}X_{,kl}
-X \nabla^2A^{(1)}
\big)
     \Big]
+\frac{1}{\tau^4}
\Big[
8A^{(1)}_{,ij}X
+12A^{(1),\,k}\nabla^{-2}X_{,kij}
\nn\\
&
+\delta_{ij}\nabla^{-2}\big(
6A^{(1),klm}\nabla^{-2}X_{,klm}
-6X_{,k}\nabla^2A^{(1),k}
\big)
+8\nabla^{-2}\big(
X^{,k}_{,i}A^{(1)}_{,\,kj}
+X^{,k}_{,j}A^{(1)}_{,\,ki}
-X_{,ij}\nabla^2A^{(1)}
\nn\\
&
-A^{(1)}_{,ij}\nabla^2X
\big)
+4\partial_i\nabla^{-2}\big(
3A^{(1),kl}_{,j}\nabla^{-2}X_{,kl}
-X_{,kj}A^{(1),k}
-2X\nabla^2A^{(1)}_{,j}
\big)
+4\partial_j\nabla^{-2}
\big(
3A^{(1),kl}_{,i}\nabla^{-2}X_{,kl}
\nn\\
&
-X_{,ki}A^{(1),\,k}
-2X\nabla^2A^{(1)}_{,i}
\big)
+4\partial_i\partial_j\nabla^{-2}
\big(
2X\nabla^2A^{(1)}
-X_{,k}A^{(1),k}
-6A^{(1),\,kl}\nabla^{-2}X_{,kl}
\big)
\nn\\
&
+6\partial_i\partial_j\nabla^{-2}\nabla^{-2}
\big(
A^{(1),klm}\nabla^{-2}X_{,klm}
-X_{,k}\nabla^2A^{(1),k}
\big)
-7\nabla^{-2}\big(
-4A^{(1),k}_{,i}A^{(1)}_{,jk}
+4A^{(1)}_{,ij}\nabla^2A^{(1)}
\nn\\
&
+A^{(1),kl}A^{(1)}_{,kl}\delta_{ij}
-\nabla^2A^{(1)}\nabla^2A^{(1)}\delta_{ij}
\big)
-7\partial_i\partial_j\nabla^{-2}\nabla^{-2}\big(
A^{(1),kl}A^{(1)}_{,kl}
-\nabla^2A^{(1)}\nabla^2A^{(1)}\big)
\Big]
\nn\\
&
+\frac{1}{\tau^3}
\Big[
-8C^{||(1)}_{,ij}X
-12C^{||(1),\,k}\nabla^{-2}X_{,kij}
+\delta_{ij}\nabla^{-2}
\big(
6X_{,k}\nabla^2C^{||(1),\,k}
-6C^{||(1),klm}\nabla^{-2}X_{,klm}
\big)
\nn\\
&
+8\nabla^{-2}
\big(
X_{,ij}\nabla^2C^{||(1)}
+C^{||(1)}_{,ij}\nabla^2 X
-X^{,k}_{,i}C^{||(1)}_{,\,kj}
-X^{,k}_{,j}C^{||(1)}_{,\,ki}
\big)
+4\partial_i\nabla^{-2}
\big(
X_{,kj}C^{||(1),k}
+2X\nabla^2C^{||(1)}_{,j}
\nn\\
&
-3C^{||(1),kl}_{,j}\nabla^{-2}X_{,kl}
\big)
+4\partial_j\nabla^{-2}
\big(
X_{,ki}C^{||(1),k}
+2X\nabla^2C^{||(1)}_{,i}
-3C^{||(1),kl}_{,i}\nabla^{-2}X_{,kl}
\big)
\nn\\
&
+4\partial_i\partial_j\nabla^{-2}
\big(
X_{,k}C^{||(1),k}
-2X\nabla^2C^{||(1)}
+6C^{||(1),kl}\nabla^{-2}X_{,kl}
\big)
+6\partial_i\partial_j\nabla^{-2}\nabla^{-2}
\big(
X_{,k}\nabla^2C^{||(1),\,k}
\nn\\
&
-C^{||(1),klm}\nabla^{-2}X_{,klm}
\big)
+8\nabla^{-2}
\big(
-2A^{(1),k}_{,j}C^{||(1)}_{,ik}
-2A^{(1),k}_{,i}C^{||(1)}_{,jk}
+2C^{||(1)}_{,ij}\nabla^2A^{(1)}
+2A^{(1)}_{,ij}\nabla^2C^{||(1)}
\nn\\
&
+A^{(1),kl}C^{||(1)}_{,kl}\delta_{ij}
-\nabla^2A^{(1)}\nabla^2C^{||(1)}\delta_{ij}
\big)
+8\partial_i\partial_j\nabla^{-2}\nabla^{-2}\big(
A^{(1),kl}C^{||(1)}_{,kl}
-\nabla^2A^{(1)}\nabla^2C^{||(1)}\big)
\Big]
\nn\\
&
+\frac{1}{\tau^2}\Big[
\nabla^{-2}\big(
4A^{(1),kl}_{,i}A^{(1)}_{,jkl}
-4A^{(1)}_{,ijk}\nabla^2A^{(1),k}
-A^{(1),klm}A^{(1)}_{,klm}\delta_{ij}
+\nabla^2A^{(1)}_{,k}\nabla^2A^{(1),k}\delta_{ij}
\big)
 \nn \\
&
+\partial_i\partial_j\nabla^{-2}\nabla^{-2}\big(
-A^{(1),klm}A^{(1)}_{,klm}
+\nabla^2A^{(1)}_{,k}\nabla^2A^{(1),k}\big)
\Big]
+\frac{1}{\tau}\nabla^{-2}\Big[
2\big(6\varphi^{,k}_{,j}A^{(1)}_{,ik}
+6\varphi^{,k}_{,i}A^{(1)}_{,jk}
\nn \\
&
-6A^{(1)}_{,ij}\nabla^2\varphi
-6\varphi_{,ij}\nabla^2A^{(1)}
-3A^{(1),kl}\varphi_{,kl}\delta_{ij}
+3\nabla^2A^{(1)}\nabla^2\varphi\delta_{ij}
-2A^{(1),kl}_{,j}C^{||(1)}_{,ikl}
-2A^{(1),kl}_{,i}C^{||(1)}_{,jkl}
\nn\\
&
+2C^{||(1)}_{,ijk}\nabla^2A^{(1),k}
+2A^{(1)}_{,ijk}\nabla^2C^{||(1),k}
+A^{(1),klm}C^{||(1)}_{,klm}\delta_{ij}
-\nabla^2A^{(1)}_{,k}\nabla^2C^{||(1),k}\delta_{ij}
\big)
\nn\\
&
+2\partial_i\partial_j\nabla^{-2}\big(
-3A^{(1),kl}\varphi_{,kl}
+3\nabla^2A^{(1)}\nabla^2\varphi
+A^{(1),klm}C^{||(1)}_{,klm}
-\nabla^2A^{(1)}_{,k}\nabla^2C^{||(1),k}\big)
\Big]
\nn\\
&
+\Big[
\nabla^{-2}\big(
-\frac{40}3\varphi^{,k}_{,j}C^{||(1)}_{,ik}
-\frac{40}3\varphi^{,k}_{,i}C^{||(1)}_{,jk}
+\frac{40}3 C^{||(1)}_{,ij}\nabla^2\varphi
+\frac{40}3\varphi_{,ij}\nabla^2C^{||(1)}
+\frac{20}3\varphi^{,kl}C^{||(1)}_{,kl}\delta_{ij}
\nn\\
&
-\frac{20}3\nabla^2\varphi\nabla^2C^{||(1)}\delta_{ij}
+4C^{||(1),kl}_{,j}C^{||(1)}_{,ikl}
-4C^{||(1)}_{,ijk}\nabla^2C^{||(1),k}
-C^{||(1),klm}C^{||(1)}_{,klm}\delta_{ij}
\nn\\
&
+\nabla^2C^{||(1),k}\nabla^2C^{||(1)}_{,k}\delta_{ij}
\big)
+\partial_i\partial_j\nabla^{-2}\nabla^{-2}\big(
\frac{20}3\varphi^{,kl}C^{||(1)}_{,kl}
-\frac{20}3\nabla^2\varphi\nabla^2C^{||(1)}
+\nabla^2C^{||(1),k}\nabla^2C^{||(1)}_{,k}
\nn\\
&
-C^{||(1),klm}C^{||(1)}_{,klm}
\big)
\Big]
+\tau \Big[
\nabla^{-2}\big(
-\frac{2}{3}A^{(1),kl}_{,j}\varphi_{,ikl}
-\frac{2}{3}A^{(1),kl}_{,i}\varphi_{,jkl}
+\frac{2}{3}A^{(1)}_{,ijk}\nabla^2\varphi^{,k}
+\frac{2}{3}\varphi_{,ijk}\nabla^2A^{(1),k}
\nn \\
&
+\frac{1}{3}A^{(1),klm}\varphi_{,klm}\delta_{ij}
-\frac{1}{3}\nabla^2A^{(1),k}\nabla^2\varphi_{,k}\delta_{ij}
\big)
+\partial_i\partial_j\nabla^{-2}\nabla^{-2}\big(
-\frac{1}{3}\nabla^2A^{(1),k}\nabla^2\varphi_{,k}
+\frac{1}{3}A^{(1),klm}\varphi_{,klm}
\big) \Big]
\nn\\
&
+\tau^2\Big[
\nabla^{-2}\big(\frac{2}{3}\varphi^{,kl}_{,j}C^{||(1)}_{,ikl}
+\frac{2}{3}\varphi^{,kl}_{,i}C^{||(1)}_{,jkl}
-\frac{2}{3}C^{||(1)}_{,ijk}\nabla^2\varphi^{,k}
-\frac{2}{3}\varphi_{,ijk}\nabla^2C^{||(1),k}
-\frac{1}{3}\varphi^{,klm}C^{||(1)}_{,klm}\delta_{ij}
\nn\\
&
+\frac{1}{3}\nabla^2\varphi^{,k}\nabla^2C^{||(1)}_{,k}\delta_{ij}
\big)
+\frac{1}{3}\partial_i\partial_j\nabla^{-2}\nabla^{-2}\big(
\nabla^2\varphi^{,k}\nabla^2C^{||(1)}_{,k}
-\varphi^{,klm}C^{||(1)}_{,klm}
\big)  \Big] .
\el
}
The  transformation formulas  (\ref {phi2tr})--(\ref {chiT2trans2})
are   lengthy because of those   terms  brought about by   $\xi^{(1)\mu}$.
Eq.(\ref {chiT2trans2}) tells  us that
the transformation of 2nd order tensor
involves only  $\xi^{(1)\mu}$,
 independent of  the 2nd-order vector field $\xi^{(2)\mu}$.

The synchronous-to-synchronous transformation of 2nd-order
          density perturbation is derived
by applying  (\ref {metricTrans2nd3})  to the 2nd order  $T^{(2)}_{00 } $,
\be \label{T2tranfs}
\bar{T}^{(2)}_{00}
 =
T^{(2)}_{00}
-2\mathcal{L}_{\xi^{(1)}} T^{(1)}_{00 }
+\mathcal{L}_{\xi^{(1)}}\l(\mathcal{L}_{\xi^{(1)}}T^{(0)}_{00 } \r)
-\mathcal{L}_{\xi^{(2)}} T^{(0)}_{00} .
\ee
Up to the 2nd order, one calculates    $T^{(2)}_{00}  = \rho^{(2)} a^2 $,
$\bar U^{(2)}_0 = -\frac{A^{(1),i } A^{(1)} _{,i}}{\tau^2 } $,
$\bar T^{(2)}_{00}= a^2\bar\rho^{(2)} +2\rho^{(0)}A^{(1),i }  A^{(1)}  _{,i}$.
Plugging these into (\ref {T2tranfs}) leads to the result
\bl \label{delt2tr}
\bar  \rho ^{(2)}_S
=&  \rho ^{(2)}_S
-2  \rho^{(1)} _{,0} \frac{A^{( 1 )}}{\tau^2}
-2   \rho^{(1)} _{,k }
    \l( -\frac{A^{(1),k}}{\tau} +C^{||(1),k}  \r)
      \nn \\
& +  54 \rho^{(0)}   \frac{A^{( 1 )}A^{( 1 )} }{\tau^6}
  -  6 \rho^{(0)}   \frac{A^{(1)}_{, \, k}}{\tau^3}
            \l( -\frac{A^{(1),k}}{\tau} +C^{||(1),k} \r)
            + 6 \rho^{(0)} \frac{A^{( 2 )}}{\tau^3},
\el
This result  can be also derived by applying (\ref{f2Trans})
to $\rho ^{(2)}$ as a scalar function,
\bl \label{delt2tr2}
\bar  \rho ^{(2)}  & =   \rho ^{(2)}
-2\mathcal{L}_{\xi^{(1)}}    \rho^{(1)}
+\mathcal{L}_{\xi^{(1)}}\l(\mathcal{L}_{\xi^{(1)}} \rho ^{(0)} \r)
-\mathcal{L}_{\xi^{(2)}} \rho ^{(0)}   .
\el
(\ref {delt2tr}) is also written in terms of  the density contrast
\bl  \label{delta2transfmt}
\bar \delta   ^{(2)}_S = & \delta  ^{(2)}_S
 + \l( \frac{ 12 }{ \tau } \delta^{(1)}
 - 2 \delta^{(1)}_{, \, 0}   \r)\frac{A^{(1)}}{\tau^2}
-2  \delta^{(1)}_{, \, k}
   \l( -\frac{A^{(1),k}}{\tau} +C^{||(1),k}  \r)
\nn \\
&
      + 54 \frac{ A^{( 1 )}  A^{( 1 )}  }{\tau^ 6}
 -  6   \frac{A^{(1)}_{, \, k}}{\tau^3}
            \l( -\frac{A^{(1),\,k}}{\tau} +C^{||(1),k} \r)
      + 6  \frac{A^{( 2 )}}{\tau^3 } \, .
\el

So far in this paper,
the  perturbations of 4-velocity
are  taken to be zero,
$U^{(1)\mu} = U^{(2)\mu}= 0 $,
for the  dust model.
When  one also requires in the new synchronous coordinate $\bar x^\mu$
the transformed 3-velocity is zero $\bar U^{(1) i } =0$, \cite{Russ1996},
one gets an  extra constraint on the transformation vector field
\be\label{A1}
 A^{(1)} ({\bf x})  _{,i}  =0,
 \ee
 i.e,  $A^{(1)} = const$. [See Eq.(\ref{A1const}) in Appendix C]

When  one further requires the transformed 2nd-order perturbed velocity
$\bar U^{(2) i } =0 $ by using  (\ref{vect2transf}),
this  leads  to
$ \mathcal{L}_{\xi^{(2)}}  U^{(0)i}  =0  $,
which gives another  constraint
\be \label{A2}
 A^{(2)} ({\bf x})  _{,i} =0,
 \ee
 i.e, $A^{(2)}= const$.
 Under the condition  (\ref{A1}) and (\ref{A2}),
the components of $\xi^{(2)\mu }$ in  (\ref{alpha2_2}),  (\ref{beta2_2}), and (\ref{d2_1})
are much simplified as the following
\bl
\alpha^{(2)}
= & \frac{ A^{(2)} }{\tau^2} \, , \\
\beta^{(2)} =  & -\frac{1}{2\tau^4}A^{(1)}A^{(1)}
        -\frac1\tau A^{(2)}  +C^{||(2)} ({\bf x})   \,, \\
d^{(2)}_i =  & C^{\perp(2)}_{\,i}({\bf x}) ,
\el
and
(\ref {phi2tr}) (\ref{chi||2trans2}),  (\ref{chiPerp2Trans}), and (\ref{chiT2trans2})
substantially reduce to
\bl  \label{phi2trA}
\bar \phi^{(2)}_S=&\phi^{(2)}_S
-\frac{2}{\tau^6}\Big[XA^{(1)}+A^{(1)}A^{(1)}\Big]
\nn\\
&
+\frac{1}{\tau^3}\Big[
-2X_{,\,k}C^{||(1),k}
-4C^{||(1),\,kl}\nabla^{-2}X_{,kl}
-\frac{40}{3}\varphi A^{(1)}
-\frac{8}{3}A^{(1)}\nabla^2C^{||(1)}
\Big]
\nn\\
&
+\frac{1}{\tau}\Big[
-\frac{2}{3} A^{(1)}\nabla^2\varphi
\Big]
+\Big[
-\frac{20}9\varphi\nabla^2C^{||(1)}
-\frac{10}3\varphi_{,\,k}C^{||(1),k}
-\frac13 C^{||(1),k}\nabla^2C^{||(1)}_{,k}
\nn\\
&
-\frac23C^{||(1)}_{,kl}C^{||(1),kl}
\Big]
+\tau^2\Big[
-\frac{1}{9}\nabla^2\varphi_{,\,k}C^{||(1),k}
-\frac{2}{9}\varphi_{,kl} C^{||(1),kl} \Big]
\nn \\
&
+ 2 \frac{ A^{( 2)}}{\tau^3 }
  + \frac{1}{3}\nabla^2 C^{||(   2)} ({\bf x}) ,
\el
\bl   \label{chi||2trA}
D_{ij}\bar\chi^{||(2)}_S
&=
D_{ij}\chi^{||(2)}_S
+\frac{1}{\tau^6}D_{ij}\Big[
12\nabla^{-2}\big(A^{(1)}X\big)
\Big]
\nn\\
&
+\frac{1}{\tau^3}
D_{ij}\Big[
12C^{||(1),\,k}\nabla^{-2}X_{,k}
-12\nabla^{-2}
\big(
\nabla^{2}C^{||(1),\,k}\nabla^{-2}X_{,k}
\big)
\nn\\
&
+18\nabla^{-2}\nabla^{-2}\big(
X^{,k}\nabla^2C^{||(1)}_{,k}
-C^{||(1),klm}\nabla^{-2}X_{,klm}
\big)
 +16   A^{(1) }  C^{||(1)}
\Big]
\nn\\
&
+\frac{1}{\tau}D_{ij}\Big[
\frac{20 }{3}\varphi A^{(1)}
+\frac{4}{3}\nabla^{-2}\big(
-4A^{(1)}\nabla^2\varphi
\big)
\Big]
\nn\\
&
+D_{ij}\Big[
2 C^{||(1),k}C^{||(1)}_{,k}
-\nabla^{-2}\big(-\frac{40}3\varphi\nabla^2C^{||(1)}
+2 C^{||(1)}_{,k}\nabla^2C^{||(1),k}\big)
\nn\\
&
+\nabla^{-2}\nabla^{-2}\big(
-20\nabla^2\varphi\nabla^2C^{||(1)}
+20\varphi^{,kl}C^{||(1)}_{,kl}
+3\nabla^2C^{||(1)}_{,k}\nabla^2C^{||(1),k}
\nn\\
&
-3C^{||(1),klm}C^{||(1)}_{,klm}
\big)
\Big]
+\tau^2D_{ij}\Big[
\frac{2}{3}C^{||(1),k}\varphi_{,\,k}
-\nabla^{-2}\big(\frac23\varphi^{,k}\nabla^2C^{||(1)}_{,k}\big)
\nn\\
&
+\nabla^{-2}\nabla^{-2}\big(
-\varphi^{,klm}C^{||(1)}_{,klm}
+\nabla^2\varphi^{,k}\nabla^2C^{||(1)}_{,k}
\big)
\Big]
 -2D_{ij}C^{||(2)}
,
\el
\bl   \label{chiperp2trA}
\bar\chi^{\perp(2)}_{S\,ij}
&=\chi^{\perp(2)}_{S\,ij}
+\frac{1}{\tau^3}
\Big[
\partial_i\nabla^{-2}
\Big(
12C^{||(1),kl}\nabla^{-2}X_{,klj}
+12\nabla^{-2}X_{,kj}\nabla^2C^{||(1),\,k}
\Big)
\nn\\
&
+\partial_i\partial_j\nabla^{-2}
\Big(
12C^{||(1),kl}\nabla^{-2}X_{,kl}
-12X_{,k}C^{||(1),\,k}
\Big)
\nn\\
&
+\partial_i\partial_j\nabla^{-2}\nabla^{-2}
\Big(
12C^{||(1),k}\nabla^2X_{,k}
-24\nabla^{-2}X_{,kl}\nabla^2C^{||(1),\,kl}
\nn\\
&
-36C^{||(1),klm}\nabla^{-2}X_{,klm}
\Big)
\Big]
\nn\\
&
+\Big[
2\partial_j\nabla^{-2}\big(
\frac{20}3\varphi^{,k}C^{||(1)}_{,ik}
-\frac{20}3\varphi_{,i}\nabla^2C^{||(1)}
-C^{||(1),kl}C^{||(1)}_{,ikl}
+C^{||(1)}_{,ik}\nabla^2C^{||(1),k}
\big)
\nn\\
&
+2\partial_i\partial_j\nabla^{-2}\nabla^{-2}\big(
-\frac{20}3\varphi^{,kl}C^{||(1)}_{,kl}
+\frac{20}3\nabla^2\varphi\nabla^2C^{||(1)}
+C^{||(1),klm}C^{||(1)}_{,klm}    \nn   \\
& -\nabla^2C^{||(1),k}\nabla^2C^{||(1)}_{,k}\big)
\Big]
+\frac{2}{3}\tau^2\Big[
\partial_j\nabla^{-2}\big(
\varphi_{,ik}\nabla^2C^{||(1),k}
-\varphi^{,kl}C^{||(1)}_{,ikl}
\big)
\nn \\
&
+\partial_i\partial_j\nabla^{-2}\nabla^{-2}\big(
\varphi^{,klm}C^{||(1)}_{,klm}
-\nabla^2\varphi^{,k}\nabla^2C^{||(1)}_{,k}\big)   \Big]
-C^{\perp(2)}_{i,j}  \nn \\
&
+\big( i \leftrightarrow j  \big)
\el
\bl  \label{chitop2trA}
\bar\chi^{\top(2)}_{S\,ij}
& =\chi^{\top(2)}_{S\,ij}
+\frac{1}{\tau^6}
\Big[
12A^{(1)}\nabla^{-2}X_{,ij}
-\partial_i\partial_j\nabla^{-2}
\Big(
12A^{(1)}X
\Big)
     \Big]
\nn\\
&
+\frac{1}{\tau^3}
\Big[
-8C^{||(1)}_{,ij}X
-12C^{||(1),\,k}\nabla^{-2}X_{,kij}
\nn\\
&
+\delta_{ij}\nabla^{-2}
\Big(
6X_{,k}\nabla^2C^{||(1),\,k}
-6C^{||(1),klm}\nabla^{-2}X_{,klm}
\Big)
\nn\\
&
+8\nabla^{-2}
\Big(
X_{,ij}\nabla^2C^{||(1)}
+C^{||(1)}_{,ij}\nabla^2 X
-X^{,k}_{,i}C^{||(1)}_{,\,kj}
-X^{,k}_{,j}C^{||(1)}_{,\,ki}
\Big)
\nn\\
&
+4\partial_i\nabla^{-2}
\Big(
X_{,kj}C^{||(1),k}
+2X\nabla^2C^{||(1)}_{,j}
-3C^{||(1),kl}_{,j}\nabla^{-2}X_{,kl}
\Big)
\nn\\
&
+4\partial_j\nabla^{-2}
\Big(
X_{,ki}C^{||(1),k}
+2X\nabla^2C^{||(1)}_{,i}
-3C^{||(1),kl}_{,i}\nabla^{-2}X_{,kl}
\Big)
\nn\\
&
+4\partial_i\partial_j\nabla^{-2}
\Big(
X_{,k}C^{||(1),k}
-2X\nabla^2C^{||(1)}
+6C^{||(1),kl}\nabla^{-2}X_{,kl}
\Big)
\nn\\
&
+6\partial_i\partial_j\nabla^{-2}\nabla^{-2}
\Big(
X_{,k}\nabla^2C^{||(1),\,k}
-C^{||(1),klm}\nabla^{-2}X_{,klm}
\Big)
\Big]
\nn\\
&
+\Big[
\nabla^{-2}\big(
-\frac{40}3\varphi^{,k}_{,j}C^{||(1)}_{,ik}
-\frac{40}3\varphi^{,k}_{,i}C^{||(1)}_{,jk}
+\frac{40}3 C^{||(1)}_{,ij}\nabla^2\varphi
+\frac{40}3\varphi_{,ij}\nabla^2C^{||(1)}
\nn\\
&
+\frac{20}3\varphi^{,kl}C^{||(1)}_{,kl}\delta_{ij}
-\frac{20}3\nabla^2\varphi\nabla^2C^{||(1)}\delta_{ij}
+4C^{||(1),kl}_{,j}C^{||(1)}_{,ikl} -4C^{||(1)}_{,ijk}\nabla^2C^{||(1),k}
\nn\\
&
-C^{||(1),klm}C^{||(1)}_{,klm}\delta_{ij}
+\nabla^2C^{||(1),k}\nabla^2C^{||(1)}_{,k}\delta_{ij}
\big)
\nn\\
&
+\partial_i\partial_j\nabla^{-2}\nabla^{-2}\big(
-\frac{20}3\nabla^2\varphi\nabla^2C^{||(1)}
+\frac{20}3\varphi^{,kl}C^{||(1)}_{,kl}
+\nabla^2C^{||(1),k}\nabla^2C^{||(1)}_{,k} \nn \\
& -C^{||(1),klm}C^{||(1)}_{,klm}
\big)
\Big]
+\tau^2\Big[
\nabla^{-2}\big(\frac{2}{3}\varphi^{,kl}_{,j}C^{||(1)}_{,ikl}
+\frac{2}{3}\varphi^{,kl}_{,i}C^{||(1)}_{,jkl}
-\frac{2}{3}C^{||(1)}_{,ijk}\nabla^2\varphi^{,k} \nn \\
&
-\frac{2}{3}\varphi_{,ijk}\nabla^2C^{||(1),k}
-\frac{1}{3}\varphi^{,klm}C^{||(1)}_{,klm}\delta_{ij}
+\frac{1}{3}\nabla^2\varphi^{,k}\nabla^2C^{||(1)}_{,k}\delta_{ij}
\big) \nn \\
&
+\frac{1}{3}\partial_i\partial_j\nabla^{-2}\nabla^{-2}\big(
-\varphi^{,klm}C^{||(1)}_{,klm}
+\nabla^2\varphi^{,k}\nabla^2C^{||(1)}_{,k}\big)  \Big] ,
\el
and   Eq.(\ref {delt2tr}) and Eq.(\ref {delta2transfmt}) reduce to
\bl \label{rho2transf}
\bar \rho^{(2)}_S   =  & \rho^{(2)}_S
     -2  \rho^{(1)'}  \frac{A^{(1)}}{\tau^2 }
     -2  \rho^{(1)}_{,k}  C^{|| (1),k}
     +    54 \frac{A^{(1)}  A^{(1)} }{\tau^6 } \rho^{(0)}
     +  6  \frac{  A^{(2)}}{\tau^3} \rho^{(0)}  \, .
\el
\bl  \label{delta2transf}
\bar \delta   ^{(2)} _S&  = \delta  ^{(2)}_S
 + ( \frac{ 12 }{ \tau } \delta^{(1)}
 - 2 \delta^{(1)}_{, \, 0}   )\frac{A^{(1)}}{\tau^2}
-2  \delta^{(1)}_{, \, k}
    C^{||(1),k}
      + 54 \frac{ A^{( 1 )}  A^{( 1 )}  }{\tau^ 6}
      + 6  \frac{A^{( 2 )}}{\tau^3 } \, .
\el

The above results of synchronous-to-synchronous transformations are general ones,
in the sense that
two vector fields $\xi^{(1)\mu}$ and $\xi^{(2)\mu}$ are involved simultaneously.
However, in applications,
certain distinctions should be made
in regard to transformations due to $\xi^{(1)\mu}$ and $\xi^{(2)\mu}$.
If one sets $\xi^{(2)\mu}=0$  \cite{Abramo1997,HwangNoh2012},
only  $\xi^{(1)\mu}  $ remains,
which ensures $\bar g^{(1)}_{00}=0$,  $\bar g^{(1)}_{0i}=0$
and the obtained 1st-order perturbations unchanged
in the fixed 1st-order synchronous coordinate,
one has no freedom
to make  $\bar g^{(2)}_{00}=0$ and  $\bar g^{(2)}_{0i}=0$ anymore,
because   $\xi^{(1)\mu}$ has been already fixed,
as specified by (\ref {xi0trans}) and (\ref{gi0}).
Thus,  effective 2nd-order transformations
from synchronous to synchronous
can not be made when   $\xi^{(2)\mu}=0$.
On the other hand,
if  the 1st-order solutions are held fixed
and only the 2nd-order  metric perturbations are transformed \cite{Gleiser1996},
one simply sets $\xi^{(1)\mu}=0$ but $\xi^{(2)\mu} \ne 0$,
so that   (\ref{alpha2_2}), (\ref{beta2_2}), and (\ref{d2_1})   reduce to
\bl \label{alpha2}
 \alpha^{(2)} (\tau,\mathbf x)  = &  \frac{A^{(2)}(\mathbf x)}{\tau^2}  \\
 \label{beta2}
\beta^{( 2)}_{,\,i}  (\tau,\mathbf x)  =  & -\frac{A^{( 2)}(\mathbf x)_{,\,i}}{\tau}
         +C^{||( 2)} (\mathbf x)_{,\,i}  \\
 \label{d2}
d^{(2)}_i  = & 0 ,
\el
and
(\ref{phi2tr}), (\ref{chi||2trans2}), (\ref{chiPerp2Trans}),
           (\ref{chiT2trans2}), and (\ref{delta2transf})
reduce to
\bl  \label{2phigauge0}
\bar \phi^{(2)}_S=& \phi^{(2)}_S
+\frac{2}{\tau^3} A^{(2)}
-\frac{1}{3\tau} \nabla^2A^{(2)}
+\frac13\nabla^2C^{||(2)} \, , \\
\label{2chi||0}
D_{ij}\bar\chi^{||(2)}_S =&
  D_{ij}\chi^{||(2)}_S
+\frac{2}{\tau}D_{ij}A^{(2)}
-2D_{ij}C^{||(2)} \,, \\
\label{2chiperp0}
\bar\chi^{\perp(2)}_{S\,ij}
= & \chi^{\perp(2)}_{S\,ij}
-\Big[ C^{\perp(2)}_{ i,j} +C^{\perp(2)}_{ j,i} \Big] \,, \\
 \label{2chitop0}
\bar\chi^{\top(2)}_{S\,ij}  = & \chi^{\top(2)}_{S\,ij} \, , \\
\bar \delta   ^{(2)} _S = & \delta   ^{(2)} _S+6   \frac{A^{(2)}} {\tau^3} \, .
   \label{2deltatr}
\el
Thus, it is seen that only  $\xi^{(2)\mu}$ is effective
in carrying out 2nd-order transformations that we consider,
because $\xi^{(1)\mu}$  has been used in obtaining the 1st-order perturbations.
This is also consistent with the fact that
one can have 4 degrees of freedom at each order of transformation.
The  gauge transformations  (\ref {2phigauge0})-- (\ref{2deltatr})
have a similar structure to the 1st-order gauge transformations
 in  (\ref {gaugetrchiT}), (\ref {phigauge}),
  (\ref {phigauge2}), and (\ref  {delta1transf}) in Appendix C.
In particular,
the gauge mode of 2nd-order vector   is   time-independent,
and the  2nd-order tensor   contains  no gauge mode.
Furthermore,  under  the condition  $\bar U^{(2) i } =0 $
(i.e, $A^{(2)}_{,\, i} =0$),
 (\ref{alpha2}) (\ref {d2})
 (\ref {2chiperp0}) (\ref {2chitop0}) (\ref {2deltatr}) remain unchanged,
whereas  (\ref{beta2}), (\ref{2phigauge0}), and  (\ref{2chi||0})
 reduce  to
\bl
\beta^{(2)}_{,\,i}  (\mathbf x) = &  C^{||(  2)} (\mathbf x)_{,\,i} \, , \\
\bar  \phi^{( 2)}_S   = & \phi^{( 2)}_S  +   2 \frac{ A^{( 2)}}{\tau^3 }
          + \frac{1}{3}\nabla^2 C^{||(   2)} ({\bf x})   , \label{phi2gauge}  \\
D_{ij}\bar\chi^{||(2)}_S
       =&    D_{ij}\chi^{||(2)}_S  -2D_{ij}C^{||(2)} (\bf x)  .  \label{Dij2gauge}
\el

Though having residual gauge freedom,
synchronous gauges have  advantages in regard to interpretation
of cosmological observations in terms of calculational results.
The RW spacetime as given by   Eq.(\ref {18q1})
in synchronous coordinate
can be written as $ds^2= - dt^2+ a^2(t) \gamma_{ij}dx^i dx^j $,
where $dt\equiv  a(\tau)d \tau $ is the cosmic time,
and is equal to the proper time   $ ds/c= \sqrt{- g_{00}  dt^2}  $
 measured by a comoving observer ($dx^i  =0$).
For  galaxies without peculiar velocity
(or, the typical peculiar velocity $v\sim 10^2$km/s $\sim 10^{-3}$c
is neglected as an approximation),
an observer   on a galaxy is a  comoving observer.
This holds for a perturbed RW spacetime
as $g^{(1)}_{00} = g^{(2)}_{00}=0$ by definition of synchronous coordinate.
Therefore, the time $\tau$ occurring  in
perturbed cosmological quantities
is  directly related to the proper time $t$ used  by the observer on our Galaxy.

On the other  hand,
in the Poisson coordinates (conformal-Newtonian coordinates) (see Appendix D)
for the perturbed RW spacetime,
the proper time of a comoving observer is
 $ds/c  =\sqrt{(1 +2\psi   )d t_P ^2} $ ,
where  $dt_P$ is the time in Poisson coordinates
and $\psi =\psi (t_P, {\bf x} )$ is the lapse function,
which is also the gravitational potential at the  position of observer.
Thus,  the proper time measured by a comoving observer on our Galaxy
is  not equal to the  time  $ dt_P  $
  appearing in the  perturbed cosmological quantities
  in Poisson gauges.
In principle,
the observer  would need to know
the potential $\psi (t_P, {\bf x} )$ at  the Galaxy,
in order to relate his  proper time to the  Poisson coordinate time $t_P$ .

\section{Conclusion}

We have conducted a comprehensive study of the 2nd-order perturbed  Einstein equation
for the Einstein-de Sitter model
in synchronous coordinates.
There are  three types of couplings of 1st-order metric perturbations:
scalar-scalar,  scalar-tensor, and    tensor-tensor,
which serve as the source for 2nd-order metric perturbations.
We have  decomposed the 2nd-order perturbed  Einstein  equation
into three sets according to the couplings.
For  the scalar-scalar  coupling   in this paper,
we  have obtained the solutions of the 2nd-order scalar, vector and  tensor perturbations
  for general initial conditions.
In particular,
the 2nd-order vector perturbations  in Eq.(\ref{chi2Sperp4})  are  produced
by the  coupling of 1st-order scalar perturbations,
even though  there is no  1st-order vector metric perturbation.
Besides, the complete expression of 2nd-order
tensor metric perturbation is given in Eq.(\ref{chi2S2})
and Eq.(\ref{tensor22}),
which  includes  some extra terms as well as the   homogeneous solution,
correcting  that in literature.

Moreover, we have  also performed a detailed study of
general synchronous-to-synchronous  2nd-order gauge transformations,
which are generated by both  a 1st-order vector field $\xi^{(1)\mu}$
and   an independent  2nd-order vector field  $\xi^{(2)\mu}$ as well.
While  $\xi^{(1)\mu}$ is actually fixed for  the usual 1st-order gauge transformation,
$\xi^{(2)\mu}$
is effective for carrying out the required 2nd-order transformations.
As a main result,
the residual gauge modes of 2nd-order metric perturbations and density contrast
have been found   explicitly,
    listed in (\ref{phi2transform0}), (\ref{chi||2transF2}), (\ref{chiPerp2TransF}), and (\ref {chiT2transF2})
    for a general RW spacetime,
and in (\ref {phi2tr}), (\ref {chi||2trans2}),
(\ref{chiPerp2Trans}),  (\ref{chiT2trans2}),  and (\ref{delta2transfmt})
for  the Einstein-de Sitter model.
We have also found that,
when one further requires that the transformed  3-velocity perturbations
are also zero,   $\bar U^{(1)i} =\bar U^{(2)i} =0$,
the vector fields $\xi^{(1)\mu} $ and $\xi^{(2)\mu} $
are consequently constrained,
and the residual gauge modes of   perturbations are substantially  reduced,
 which are   listed in (\ref{phi2trA}), (\ref{chi||2trA}),
(\ref {chiperp2trA}), (\ref{chitop2trA}),
and   (\ref {delta2transf}).
Furthermore,
if we  fix  the 1st-order solutions of perturbations ($\xi^{(1)\mu}=0$)
and  transform only the 2nd-order  perturbations  by $\xi^{(2)\mu} $,
all the things become  very simple
and the resulting  formulas are listed in (\ref {alpha2}) --   (\ref {Dij2gauge}),
which  have   exactly the same  structure as  the 1st-order transformations.
In particular,
the 2nd-order vector  contains
a time-independent, residual gauge mode,
and the 2nd-order tensor contains no residual gauge mode.

There are several related issues that can be explored in future research.
Firstly,
we can extend the above method to the case of
the scalar-tensor and tensor-tensor types of couplings of 1st-order metric perturbations
as the source for the 2nd-order metric perturbations.
This will be more involved and  will be presented in a subsequent paper.
Next, one can extend the work to other stages of expansion,
such as  the radiation  and inflationary stages,
and the  calculations would be  more involved.
For these earlier stages,
the scalar-tensor  and tensor-tensor couplings
will be more significant than for the  matter stage.
Besides,
based on our results of the 2nd-order perturbations with scalar-scalar coupling,
a detailed computation of the evolution
during the matter stage in a realistic Big-Bang cosmology
will be  a whole new project in the future.

\

\textbf{Acknowledgements}

Yang  Zhang is supported by NSFC Grant No. 11275187,
NSFC 11421303, 11675165, 11633001
SRFDP, and CAS,
the Strategic Priority Research Program
``The Emergence of Cosmological Structures"
of the Chinese Academy of Sciences, Grant No. XDB09000000.

\newpage

\appendix

\numberwithin{equation}{section}

\section{The 1st order perturbations}

The formulas of the Ricci tensor are
$R _{ \mu\nu} =\Gamma^{\alpha}_{ \mu\nu,\alpha}-\Gamma^{\alpha}_{\nu\alpha,\mu}
+\Gamma^{\alpha}_{\lambda\alpha}\Gamma^{\lambda}_{\mu\nu}
-\Gamma^{\alpha}_{\lambda\nu}\Gamma^{\lambda}_{\alpha\mu}$,
and
$\Gamma^{\alpha}_{\beta\gamma}=
\frac{1}{2}g^{\alpha\rho}(g_{\rho\gamma,\beta}+g_{\beta\rho, \gamma}
- g_{\beta\gamma,\rho})$.
In the 0th order,
$R^{(0)}_{00}=-\frac{3a''}{a}+3\l(\frac{a'}{a}\r)^2$,
$R^{(0)}_{ij}=\delta_{ij} \l[\frac{a''}{a}+ (\frac{a'}{a} )^2\r]$,
$R^{(0)}=\frac{6}{a^2} \frac{a''}{a}$.
The 1st-order Ricci tensor
$R^{(1) }_{00}=   3\phi^{(1)''} +3\frac{a'}{a}\phi^{(1)'} $,
$R^{(1)}_{0i}= 2\phi^{(1)' }_{,i}+ \frac{1}{2} D_{ij}\chi^{||(1)',j  }$,
\bl \label{cnm11}
R^{(1)}_{ij}=&\phi^{(1) }_{,ij}- \l( \phi^{(1)''} + 5\frac{a'}{a}\phi^{(1)'}
+2(\frac{a''}{a} +2(\frac{a'}{a})^2 ) \phi^{(1)} -\nabla^2 \phi^{(1) } \r)\delta_{ij}
 \nn\\
&+\frac{1}{2} D_{ij}\chi^{||(1)''}+\frac{a'}{a}D_{ij}\chi^{||(1)'}
+(\frac{a''}{a}
+(\frac{a'}{a})^2 ) D_{ij}\chi^{||(1)}
-\frac{1}{2}\nabla^2D_{ij}\chi^{||(1) }
+  D^k_i\chi^{||(1)}_{,jk}   \nn\\
&
 +\frac{1}{2} \chi^{\top(1)''}_{ij}
 +\frac{a'}{a} \chi^{\top(1)'}_{ij}
 +(\frac{a''}{a}
 +(\frac{a'}{a})^2 )\chi^{\top(1) }_{ij}
 - \frac{1}{2}\nabla^2\chi^{\top(1) }_{ij} , \nn
\el
and
$R^{(1)}   =  \frac{1}{a^2}(-6\phi^{(1)''} -18\frac{a'}{a}\phi^{(1)'}
          +4\nabla^2\phi^{(1) }+D_{ij}\chi^{||(1),ij  }) $.

The  $(0i)$ component of the 1st-order perturbed Einstein equation is
\be \label{cnmi19}
 G^{(1)}_{0i}\equiv  R^{(1)}_{0i}=0,
\ee
which  leads  the 1st-order momentum constraint
\be \label{momentconstr1}
 \phi^{(1)'}_{,i}  +\frac{1}{6}\nabla^2 \chi^{\parallel(1)'}_{,i} = 0.
\ee
This  relates the two   scalar modes,
and  can be written as
 \be \label{impli}
 \phi^{(1)'}  +\frac{1}{6}\nabla^2 \chi^{\parallel(1)'}  = 0,
\ee
after dropping a space-independent constant.
Integration of (\ref{impli}) gives
\be   \label{eq17}
\phi^{(1)} +\frac{1}{6}\nabla^2 \chi^{\parallel(1)}
    =\phi^{(1)}_0 +\frac{1}{6}\nabla^2 \chi^{\parallel(1)}_0,
\ee
where  $\phi^{(1)}_0$  and $\chi^{\parallel(1)}_0$
denote the initial values of the two scalar modes, respectively.

The $(00)$ component of 1st-order perturbed Einstein equation is
\be  \label{1st00}
 G^{(1)}_{00}\equiv
  R_{00}^{(1)}  -\frac{1}{2}g_{00}^{(0)}R^{(1)}
 = 8\pi G \rho^{(0)}   a^2  \delta^{(1)} ,
\ee
which
leads to the 1st-order energy constraint
\be\label{G001eq}
-6\frac{a'}{a}\phi^{(1)'}
+2 ( \nabla^2\phi^{(1) }+\frac{1}{6} \nabla^2\nabla^2 \chi^{\parallel(1)})
=8\pi G \rho^{(0)}   a^2    \delta^{(1)}.
\ee
Using    (\ref{impli}),  (\ref{eq21}) into the above  gives
\be   \label{eq18}
\nabla^2 \left(\frac{2}{\tau}\chi^{\parallel(1)'}
    +\frac{6}{\tau^2}(\chi^{\parallel(1)}-\chi^{\parallel(1)}_0)
    +2\phi^{(1)}_0+\frac{1}{3}\nabla^2\chi^{\parallel(1)}_0  \right)
    =\frac{12}{\tau^2}\delta_0^{(1)}
\ee

The $(ij)$ component of 1st-order perturbed Einstein equation is:
\be\label{cnm29}
G^{(1)}_{ij}  \equiv
R^{(1)}_{ij}  -\frac{1}{2} \delta_{ij}a^2 R^{(1)}  -\frac{1}{2}a^2\gamma^{(1)}_{ij}  R^{(0)}
 =0.
\ee
which   gives the evolution equation
\be  \label{cnm31}
\begin{split}
&2 \phi^{(1)\, ''} \delta_{ij}
+\frac{8}{\tau}\phi^{(1)'}\delta_{ij} +\phi^{(1)}_{,ij}
-\nabla^2  \phi^{(1) }\delta_{ij}           \\
 &+\frac{1}{2}D_{ij}\chi^{\parallel(1)''}
 +\frac{2}{\tau} D_{ij}\chi^{\parallel(1)'}
 +\frac{1}{6}\nabla^2D_{ij}\chi^{\parallel(1)}
 -\frac{1}{9}\delta_{ij}\nabla^2\nabla^2\chi^{\parallel(1)}          \\
&  +\frac{1}{2} \chi^{\top(1)''}_{ij}
+\frac{2}{\tau} \chi^{\top(1)'}_{ij}
- \frac{1}{2}\nabla^2\chi^{\top(1) }_{ij}  =0,
\end{split}
\ee
  involving  all the metric perturbations.
The traceless part of Eq.(\ref{cnm31}) is
\ba \label{wedsl}
&&\left( D_{ij}\chi^{||(1)''} +\frac{4}{\tau}D_{ij}\chi^{||(1)'}
           +\frac{1}{3} \nabla^2D_{ij}\chi^{||(1) } \right)
+\left(\chi^{\top(1)''}_{ij}
+\frac{4}{\tau}\chi^{\top(1)'}_{ij}
-\nabla^2\chi^{\top(1)}_{ij}\right) \nn \\
& &  +2\left(\partial_i\partial_j\phi^{(1)}
   -\frac{1}{3}\delta_{ij}\nabla^2\phi^{(1)}\right)=0.
\ea
The transverse part of (\ref{wedsl})  directly gives
the hyperbolic, partial differential equation of 1st-order tensor
\be  \label{eq20}
\chi^{\top(1)''}_{ij}+\frac{4}{\tau}\chi^{\top(1)'}_{ij}
-\nabla^2\chi^{\top(1)}_{ij}=0,
\ee
which describes gravitational wave propagating at the speed of light.
 Eq.(\ref {eq20}), Eq.(\ref {momentconstr1}), and Eq.(\ref {eq18})
show that,  at 1st order,  the tensor and the scalar are independent.
The   solution of (\ref{eq20})
is given by (\ref{Fourier}) and (\ref{GWmode}).

Taking the trace of (\ref{cnm31}) gives
\be
6\phi^{(1)\, ''}
 + \frac{24 }{\tau }\phi^{(1)\, '}
 -2 (\nabla^2\phi^{(1) }
 + \frac{1}{6} \nabla^2 \nabla^2  \chi^{||(1)})     =0 ,
\ee
which, by the   momentum constraint (\ref{eq17}),
is also written as  the    equation of $\phi^{(1)}$
\be\label{init}
\phi^{(1)\, ''}
+ \frac{4}{\tau}\phi^{(1)'}
 =\frac{1}{3}  (  \nabla^2  \phi^{(1)}_0
    +\frac{1}{6} \nabla^2  \nabla^2 \chi^{\parallel(1)}_0 ).
\ee
By  (\ref{impli}),
it can be also written as the   equation of $  \chi^{\parallel(1)}$:
\be \label{ggtw}
  \chi^{\parallel(1)\, ''}
+ \frac{4}{\tau}  \chi^{\parallel(1)'} =
  -2(  \phi^{(1)}_0 +\frac{1}{6}   \nabla^2 \chi^{\parallel(1)}_0)  .
\ee
Note that the evolution equations (\ref {init}) (\ref {ggtw})
are not hyperbolic differential equations,
thus, the scalar perturbations $\phi^{(1) }$ and $ \chi^{\parallel(1)}$
 just  follow  where the density perturbation is distributed,
and  do not propagate at speed of light,
in contrast to the tensor $\chi^{\top(1)}_{ij}$.
Combining    (\ref{eq21}) and    (\ref{eq18}) yields
 the equation of the  1st-order  density contrast
\be \label{eqdelt}
\delta^{(1)\, ''}   +\frac{2}{\tau}\delta^{(1)\, '} -\frac{6}{\tau^2}\delta^{(1)}=0 .
\ee
Note that Eq.(\ref {eqdelt}) has no sound speed term
because  the pressure is zero for the dust model.
The general solution of (\ref {eqdelt}) consists of
a growing mode  $ \propto  \tau^2$ and a decaying mode $\tau^{-3}$,
\be \label{deltasol}
\delta^{(1)} = \delta_{0g}^{(1)} \frac{\tau^2}{\tau_0^2}  +\frac{3 X}{\tau^3}.
\ee
where $X= X({\bf x})$
 represents  the decaying mode of the  density contrast.
$\delta_{0}^{(1)}\equiv  \delta_{0g}^{(1)}+ \frac{3 X}{\tau^3_0}$
will denote the initial value of $\delta^{(1)}$.
Combining   (\ref{delt1})  and (\ref{eqdelt}) leads to
\be  \label{eq23}
\chi^{\parallel(1)''}+\frac{2}{\tau}\chi^{\parallel(1)'}
-\frac{6}{\tau^2}\chi^{\parallel(1)}=0,
\ee
which is the same equation as (\ref {eqdelt}).
Its solution is taken as
\be\label{yxe}
\nabla^2 \chi^{\parallel(1)}  =-2\frac{\delta_{0g}^{(1)}}{\tau_0^2}\tau^2
   -\frac{6 X}{\tau^3}  .
\ee
 $\phi^{(1)}$ satisfies a similar equation to (\ref{eq23})
with the inhomogeneous terms
$-\frac{6}{\tau^2} (\phi^{(1)}_0 +\frac16 \nabla^2 \chi^{\parallel(1)}_0)$.
For convenience,
introduce  the gravitational potential $\varphi$
defined by
\be\label{poisson}
\nabla^2 \varphi({\bf x}) =  \frac{6}{\tau_0^2}\delta_{0g}^{(1)}({\bf x}) .
\ee
Then  one obtains the solution
\bl \label{q2a}
\phi^{(1)} &  =   \frac{5}{3}  \varphi + \frac{\tau^2}{18}  \nabla^2 \varphi
 +\frac{ X}{\tau^3} \\
D_{ij} \chi^{\parallel(1)}
  & =   -  \frac{\tau^2}{3} (\varphi_{,ij} - \frac{1}{3} \delta_{ij} \nabla^2 \varphi )
     { -\frac{6\nabla^{-2} D_{ij} X}{\tau^3} }
     \label{q1a}
\el
by   (\ref{eq17}) and (\ref{ggtw}).
We shall keep the decaying terms
$\propto X/\tau^3$ in  (\ref {deltasol}), (\ref {q2a}), and (\ref {q1a}).
The   scalar perturbations
$\phi^{(1)} $ and $D_{ij} \chi^{\parallel(1)}$ are independent dynamic fields,
but the two fields and their first time derivatives
are related through the energy, and momentum constraints.
This leads to the fact that
the growing mode of $D_{ij}  \chi^{\parallel(1)}$ is
related to that of $\phi^{(1)} $ and, respectively,
so is  the decaying mode.
Thus, there are only two unknown functions $\varphi$ and $X$
in the solutions (\ref{q1a}) and (\ref{q2a}) which  will be determined
by the initial condition at $\tau_0$.
We remark that the solutions  (\ref{q1a}) and (\ref{q2a}) are consistent
with each other in a fixed gauge.
If the time-independent term $\frac{5}{3}\varphi $ in Eq.(\ref{q2a})
  was  discarded as   a gauge term,
$D_{ij} \chi^{\parallel(1)}$  of (\ref{q1a})
would acquire a term $10 D_{ij}( \nabla^{-2} \varphi ) $ simultaneously,
according to the gauge transformation (\ref{gaugemodes}) and
(\ref {phigauge3})
 of  Appendix C.

\subsection{ The energy density contrast}

For the matter source of gravity,
the equation of dust is   determined by
the conservation of  energy-momentum tensor,
by which the perturbations of density will be related the metric perturbations
as the following.
Define the   density contract
\be\label{the20}
\delta=\frac{\rho- \rho^{(0)} }{  \rho^{(0)} },
\ee
where $\rho= \rho({\bf x},\tau)$ is the mass density,
$\rho^{(0)}$ is its mean density.
From the energy conservation    $T^{0\nu}\,_{; \, \nu}=0$,
one obtains the solution $\delta$ in terms of metric perturbations
\begin{equation}  \label{eq13}
\delta(\tau, {\bf x} )
=(1+\delta_0({\bf x}))
\left[\frac{\gamma(\tau, {\bf x} )} {\gamma_0({\bf x})} \right]^{-\frac{1}{2}}-1,
\end{equation}
where $\delta_0$
 and $\gamma_0$ are the initial values,
and
\be\label{gamma}
\gamma \equiv \det(\gamma_{ij})
=1+  \gamma^{(1)i}_i  +\frac{1}{2} \gamma^{(2)i}_i
   + \frac{1}{2} \gamma^{(1)i}_{i} \gamma^{(1)j}_{j}
   -\frac{1}{2} \gamma^{(1)ij} \gamma^{(1)}_{ij} .
\ee
Expanding  (\ref {eq13}) up to 2nd order gives
\bl\label{delta2}
\delta =&-\frac{1}{2}\gamma^{(1)i}_i+\frac{1}{2}\gamma^{(1)i}_{0i}
+\delta _0^{(1)}
+\frac12\delta _0^{(2)}
+\frac{1}{4}\gamma^{(2)i}_{0\,i}
-\frac{1}{4}\gamma^{(2)i}_i
+\frac{1}{8}(\gamma^{(1)i}_{i})^2
+\frac{1}{8}(\gamma^{(1)i}_{0i})^2 \nn \\
& -\frac{1}{4}\gamma^{(1)i}_i\gamma^{(1)j}_{0j}
 +\frac{1}{4}\gamma^{(1)ij}\gamma^{(1)}_{ij}
-\frac{1}{4}\gamma^{(1)ij}_0\gamma^{(1)}_{0ij}
-\frac{1}{2}\gamma^{(1)i}_i\delta_0^{(1)}
+\frac{1}{2}\gamma^{(1)i}_{0i}\delta_0^{(1)} .
\el
where $\delta _0 \equiv  \delta _0^{(1)} +\frac12\delta _0^{(2)}$.
From (\ref{delta2}) one reads off the 1st-order density contrast
\be \label{deltaphi}
\delta^{(1)}  =\delta_0^{(1)}  +  \frac{1}{2}        (6\phi^{(1)} - 6\phi^{(1)}_0 )
\ee
which, by  Eq.(\ref{eq17}),   can be also written as
\be  \label{eq21}
\delta^{(1)}=
  \delta_0^{(1)} -\frac{1}{2} \nabla^2 (\chi^{\parallel(1)}-\chi^{\parallel(1)}_0).
\ee
One can  use the residual gauge freedom
to take $\nabla^2  \chi^{\parallel(1)}_0= -2\delta_0^{(1)}$
(see Eq.(\ref{eq22}) in Appendix C),
so that (\ref {eq21}) reduces to
\be \label{delt1}
\delta^{(1)}  = - \frac{1}{2} \nabla^2  \chi^{\parallel(1)}.
\ee
From (\ref{delta2}) one reads off the 2nd-order density contrast
\be  \label{delta2nd}
\begin{split}
\delta^{(2)}
= &\delta^{(2)}_0
+\frac{1}{2} \gamma^{(2)i}_{0\,i}
-\frac{1}{2} \gamma^{(2)i}_i
+\frac{1}{4} (\gamma^{(1)i}_{i})^2
+\frac{1}{4}(\gamma^{(1)i}_{0\, i})^2
-\frac{1}{2}\gamma^{(1)i}_i\gamma^{(1)j}_{0j} \\
 &+\frac{1}{2}\gamma^{(1)ij}\gamma^{(1)}_{ij}
-\frac{1}{2}\gamma^{(1)ij}_0\gamma^{(1)}_{0ij}
-\gamma^{(1)i}_i\delta_0^{(1)}
+\gamma^{(1)i}_{0\, i}\delta_0^{(1)} \, .
\end{split}
\ee
which depends on 2nd-order  perturbations
only through $\gamma^{(2)i}_i =-6 \phi^{(2)}$,
independent of  the 2nd-order tensor.
But $\delta^{(2)}$ depends on
 the 1st-order tensor via the term
$\frac{1}{4}\gamma^{(1)ij}\gamma^{(1)}_{ij}= \frac{1}{4}
(  2 \chi^{\top(1)}_{ij}D^{ij}\chi^{\parallel(1)}
 + \chi^{\top(1)}_{ij}\chi^{\top(1)ij} ) $.

\section{ Expressions of 2nd  Order Perturbed Ricci  Tensors}

By Eq.(\ref{18q1}) and Eq.(\ref{eq1}),
we  calculate  the  2nd-order  Ricci and Einstein tensors.
The 2nd-order Ricci tensors are
 \bl\label{cnm12}
R^{(2) }_{00}=& \frac{3a'}{2a} \phi^{(2)'}  +\frac{3}{2} \phi^{(2)''} +6\frac{a'}{a}\phi^{(1) }  \phi^{(1) '}
+6 \phi^{(1) }  \phi^{(1)''} +3\phi^{(1) '}\phi^{(1) '}
+\frac{1}{2}D^{ij}\chi^{||(1)}D_{ij}\chi^{||(1)''}
\nn\\
&
+\frac{1}{4}D^{ij}\chi^{||(1)'}D_{ij}\chi^{||(1)'}
+\frac{a'}{2a}D^{ij}\chi^{||(1)}D_{ij}\chi^{||(1)'}+\frac{1}{2} \chi^{\top(1)ij} \chi^{\top(1)''}_{ij}+\frac{1}{4} \chi^{\top(1)'ij} \chi^{\top (1)'}_{ij}
\nn\\
&
+\frac{a'}{2a} \chi^{\top(1)ij} \chi^{\top (1)'}_{ij}+\frac{1}{2}\chi^{\top(1)ij}D_{ij}\chi^{||(1)''}
+\frac{1}{2}\chi^{\top(1)'ij}D_{ij}\chi^{||(1)'}
+\frac{a'}{2a}\chi^{\top(1)ij}D_{ij}\chi^{||(1)'}
\nn\\&
+\frac{1}{2} \chi^{\top(1)''}_{ij}D^{ij}\chi^{||(1)}
 +\frac{a'}{2a}\chi^{\top (1)'}_{ij}D^{ij}\chi^{||(1)} ,
\el
\bl  \label{cnm13}
R^{(2) }_{0i}=&  \phi^{(2)'} _{,\,i}
+\frac{1}{4}D_{ij}\chi^{||(2)',\,j}
+\frac{1}{4}\chi^{\perp(2)',\,j}_{ij}
+4\phi^{(1)'} \phi^{(1) } _{,\,i}+4\phi^{(1) } \phi^{(1) '} _{,\,i}
+\phi^{(1) }D_{ij}\chi^{||(1)',\,j}
\nn\\
&
+\phi^{(1)'}D_{ij}\chi^{||(1) ,\,j}
-\frac{1}{2} \phi^{(1) ,\,j }D_{ij}\chi^{||(1)' }
+\phi^{(1)' ,\,j }D_{ij}\chi^{||(1) }
-\frac{1}{2} \phi^{(1) ,\,j }\chi^{\top(1)' }_{ij}
\nn\\
&
    +\phi^{(1)' ,\,j } \chi^{\top(1) }_{ij}
-\frac{1}{2} D^k_{j}\chi^{||(1),\,j}D_{ik}\chi^{||(1)' }
-\frac{1}{2} D^k_{j}\chi^{||(1)  }D_{ik}\chi^{||(1)',\,j}
  + \frac{1}{2} D^{jk}\chi^{||(1)'}_{,\,i}D_{jk}\chi^{||(1)  }
   \nn\\
&  + \frac{1}{4} D^{jk}\chi^{||(1)' }D_{jk}\chi^{||(1) }_{,\,i}
 -\frac{1}{2}\chi^{\top(1) k }_{j}D_{ik}\chi^{||(1)',\,j}
+ \frac{1}{2}\chi^{\top(1)j k} D_{jk}\chi^{||(1)' }_{,\,i}
    + \frac{1}{4} \chi^{\top(1)  jk}_{,\,i} D_{jk}\chi^{||(1)'}
    \nn\\
&    -\frac{1}{2} \chi^{\top(1)',\,j}_{ik}D^k_{j}\chi^{||(1)}
    + \frac{1}{2}\chi^{\top(1)' }_{jk,  \, i}D^{jk}\chi^{||(1)}
    + \frac{1}{4}\chi^{\top(1)'}_{jk }D^{jk}\chi^{||(1) }_{,\,i}
-\frac{1}{2} \chi^{\top (1)' }_{ik}D^k_{j}\chi^{||(1),\,j}
\nn\\
& -\frac{1}{2} \chi^{\top(1) k }_{j} \chi^{\top(1)',\,j}_{ik}
+ \frac{1}{2}\chi^{\top(1)j k} \chi^{\top(1)' }_{jk,i}
          + \frac{1}{4}  \chi^{\top(1)  jk}_{ ,\,i} \chi^{\top(1)'}_{jk },
\el
\bl\label{cnm14}
R_{ij}^{(2)}=&
\delta_{ij}\Big[
-\frac52\frac{{a'}}{a}\phi^{(2)'}
-( \frac{a'}a )^2 \phi^{(2)}
-\frac{a''}{a}\phi^{(2)}
-\frac{1}{2}\phi^{(2)''}
+\frac{1}{2}\nabla^2\phi^{(2)}
+(\phi^{(1)'})^2
+\phi^{(1),\,k}\phi^{(1)}_{,\,k}
\nn\\
&
+2\phi^{(1)}\nabla^2\phi^{(1)}
-\phi^{(1)}_{,\,k}D^{lk}\chi^{||(1)}_{,\,l}
-\phi^{(1)}_{,\,kl}D^{kl}\chi^{||(1)}
-\phi^{(1)}_{,\,kl}\chi^{\top(1)kl}
-\frac{a'}{2a}D^{kl}\chi^{||(1)}D_{kl}\chi^{||(1)'}
\nn\\
&-\frac{a'}{2a}\chi^{\top(1)'}_{kl}D^{kl}\chi^{||(1)}
-\frac{a'}{2a}\chi^{\top(1)kl}D_{kl}\chi^{||(1)'}
-\frac{a'}{2a}\chi^{\top(1)kl}\chi^{\top(1)'}_{kl}
\Big] +\frac{1}{2}\phi^{(2)}_{,\,ij}    \nn
\\
&+\frac{1}{4}D_{ij}\chi^{||(2)''}
 +\frac{1}{2}\l[(\frac{a'}a )^2
    +\frac{a''}a \r]D_{ij}\chi^{||(2)}
+\frac{a'}{2a} D_{ij}\chi^{||(2)'}
+\frac{1}{4}D^k_j\chi^{||(2)}_{,\,ki}
+\frac{1}{4}D^k_i\chi^{||(2)}_{,\,kj}
\nn\\
&
-\frac{1}{4}\nabla^2D_{ij}\chi^{||(2)}
+\frac{1}{4}\chi^{\perp(2)''}_{ij}
+\frac{1}{2}\l[(\frac{a'}a )^2
    +\frac{a''}a \r]\chi^{\perp(2)}_{ij}
+\frac{a'}{2a} \chi^{\perp(2)'}_{ij}
+\frac{1}{4}\chi^{\perp(2),\,k}_{kj,\,i}
\nn\\
&
+\frac{1}{4}\chi^{\perp(2),\,k}_{ki,\,j}
-\frac{1}{4}\nabla^2\chi^{\perp(2)}_{ij}
+\frac{1}{4}\chi^{\top(2)''}_{ij}
+\frac{1}{2}\l[(\frac{a'}a )^2
    +\frac{a''}a \r]\chi^{\top(2)}_{ij}
+\frac{a'}{2a} \chi^{\top(2)'}_{ij}
-\frac{1}{4}\nabla^2\chi^{\top(2)}_{ij}
\nn\\
&
+3\phi^{(1)}_{,\,i}\phi^{(1)}_{,\,j}
+2\phi^{(1)}\phi^{(1)}_{,\,ij}
-\frac{3a'}a \phi^{(1)'}D_{ij}\chi^{||(1)}
+\frac{1}{2}\phi^{(1)'}D_{ij}\chi^{||(1)'}
\nn\\
&
+\frac{1}{2}\phi^{(1)}_{,\,k}D^k_j\chi^{||(1)}_{,\,i}
+\frac{1}{2}\phi^{(1)}_{,\,k}D^k_i\chi^{||(1)}_{,\,j}
-\frac32\phi^{(1)}_{,\,k}D_{ij}\chi^{||(1),\,k}
+\phi^{(1)}D^k_j\chi^{||(1)}_{,\,ik}
+\phi^{(1)}D^k_i\chi^{||(1)}_{,\,jk}
\nn
\\
&
-\phi^{(1)}\nabla^2D_{ij}\chi^{||(1)}
+\phi^{(1)}_{,\,i}D^k_j\chi^{||(1)}_{,\,k}
+\phi^{(1)}_{,\,j}D^k_i\chi^{||(1)}_{,\,k}
+\phi^{(1)}_{,\,ki}D^k_j\chi^{||(1)}
+\phi^{(1)}_{,\,kj}D^k_i\chi^{||(1)}
\nn\\
&
-\frac{3a'}a \phi^{(1)'}\chi^{\top(1)}_{ij}
+\frac{1}{2}\phi^{(1)'}\chi^{\top(1)'}_{ij}
+\frac{1}{2}\phi^{(1)}_{,\,k}\chi^{\top(1)k}_{j,\,i}
+\frac{1}{2}\phi^{(1)}_{,\,k}\chi^{\top(1)k}_{i,\,j}
-\frac32\phi^{(1)}_{,\,k}\chi^{\top(1),\,k}_{ij}
\nn
\\
&
-\phi^{(1)}\nabla^2\chi^{\top(1)}_{ij}
+\phi^{(1)}_{,\,ki}\chi^{\top(1)k}_{j}
+\phi^{(1)}_{,\,kj}\chi^{\top(1)k}_{i}
-\frac{1}{2}D^k_i\chi^{||(1)'}D_{kj}\chi^{||(1)'}
-\frac{1}{2}D^{kl}\chi^{||(1)}_{,\,l}D_{kj}\chi^{||(1)}_{,\,i}
\nn\\
&
-\frac{1}{2}D^{kl}\chi^{||(1)}_{,\,l}D_{ki}\chi^{||(1)}_{,\,j}
+\frac{1}{2}D^{lk}\chi^{||(1)}_{,\,l}D_{ij}\chi^{||(1)}_{,\,k}
-\frac{1}{2}D^{kl}\chi^{||(1)}D_{lj}\chi^{||(1)}_{,\,ik}
-\frac{1}{2}D^{kl}\chi^{||(1)}D_{li}\chi^{||(1)}_{,\,jk}
\nn
\\
&
+\frac{1}{2}D^{kl}\chi^{||(1)}D_{ij}\chi^{||(1)}_{,\,kl}
+\frac{1}{2}D^{kl}\chi^{||(1)}D_{kl}\chi^{||(1)}_{,\,ij}
+\frac{1}{4} D^{kl}\chi^{||(1)}_{,\,i}D_{kl}\chi^{||(1)}_{,\,j}
-\frac{1}{2}D^l_i\chi^{||(1)}_{,\,k}D^k_j\chi^{||(1)}_{,\,l}
\nn\\
&
+\frac{1}{2}D^k_i\chi^{||(1)}_{,\,l}D_{kj}\chi^{||(1),\,l}
-\frac{1}{2}\chi^{\top(1)'}_{kj}D^k_i\chi^{||(1)'}
-\frac{1}{2}\chi^{\top(1)'k}_{i}D_{kj}\chi^{||(1)'}
-\frac{1}{2}\chi^{\top(1)}_{kj,\,i}D^{kl}\chi^{||(1)}_{,\,l}
\nn\\
&
-\frac{1}{2}\chi^{\top(1)}_{ki,\,j}D^{kl}\chi^{||(1)}_{,\,l}
+\frac{1}{2}\chi^{\top(1)}_{ij,\,k}D^{lk}\chi^{||(1)}_{,\,l}
-\frac{1}{2}\chi^{\top(1)}_{lj,\,ik}D^{kl}\chi^{||(1)}
-\frac{1}{2}\chi^{\top(1)kl}D_{lj}\chi^{||(1)}_{,\,ik}
\nn
\\
&
-\frac{1}{2}\chi^{\top(1)}_{li,\,jk}D^{kl}\chi^{||(1)}
-\frac{1}{2}\chi^{\top(1)kl}D_{li}\chi^{||(1)}_{,\,jk}
+\frac{1}{2}\chi^{\top(1)}_{ij,\,kl}D^{kl}\chi^{||(1)}
+\frac{1}{2}\chi^{\top(1)kl}D_{ij}\chi^{||(1)}_{,\,kl}
\nn
\\
&
+\frac{1}{2}\chi^{\top(1)}_{kl,\,ij}D^{kl}\chi^{||(1)}
+\frac{1}{2}\chi^{\top(1)kl}D_{kl}\chi^{||(1)}_{,\,ij}
+\frac{1}{4}\chi^{\top(1)}_{kl,\,j}D^{kl}\chi^{||(1)}_{,\,i}
+\frac{1}{4}\chi^{\top(1)kl}_{,\,i}D_{kl}\chi^{||(1)}_{,\,j}
\nn
\\
&
-\frac{1}{2}\chi^{\top(1)k}_{j,\,l}D^l_i\chi^{||(1)}_{,\,k}
-\frac{1}{2}\chi^{\top(1)l}_{i,\,k}D^k_j\chi^{||(1)}_{,\,l}
+\frac{1}{2}\chi^{\top(1),\,l}_{kj}D^k_i\chi^{||(1)}_{,\,l}
+\frac{1}{2}\chi^{\top(1)k}_{i,\,l}D_{kj}\chi^{||(1),\,l}
\nn\\
&
-\frac{1}{2}\chi^{\top(1)'k}_{i}\chi^{\top(1)'}_{kj}
-\frac{1}{2}\chi^{\top(1)kl}\chi^{\top(1)}_{lj,\,ik}
-\frac{1}{2}\chi^{\top(1)kl}\chi^{\top(1)}_{li,\,jk}
+\frac{1}{2}\chi^{\top(1)kl}\chi^\top{(1)}_{ij,\,kl}
\nn
\\
&
+\frac{1}{2}\chi^{\top(1)kl}\chi^{\top(1)}_{kl,\,ij}
+\frac{1}{4}\chi^{\top(1)kl}_{,\,i}\chi^{\top(1)}_{kl,\,j}
-\frac{1}{2}\chi^{\top(1)l}_{i,\,k}\chi^{\top(1)k}_{j,\,l}
+\frac{1}{2}\chi^{\top(1)k}_{i,\,l}\chi^{\top(1),\,l}_{kj}
\el
and the 2nd-order Ricci scalar is
\bl\label{cnm18}
R^{(2) } =&\frac{1}{a^2}\Bigg[
2\nabla^2\phi^{(2)}
-9\frac{a'}a \phi^{(2)'}
-3\phi^{(2)''}
+\frac12D^{kl}\chi^{||(2)}_{,\,kl}
-12\phi^{(1)}\phi^{(1)''}
-36\frac{a'}a\phi^{(1)'}\phi^{(1)}
\nn
\\
&
+6\phi^{(1)}_{,\,k}\phi^{(1),\,k}
+16\phi^{(1)}\nabla^2\phi^{(1)}
+4\phi^{(1)}D^{kl}\chi^{||(1)}_{,\,kl}
-2\phi^{(1)}_{,\,kl}D^{kl}\chi^{||(1)}
-2\phi^{(1)}_{,\,kl}\chi^{\top(1)kl}
\nn\\
    &
-D^{kl}\chi^{||(1)}D_{kl}\chi^{||(1)''}
-\frac34D^{kl}\chi^{||(1)'}D_{kl}\chi^{||(1)'}
    -3\frac{a'}a D^{kl}\chi^{||(1)}D_{kl}\chi^{||(1)'}
    \nn
    \\
    &
    -2D_{ml}\chi^{||(1),\,l}_{,\,k}D^{km}\chi^{||(1)}
+D^{km}\chi^{||(1)}\nabla^2D_{km}\chi^{||(1)}
-D^{km}\chi^{||(1)}_{,\,k}D_{ml}\chi^{||(1),\,l}
\nn\\
&
    +\frac34D^{km}\chi^{||(1)}_{,\,l}D_{km}\chi^{||(1),\,l}
-\frac12D^{km}\chi^{||(1)}_{,\,l}D^l_k\chi^{||(1)}_{,\,m}
-D^{kl}\chi^{||(1)}\chi^{\top(1)''}_{kl}
-\chi^{\top(1)kl}D_{kl}\chi^{||(1)''}
\nn
\\
&
-\frac32\chi^{\top(1)'}_{kl}D^{kl}\chi^{||(1)'}
-3\frac{a'}a \chi^{\top(1)'}_{kl}D^{kl}\chi^{||(1)}
-3\frac{a'}a \chi^{\top(1)kl}D_{kl}\chi^{||(1)'}
-2\chi^{\top(1)km}D_{ml}\chi^{||(1),\,l}_{,\,k}
    \nn\\
    &
+\chi^{\top(1)km}\nabla^2D_{km}\chi^{||(1)}
+D^{km}\chi^{||(1)}\nabla^2\chi^{\top(1)}_{km}
+\frac32\chi^{\top(1),\,l}_{km}D^{km}\chi^{||(1)}_{,\,l}
-\chi^{\top(1)l}_{k,\,m}D^{km}\chi^{||(1)}_{,\,l}
    \nn
    \\
    &
-\chi^{\top(1)kl}\chi^{\top(1)''}_{kl}
-\frac34\chi^{\top(1)'kl}\chi^{\top(1)'}_{kl}
    -3\frac{a'}a \chi^{\top(1)kl}\chi^{\top(1)'}_{kl}
+\chi^{\top(1)km}\nabla^2\chi^{\top(1)}_{km}
    \nn
    \\
    &
    +\frac34\chi^{\top(1)km}_{,\,l}\chi^{\top(1),\,l}_{km}
    -\frac12\chi^{\top(1)km}_{,\,l}\chi^{\top(1)l}_{k,\,m}
\Bigg] .
\el
The 2nd-order perturbed Einstein tensors are
\bl \label{G002}
G_{00}^{(2)} \equiv &
         R_{00}^{(2)} -\frac{1}{2}g^{(0)}_{00}R^{(2)} \nn \\
 = &
\nabla^2\phi^{(2)}
-\frac{3a'}{a} \phi^{(2)'}
+\frac{1}{4}D^{kl}\chi^{||(2)}_{,\,kl}
\nn\\
&
-12\frac{a'}a\phi^{(1)'}\phi^{(1)}
+3\phi^{(1) '}\phi^{(1) '}
+3\phi^{(1)}_{,\,k}\phi^{(1),\,k}
+8\phi^{(1)}\nabla^2\phi^{(1)}
+2\phi^{(1)}D^{kl}\chi^{||(1)}_{,\,kl}
\nn\\
    &
-\phi^{(1)}_{,\,kl}D^{kl}\chi^{||(1)}
-\frac{1}{8} D^{kl}\chi^{||(1)'}D_{kl}\chi^{||(1)'}
-\frac{a'}a D^{kl}\chi^{||(1)}D_{kl}\chi^{||(1)'}
            \nn\\
            &
-D_{ml}\chi^{||(1),\,l}_{,\,k}D^{km}\chi^{||(1)}
+\frac{1}{2}D^{km}\chi^{||(1)}\nabla^2D_{km}\chi^{||(1)}
-\frac{1}{2}D^{km}\chi^{||(1)}_{,\,k}D_{ml}\chi^{||(1),\,l}
\nn\\
&
    +\frac{3}{8}D^{km}\chi^{||(1)}_{,\,l}D_{km}\chi^{||(1),\,l}
-\frac{1}{4}D^{km}\chi^{||(1)}_{,\,l}D^l_k\chi^{||(1)}_{,\,m}
-\phi^{(1)}_{,\,kl}\chi^{\top(1)kl}
\nn\\
&
-\frac{1}{4}\chi^{\top(1)'}_{kl}D^{kl}\chi^{||(1)'}
-\frac{a'}a \chi^{\top(1)kl}D_{kl}\chi^{||(1)'}
-\frac{a'}a \chi^{\top(1)'}_{kl}D^{kl}\chi^{||(1)}
    \nn\\
    &
-\chi^{\top(1)km}D_{ml}\chi^{||(1),\,l}_{,\,k}
+\frac{1}{2}\chi^{\top(1)km}\nabla^2D_{km}\chi^{||(1)}
+\frac{1}{2}D^{km}\chi^{||(1)}\nabla^2\chi^{\top(1)}_{km}
    \nn\\
    &
+\frac{3}{4}\chi^{\top(1),\,l}_{km}D^{km}\chi^{||(1)}_{,\,l}
-\frac{1}{2}\chi^{\top(1)l}_{k,\,m}D^{km}\chi^{||(1)}_{,\,l}
-\frac{1}{8}\chi^{\top(1)'kl}\chi^{\top(1)'}_{kl}
    \nn
    \\
    &
 -\frac{a'}a \chi^{\top(1)kl}\chi^{\top(1)'}_{kl}
+\frac{1}{2}\chi^{\top(1)km}\nabla^2\chi^{\top(1)}_{km}
+\frac{3}{8}\chi^{\top(1)km}_{,\,l}\chi^{\top(1),\,l}_{km}
       -\frac{1}{4}\chi^{\top(1)km}_{,\,l}\chi^{\top(1)l}_{k,\,m},
\el
\be
 G_{0i}^{(2)} \equiv   R_{0i}^{(2)},
\ee
{\large
\bl\label{EinsteinTensor2f}
G^{(2)}_{ij}=&
\delta_{ij}\Big[
-\frac{1}{2}\nabla^2\phi^{(2)}
+\phi^{(2)''}
+2\frac{{a'}}{a}\phi^{(2)'}
+\l[2\frac{a''}{a}-(\frac{a'}{a})^2\r]\phi^{(2)}
-\frac{1}{4}D^{kl}\chi^{||(2)}_{,\,kl}
-2\phi^{(1),\,k}\phi^{(1)}_{,\,k}
\nn\\
&
-2\phi^{(1)}\nabla^2\phi^{(1)}
+\l(\phi^{(1)'}\r)^2
-\phi^{(1)}_{,\,k}D^{lk}\chi^{||(1)}_{,\,l}
-\phi^{(1)}D^{kl}\chi^{||(1)}_{,\,kl}
+\frac{a'}{a}D^{kl}\chi^{||(1)}D_{kl}\chi^{||(1)'}
\nn\\
&
+D_{ml}\chi^{||(1),\,l}_{,\,k}D^{km}\chi^{||(1)}
-\frac{1}{2} D^{km}\chi^{||(1)}\nabla^2D_{km}\chi^{||(1)}
-\frac38D^{km}\chi^{||(1),\,l}D_{km}\chi^{||(1)}_{,\,l}
+\frac{1}{4}D^{km}\chi^{||(1),\,l}D_{kl}\chi^{||(1)}_{,\,m}
\nn
\\
&
+\frac{1}{2}D^{kl}\chi^{||(1)}_{,\,l}D^m_k\chi^{||(1)}_{,\,m}
+\frac38D^{kl}\chi^{||(1)'}D_{kl}\chi^{||(1)'}
+\frac{1}{2}D^{kl}\chi^{||(1)}D_{kl}\chi^{||(1)''}
+\frac{a'}{a}\chi^{\top(1)kl}D_{kl}\chi^{||(1)'}
\nn\\
&
+\frac{a'}{a}D^{kl}\chi^{||(1)}\chi^{\top(1)'}_{kl}
+\chi^{\top(1)km}D_{ml}\chi^{||(1),\,l}_{,\,k}
-\frac{1}{2}\chi^{\top(1)km}\nabla^2D_{km}\chi^{||(1)}
-\frac{1}{2} D^{km}\chi^{||(1)}\nabla^2\chi^{\top(1)}_{km}
\nn
\\
&
-\frac{3}{4}D^{km}\chi^{||(1),\,l}\chi^{\top(1)}_{km,\,l}
+\frac{1}{2}D^{km}\chi^{||(1),\,l}\chi^{\top(1)}_{kl,\,m}
+\frac{3}{4}\chi^{\top(1)'kl}D_{kl}\chi^{||(1)'}
+\frac{1}{2}\chi^{\top(1)''}_{kl}D^{kl}\chi^{||(1)}
\nn\\
&
+\frac{1}{2}\chi^{\top(1)kl}D_{kl}\chi^{||(1)''}
+\frac{a'}{a}\chi^{\top(1)kl}\chi^{\top(1)'}_{kl}
-\frac{1}{2}\chi^{\top(1)km}\nabla^2\chi^{\top(1)}_{km}
-\frac38\chi^{\top(1)km,\,l}\chi^{\top(1)}_{km,\,l}
\nn\\
&
+\frac{1}{4}\chi^{\top(1)km,\,l}\chi^{\top(1)}_{kl,\,m}
+\frac38\chi^{\top(1)'kl}\chi^{\top(1)'}_{kl}
+\frac{1}{2}\chi^{\top(1)kl}\chi^{\top(1)''}_{kl}
\Big]
+\frac{1}{2}\phi^{(2)}_{,\,ij}
+\frac{a'}{2a} D_{ij}\chi^{||(2)'}
+\frac{1}{4}D^k_j\chi^{||(2)}_{,\,ki}
\nn\\
&
+\frac{1}{4}D^k_i\chi^{||(2)}_{,\,kj}
-\frac{1}{4}\nabla^2D_{ij}\chi^{||(2)}
+\frac{1}{4}D_{ij}\chi^{||(2)''}
 +\frac{1}{2}\l[(\frac{a'}a )^2
    -2\frac{a''}a \r]D_{ij}\chi^{||(2)}
+\frac{a'}{2a} \chi^{\perp(2)'}_{ij}
+\frac{1}{4}\chi^{\perp(2)k}_{j,\,ki}
\nn\\
&
+\frac{1}{4}\chi^{\perp(2)k}_{i,\,kj}
-\frac{1}{4}\nabla^2\chi^{\perp(2)}_{ij}
+\frac{1}{4}\chi^{\perp(2)''}_{ij}
+\frac{1}{2}\l[(\frac{a'}a )^2
    -2\frac{a''}a \r]\chi^{\perp(2)}_{ij}
+\frac{a'}{2a} \chi^{\top(2)'}_{ij}
-\frac{1}{4}\nabla^2\chi^{\top(2)}_{ij}
\nn\\
&
+\frac{1}{4}\chi^{\top(2)''}_{ij}
+\frac{1}{2}\l[(\frac{a'}a )^2
    -2\frac{a''}a \r]\chi^{\top(2)}_{ij}
+3\phi^{(1)}_{,\,i}\phi^{(1)}_{,\,j}
+2\phi^{(1)}\phi^{(1)}_{,\,ij}
+\frac{1}{2}\phi^{(1)'}D_{ij}\chi^{||(1)'}
\nn\\
&
+\frac{1}{2}\phi^{(1)}_{,\,k}D^k_j\chi^{||(1)}_{,\,i}
+\frac{1}{2}\phi^{(1)}_{,\,k}D^k_i\chi^{||(1)}_{,\,j}
-\frac32\phi^{(1)}_{,\,k}D_{ij}\chi^{||(1),\,k}
+\phi^{(1)}D^k_j\chi^{||(1)}_{,\,ik}
+\phi^{(1)}D^k_i\chi^{||(1)}_{,\,jk}
\nn
\\
&
-\phi^{(1)}\nabla^2D_{ij}\chi^{||(1)}
-2 D_{ij}\chi^{||(1)}\nabla^2\phi^{(1)}
+\phi^{(1)}_{,\,i}D^k_j\chi^{||(1)}_{,\,k}
+\phi^{(1)}_{,\,j}D^k_i\chi^{||(1)}_{,\,k}
+\phi^{(1)}_{,\,ki}D^k_j\chi^{||(1)}
\nn\\
&
+\phi^{(1)}_{,\,kj}D^k_i\chi^{||(1)}_{}
+3\phi^{(1)''}D_{ij}\chi^{||(1)}
+6\frac{a'}a\phi^{(1)'}D_{ij}\chi^{||(1)}
+\frac{1}{2}\phi^{(1)'}\chi^{\top(1)'}_{ij}
+\frac{1}{2}\phi^{(1)}_{,\,k}\chi^{\top(1)k}_{j,\,i}
\nn\\
&
+\frac{1}{2}\phi^{(1)}_{,\,k}\chi^{\top(1)k}_{i,\,j}
-\frac32\phi^{(1)}_{,\,k}\chi^{\top(1),\,k}_{ij}
-\phi^{(1)}\nabla^2\chi^{\top(1)}_{ij}
-2\chi^{\top(1)}_{ij}\nabla^2\phi^{(1)}
+\phi^{(1)}_{,\,ki}\chi^{\top(1)k}_{j}
\nn
\\
&
+\phi^{(1)}_{,\,kj}\chi^{\top(1)k}_{i}
+3\phi^{(1)''}\chi^{\top(1)}_{ij}
+6\frac{a'}a \phi^{(1)'}\chi^{\top(1)}_{ij}
-\frac{1}{2}D^k_i\chi^{||(1)'}D_{kj}\chi^{||(1)'}
-\frac{1}{2}D^{kl}\chi^{||(1)}_{,\,l}D_{kj}\chi^{||(1)}_{,\,i}
\nn\\
&
-\frac{1}{2}D^{kl}\chi^{||(1)}_{,\,l}D_{ki}\chi^{||(1)}_{,\,j}
+\frac{1}{2}D^{lk}\chi^{||(1)}_{,\,l}D_{ij}\chi^{||(1)}_{,\,k}
-\frac{1}{2}D^{kl}\chi^{||(1)}D_{lj}\chi^{||(1)}_{,\,ik}
-\frac{1}{2}D^{kl}\chi^{||(1)}D_{li}\chi^{||(1)}_{,\,jk}
\nn
\\
&
+\frac{1}{2}D^{kl}\chi^{||(1)}D_{ij}\chi^{||(1)}_{,\,kl}
+\frac{1}{2}D^{kl}\chi^{||(1)}D_{kl}\chi^{||(1)}_{,\,ij}
+\frac{1}{4}D^{kl}\chi^{||(1)}_{,\,i}D_{kl}\chi^{||(1)}_{,\,j}
-\frac{1}{2}D_{kl}\chi^{||(1),\,kl}D_{ij}\chi^{||(1)}
\nn
\\
&
-\frac{1}{2}D^l_i\chi^{||(1)}_{,\,k}D^k_j\chi^{||(1)}_{,\,l}
+\frac{1}{2}D^k_i\chi^{||(1)}_{,\,l}D_{kj}\chi^{||(1),\,l}
-\frac{1}{2}\chi^{\top(1)'}_{kj}D^k_i\chi^{||(1)'}
-\frac{1}{2}\chi^{\top(1)'k}_{i}D_{kj}\chi^{||(1)'}
\nn\\
&
-\frac{1}{2}\chi^{\top(1)}_{kj,\,i}D^{kl}\chi^{||(1)}_{,\,l}
-\frac{1}{2}\chi^{\top(1)}_{ki,\,j}D^{kl}\chi^{||(1)}_{,\,l}
+\frac{1}{2}\chi^{\top(1)}_{ij,\,k}D^{lk}\chi^{||(1)}_{,\,l}
-\frac{1}{2}\chi^{\top(1)}_{lj,\,ik}D^{kl}\chi^{||(1)}
\nn
\\
&
-\frac{1}{2}\chi^{\top(1)kl}D_{lj}\chi^{||(1)}_{,\,ik}
-\frac{1}{2}\chi^{\top(1)}_{li,\,jk}D^{kl}\chi^{||(1)}
-\frac{1}{2}\chi^{\top(1)kl}D_{li}\chi^{||(1)}_{,\,jk}
+\frac{1}{2}\chi^{\top(1)}_{ij,\,kl}D^{kl}\chi^{||(1)}
\nn
\\
&
+\frac{1}{2}\chi^{\top(1)kl}D_{ij}\chi^{||(1)}_{,\,kl}
+\frac{1}{2}\chi^{\top(1)}_{kl,\,ij}D^{kl}\chi^{||(1)}
+\frac{1}{2}\chi^{\top(1)kl}D_{kl}\chi^{||(1)}_{,\,ij}
+\frac{1}{4}\chi^{\top(1)}_{kl,\,j}D^{kl}\chi^{||(1)}_{,\,i}
\nn\\
&
+\frac{1}{4}\chi^{\top(1)kl}_{,\,i}D_{kl}\chi^{||(1)}_{,\,j}
-\frac{1}{2}\chi^{\top(1)}_{ij}D_{kl}\chi^{||(1),\,kl}
-\frac{1}{2}\chi^{\top(1)k}_{j,\,l}D^l_i\chi^{||(1)}_{,\,k}
-\frac{1}{2}\chi^{\top(1)l}_{i,\,k}D^k_j\chi^{||(1)}_{,\,l}
\nn\\
&
+\frac{1}{2}\chi^{\top(1),\,l}_{kj}D^k_i\chi^{||(1)}_{,\,l}
+\frac{1}{2}\chi^{\top(1)k}_{i,\,l}D_{kj}\chi^{||(1),\,l}
-\frac{1}{2}\chi^{\top(1)'k}_{i}\chi^{\top(1)'}_{kj}
-\frac{1}{2}\chi^{\top(1)kl}\chi^{\top(1)}_{lj,\,ik}
-\frac{1}{2}\chi^{\top(1)kl}\chi^{\top(1)}_{li,\,jk}
\nn
\\
&
+\frac{1}{2}\chi^{\top(1)kl}\chi^{\top(1)}_{ij,\,kl}
+\frac{1}{2}\chi^{\top(1)kl}\chi^{\top(1)}_{kl,\,ij}
+\frac{1}{4}\chi^{\top(1)kl}_{,\,i}\chi^{\top(1)}_{kl,\,j}
-\frac{1}{2}\chi^{\top(1)l}_{i,\,k}\chi^{\top(1)k}_{j,\,l}
+\frac{1}{2}\chi^{\top(1)k}_{i,\,l}\chi^{\top(1),\,l}_{kj} \ .
\el
}
and its  trace is given by
\bl\label{EinsteinTensorTrace}
\delta^{kl} G^{(2)}_{kl}=&
-\nabla^2\phi^{(2)}
+3\phi^{(2)''}
+6\frac{a'}a \phi^{(2)'}
-\frac14D^{kl}\chi^{||(2)}_{,\,kl}
-3\phi^{(1),\,k}\phi^{(1)}_{,\,k}
-4\phi^{(1)}\nabla^2\phi^{(1)}
\nn
\\
&
+3\l(\phi^{(1)'}\r)^2
-\phi^{(1)}D^{kl}\chi^{||(1)}_{,\,kl}
+2\phi^{(1)}_{,\,kl}D^{kl}\chi^{||(1)}
+2\phi^{(1)}_{,\,kl}\chi^{\top(1)kl}
+3\frac{a'}aD^{kl}\chi^{||(1)}D_{kl}\chi^{||(1)'}
\nn\\
&
+2D_{ml}\chi^{||(1),\,l}_{,\,k}D^{km}\chi^{||(1)}
-D^{km}\chi^{||(1)}\nabla^2D_{km}\chi^{||(1)}
-\frac38D^{km}\chi^{||(1)}_{,\,l}D_{km}\chi^{||(1),\,l}
\nn\\
&
+\frac12D^{kl}\chi^{||(1)}_{,\,l}D^m_k\chi^{||(1)}_{,\,m}
+\frac58D^{kl}\chi^{||(1)'}D_{kl}\chi^{||(1)'}
+\frac32D^{kl}\chi^{||(1)}D_{kl}\chi^{||(1)''}
\nn\\
&
+\frac14D^{km}\chi^{||(1),\,l}D_{lm}\chi^{||(1)}_{,\,k}
+3\frac{a'}a\chi^{\top(1)'}_{kl}D^{kl}\chi^{||(1)}
+3\frac{a'}a\chi^{\top(1)kl}D_{kl}\chi^{||(1)'}
\nn\\
&
+2\chi^{\top(1)km}D_{ml}\chi^{||(1),\,l}_{,\,k}
-\chi^{\top(1)km}\nabla^2D_{km}\chi^{||(1)}
-D^{km}\chi^{||(1)}\nabla^2\chi^{\top(1)}_{km}
\nn\\
&
-\frac34\chi^{\top(1)km}_{,\,l}D_{km}\chi^{||(1),\,l}
+\frac54\chi^{\top(1)'kl}D_{kl}\chi^{||(1)'}
+\frac32\chi^{\top(1)''}_{kl}D^{kl}\chi^{||(1)}
+\frac32\chi^{\top(1)kl}D_{kl}\chi^{||(1)''}
\nn\\
&
+\frac{1}{2}\chi^{\top(1)}_{lm,\,k}D^{km}\chi^{||(1),\,l}
+3\frac{a'}a\chi^{\top(1)kl}\chi^{\top(1)'}_{kl}
-\chi^{\top(1)km}\nabla^2\chi^{\top(1)}_{km}
-\frac38\chi^{\top(1)km}_{,\,l}\chi^{\top(1),\,l}_{km}
\nn\\
&+\frac58\chi^{\top(1)'kl}\chi^{\top(1)'}_{kl}
+\frac32\chi^{\top(1)kl}\chi^{\top(1)''}_{kl}
+\frac14\chi^{\top(1)km,\,l}\chi^{\top(1)}_{lm,\,k} \ .
\el

\section{ Gauge Transformations  from Synchronous to Synchronous}

First we give the formulas  for the gauge transformation between two general coordinates.
Consider  a general   coordinate transformation   \cite{Weinberg1972, Matarrese98}
\be \label{xmutransf}
x^\mu \rightarrow  \bar x^\mu = x^\mu +\xi^{(1)\mu}
+ \frac{1}{2}\xi^{(1)\mu}_{,\alpha}\xi^{(1)\alpha}
+ \frac{1}{2}\xi ^{(2)\mu},
\ee
where  $\xi^{(1)\mu } $ is a  1st-order vector field,
and  $\xi ^{(2)\mu}$ is an independent vector field  whose magnitude is
  of 2nd order.
The corresponding   transformations of metric \cite{Weinberg1972,Russ1996}
\be\label{trsg}
g_{\mu\nu}(x)=
\frac{ \partial \bar x ^{\alpha}}{ \partial  x^\mu}
\frac{ \partial \bar x ^{\beta }}{ \partial  x^\nu}
\bar g_{\alpha \beta }(\bar x) .
\ee
The metric is written as
$g _{\mu\nu} =  g^{(0)}_{\mu\nu}
+ g^{(1)}_{\mu\nu}
+\frac{1}{2}g_{\mu\nu}^{(2)}$   to the 2nd order,
and similar for $\bar g _{\mu\nu}$.
Eq.(\ref{trsg}) leads to the following transformations  to each  order,
\be\label{metricTrans0th}
g^{(0)}_{\mu\nu} (x)= \bar g^{(0)}_{\mu\nu}(x),
\ee
\be  \label{metricTrans1st}
\bar g^{(1)}_{\mu  \nu  }(  x)
= g^{(1)}_{\mu  \nu  }(  x)
- \mathcal{L}_{\xi^{(1)}} g^{(0)}_{\mu\nu }(x)
,
\ee
\be\label{metricTrans2nd3}
\bar{g}^{(2)}_{\mu\nu }(x)
 =
g^{(2)}_{\mu\nu }(x)
-2\mathcal{L}_{\xi^{(1)}} g^{(1)}_{\mu\nu }(x)
+\mathcal{L}_{\xi^{(1)}}\l(\mathcal{L}_{\xi^{(1)}}g^{(0)}_{\mu\nu }(x)\r)
-\mathcal{L}_{\xi^{(2)}} g^{(0)}_{\mu\nu }(x).
\ee
where the Lie derivative along $\xi^{(1)\mu}$ is defined as
\be\label{LieDerive1}
\mathcal{L}_{\xi^{(1)}} g^{(0)}_{\mu\nu }
\equiv
g^{(0)}_{\mu\nu,\alpha }\xi^{(1)\alpha}
+g^{(0)}_{\mu\alpha} \xi^{(1)\alpha}_{ ,\,\nu}
+g^{(0)}_{\nu\alpha} \xi^{(1)\alpha}_{ ,\,\mu}
\,  ,
\ee
and others  are similarly defined.
It is   checked that,
under the  transformation (\ref{xmutransf}) and (\ref{trsg}),
the spacetime line  element remains invariant,
\be\label{spacetimeInterval}
ds^2=\bar g_{\mu\nu}(\bar x)d\bar x^\mu d\bar x^\nu
=g_{\mu\nu}(x)d x^\mu d x^\nu \, .
\ee
The formulas
(\ref{metricTrans0th} ), (\ref{metricTrans1st} ), and (\ref{metricTrans2nd3})
also apply to the energy-momentum tensor  $T_{\mu\nu} = \rho U^\mu U^\nu$.
Similarly,
under the  coordinate transformation (\ref{xmutransf}),
a scalar function transforms as $f(x)=\bar f(\bar x)$.
By writing  $f(x)=f^{(0)}(x)+f^{(1)}(x)+\frac{1}{2}f^{(2)}(x)$,
one has
\be\label{f0Trans}
\bar f^{(0)}(x)=f^{(0)}(x),
\ee
\be\label{f1Trans}
\bar f^{(1)}(x)=f^{(1)}(x)-\mathcal L_{\xi^{(1)}} f^{(0)}(x),
\ee
\be\label{f2Trans}
\bar f^{(2)}(x)=
f^{(2)}(x)
-2\mathcal{L}_{\xi^{(1)}} f^{(1)}(x)
+\mathcal{L}_{\xi^{(1)}}\l(\mathcal{L}_{\xi^{(1)}} f^{(0)}(x)\r)
-\mathcal{L}_{\xi^{(2)}}f^{(0)}(x),
\ee
where
\be
\mathcal L_{\xi}f\equiv f_{,\, \alpha}\xi^{\alpha} \,  .
\ee
Under   (\ref{xmutransf}),
 a 4-vector $Z^\mu$ transforms as
$\bar Z^\mu(\bar x) = ({\partial\bar x^\mu}/{\partial x^\alpha}) Z^\alpha(x)$.
Writing  $Z^\mu(x)  =Z^{(0)\mu}(x)+Z^{(1)\mu}(x)+\frac{1}{2}Z^{(2)\mu}(x)$,
one has
\be
\bar Z^{(0)\mu}(x)=Z^{(0)\mu}(x),
\ee
\be \label{vect1transf}
\bar Z^{(1)\mu}(x)=Z^{(1)\mu}(x)-\mathcal L_{\xi^{(1)}} Z^{(0)\mu}(x),
\ee
\be \label{vect2transf}
\bar Z^{(2)\mu}(x)=
Z^{(2)\mu}(x)
-2\mathcal{L}_{\xi^{(1)}} Z^{(1)\mu}(x)
+\mathcal{L}_{\xi^{(1)}}\l(\mathcal{L}_{\xi^{(1)}} Z^{(0)\mu}(x)\r)
-\mathcal{L}_{\xi^{(2)}}Z^{(0)\mu}(x),
\ee
where
\be
\mathcal L_{\xi}Z^\mu
\equiv
Z^\mu_{,\alpha}\xi^\alpha-\xi^\mu_{,\alpha}Z^\alpha,
\ee
\be
\mathcal L_{\xi}Z_\mu \equiv
Z_{\mu{,\alpha}}  \xi^\alpha + \xi_{ \mu{,\alpha}}   Z^\alpha.
\ee
The components of the  vector fields  $\xi^{(1)\mu}$ and $\xi^{(2)\mu}$
can be denoted    by the  parameters
\bl
&\xi^{(1)0}=\alpha^{(1)},
 \,\,\,\,\,\,\,\, \xi^{(1)i}=\partial^i\beta^{(1)}+d^{ (1)i} , \label{xi_r}
\\
&\xi^{(2)0}=\alpha^{(2)},
 \,\,\,\,\,\,\,\,  \xi^{(2)i}=\partial^i\beta^{(2)}+d^{ (2)i} , \label{xi_2}
\el
with the constraints $\partial_i d^{(1)i}=0$
and  $\partial_i d^{(2)i}=0$.

We first apply the above formulas
to the 1st-order gauge transformations
between synchronous coordinates.
By (\ref {metricTrans1st}),
requiring  $ \bar g_{00}^{(1)}(x) = g_{00}^{(1)}(x)= 0$ leads to
$\mathcal{L}_{\xi^{(1)}} g^{(0)}_{00 } =0$,
and its solution is
\be \label{xi0trans}
\xi^{(1)0}(\tau, {\bf x}) = \frac{A^{(1)} ({\bf x}) }{a(\tau)}.
\ee
Requiring  $ \bar g^{(1)}_{0i}(x) =  g^{(1)} _{0i}(x)= 0 $
leads to $ \mathcal{L}_{\xi^{(1)}} g^{(0)}_{0i } = 0$, ie,
\be \label{Liexi1}
  -\xi^{(1)0}_{\, ,i} +    \xi^{(1)  }_{i\, ,0}   =0,
\ee
and its solution is
\be  \label{gi0}
 \xi^{(1)}_i (\tau, {\bf x})=
 A^{(1)}({\bf x})_{,i} \int^\tau \frac{d\tau' }{a(\tau')} + C^{(1)}_i(\bf x)  ,
\ee
where  $A^{(1)} ({\bf x}) $ is an arbitrary function,
and $C^{(1)}_i ({\bf x}) $ is  an arbitrary  3-vector.
$A^{(1)} ({\bf x}) $ and $C^{(1)}_i ({\bf x}) $ together
represent 4 degrees of freedom of 1st-order gauge transformation
under consideration.
One  can  write $C^{(1)}_i ({\bf x}) $ as
\be \label{decompC}
C ^{(1)}_{i}  = C\, ^{ \parallel(1)}_{ i}
+  C\, ^{ \bot(1) }_{i}
\ee
where the longitudinal part
 $ C^{\parallel (1)}_{i } = C^{||(1) }_{,\,i} $
 with $C^{||(1)} $ being an arbitrary function,
and    $C^{\perp(1)}_{i} $ is a transverse part so that $C^{\perp(1),\, i}_{i}=0 $.
One can  set  $C^{\perp(1)}_{i}=0$,
since the   1st-order vector metric perturbation is zero
 for   an irrotational dust  in this paper.

Now we explain why a 1st-order vector metric perturbation
$\chi^{\perp(1)}_{ \,ij}$  is a residual gauge mode and can be set $0$.
Assume that it exists as the following
\[ \chi^{\perp(1)}_{ij}=
  v^{(1)}_{i,j}+  v^{(1)}_{j,i}
\]
where the vector satisfies
$\partial^i v^{(1)}_{i}=0$.
It is easy to give the momentum constraint:
$\frac{1}{4} \chi^{\perp(1)',j  }_{ij} =0$,
which leads to $ \frac{1}{4} \nabla^2v^{(1)'}_{i}=0$
and the solution
$v^{ (1)}_{i} = v_i({\bf x}) +e_i(\tau)$,
where $\partial^i v_i ({\bf x}) =0$
and $e_i(\tau)$ is an arbitrary vector depending on $\tau$ only.
By a residual gauge transformation  using
$ \xi^{(1)}_i$ of (\ref{gi0}) with the parameter $C^{\perp(1)}_i$ only,
$\chi^{\perp(1)}_{ \,ij}$ is changed to
$\bar\chi^{\perp(1)}_{ \,ij}
=(v _{i,j}+  v _{j,i})
-( C^{\perp(1)}_{ i,j} +C^{\perp(1)}_{ j,i} )$.
Setting  $C^{\perp(1)}_{ i} =v _i $,
 one gets  $\bar\chi^{\perp(1)}_{ \,ij}=0$.
Hence,  the 1st-order vector  is a gauge mode and can be eliminated.

Under the  transformation of (\ref{xi0trans})   and  (\ref{gi0}),
 the  metric perturbation   changes to
\be\label{gaugetrphi}
\bar  \phi^{(1)} =
  \phi^{(1)}
  + \frac{a'}{a}  \frac{ A^{(1)}}{a}
  +\frac{1}{3}\nabla^2 A^{(1)} \int^\tau \frac{d\tau' }{a(\tau')}
   + \frac{1}{3}\nabla^2 C^{||(1)}    ,
\ee
\be  \label{gaugetrchi}
\bar   \chi^{\parallel(1)} =
 \chi^{\parallel(1)}
  - 2 A^{(1)} \int^\tau \frac{d\tau' }{a(\tau')} -2C^{||(1)}    ,
\ee
\be  \label{gaugetrchiT}
\bar \chi^{\top (1)}_{ij} = \chi^{\top (1)}_{ij}.
\ee
The  results (\ref{xi0trans}), (\ref{gi0}), (\ref{gaugetrphi}),
           (\ref{gaugetrchi}), and (\ref{gaugetrchiT})
are  valid for a general scale factor  $a(\tau)$.
For the MD era with  $a(\tau)\propto \tau^2$,
(\ref{xi0trans}) and  (\ref{gi0}) reduce to
\be\label{alpha1}
\alpha^{(1)} (\tau,\mathbf x) = \frac{A^{(1)}(\mathbf x)}{\tau^2} \, ,
\ee
\be\label{beta1}
\beta^{(1)}_{,\,i}  (\tau,\mathbf x)
= -\frac{A^{(1)}(\mathbf x)_{,\,i}}{\tau}
     +C^{||(1)} (\mathbf x)_{,\,i}
\ee
\be\label{d1}
d^{(1)}_i=0 \, ,
\ee
and (\ref{gaugetrphi}) and (\ref{gaugetrchi}) become
\ba \label{phigauge}
&& \bar  \phi^{(1)}  =
   \phi^{(1)} +    2  \frac{ A^{(1)} ({\bf x}) }{\tau^3 }
   -  \frac{ \nabla^2 A^{(1)}  ({\bf x}) }{3\tau}
   + \frac{1}{3}\nabla^2 C^{||(1)} ({\bf x})   , \\
&& \bar  D_{ij}  \chi^{\parallel(1)} =
   D_{ij}  \chi^{\parallel(1)}  +
 \frac{2  D_{ij} A^{(1)} ({\bf x})}{ \tau}
    -2  D_{ij} C^{||(1)}({\bf x}) . \label{phigauge2}
\ea
From (\ref {delt1}) one sees that
for a given  initial density contrast  $\delta_0^{(1)}(\bf x)$,
one can always choose $C^{||(1)}(\bf x)$ to satisfy
\be  \label{eq22}
\nabla^2 \bar \chi^{\parallel(1)}_0({\bf x})=
 -2 \delta_0^{(1)}(\bf x).
\ee
The synchronous-to-synchronous transformation of 1st-order density perturbation
can be also derived from the following
\be \label{Tmunutransf}
\bar T^{(1)}_{00 }(  x)
        =  T^{(1)}_{00 }(  x) - \mathcal{L}_{\xi^{(1)}} T^{(0)}_{00}(x) ,
\ee
as an application of (\ref  {metricTrans1st}).
Up to the 1st order,  one has
$T^{(1)}_{00}  = \rho^{(1)} a^2 $, $\bar U^{(1)}_0 =0$,
   $ \bar T^{(1)}_{00} = \bar \rho^{(1)} a^2 $,
 and  $\mathcal{L}_{\xi^{(1)}} T^{(0)}_{00 }
=  -6  a^2 \rho^{(0)}  \frac{A^{(1)}}{\tau^3}$,
leading to the result
\be  \label{deltarho}
\bar   \rho ^{(1)}
 =   \rho ^{(1)}   + 6 \rho^{(0)}   \frac{A^{(1)}}{\tau^3}.
\ee
This result also follows from
$\bar \rho ^{(1)} = \rho ^{(1)} - \mathcal{L}_{\xi^{(1)}} \rho^{(0)}$
 as an application of  (\ref{f1Trans}).
In terms of the 1st-order density contrast,
(\ref {deltarho}) gives
\be \label{delta1transf}
\bar \delta   ^{(1)}  = \delta   ^{(1)} +6   \frac{A^{(1)}} {\tau^3} \, .
\ee
Thus,  the residual gauge mode  of $ \delta   ^{(1)} $ is $\tau^{-3}$.

In this paper on the  dust model,
the 4-velocity   $U^{(0)\mu} =(a^{-1},0,0,0)$,
and its 1st-order perturbation $U^{(1)\mu} =0 $  has been  taken.
When  one further  requires the transformed $\bar U^{(1) i } =0$ \cite{Russ1996},
the formula   (\ref{vect1transf}) leads to
\be \label{A1const}
   A^{(1)}({\bf x})_{,i}  =0,
\ee
i.e,  $A^{(1)} = const$,
so that   (\ref{alpha1}),   (\ref{beta1}),  (\ref{phigauge}),
and (\ref{phigauge2})
  reduce to
\bl  \label{xitr}
\alpha^{(1)} (\tau ) & = \frac{A^{(1)}   }{\tau^2}, \\
 \label{beta1s}
\beta^{(1)}_{, i}  (\mathbf x) & =   C^{||(1)}_{, i} (\mathbf x) \, ,
\el
\bl  \label{gaugemodes}
 \bar   \phi^{(1)} & =  \phi^{(1)} +     2  \frac{ A^{(1)}}{\tau^3 }
     + \frac{1}{3}\nabla^2 C^{||(1)} ({\bf x})   , \\
 \bar D_{ij}  \chi^{\parallel(1)} & =  D_{ij} \chi^{\parallel(1)}
     -2  D_{ij}  C^{||(1)} ({\bf x}) . \label{phigauge3}
\el

If starting with a   velocity $U^{(1)\mu} \ne 0$ in a given synchronous coordinate,
we can make a coordinate transform to render the velocity vanish
at $\bar U^{(1)\mu} = 0$;
however, at the same time,  the new coordinate  is  no longer synchronous in general.
This is explained as follows:
by (\ref{vect1transf}),
the vector $U^{(1)\mu}$ transforms  as
\be\label{vect1transf2}
\bar U^{(1)0} = U^{(1)0}
-[U^{(0)0}_{,0}\xi^{(1)0}
-\xi^{(1)0}_{,0}U^{(0)0}]
,
~~~~~~~
\bar U^{(1)i} = U^{(1)i}
+\xi^{(1)i}_{,0}U^{(0)0},
\nonumber
\ee
under a general $\xi^{(1)\nu}$.
Requiring $\bar U^{(1)0} =0,\,\,\bar U^{(1)i} =0$ in the new coordinate leads to
\[
 U^{(1)0} -[U^{(0)0}_{,0}\xi^{(1)0} -\xi^{(1)0}_{,0}U^{(0)0}]=0, \,\,\,
 U^{(1)i}+\xi^{(1)i}_{,0}U^{(0)0} =0.
\]
The solution of the transformation vector is
\ba\label{xxii}
&&\xi^{(1)0}= \frac{H(\bf x)}{a(\tau)}
     - \frac{1}{a(\tau)} \int^\tau a^2(\tau') U^{(1)0}d\tau',\nonumber \\
&&\xi^{(1)i}= -\int^\tau   a(\tau')U^{(1)i} d \tau' +E^i({\bf x}),
\ea
where $H({\bf x})$ is an arbitrary function and $E^i({\bf x})$ is an arbitrary vector.
For this solution to satisfy
the synchronous-to-synchronous conditions (\ref{xi0trans})
and (\ref{gi0}),
the given velocity must be of the following special form
\be\label{U1Condition1}
 U^{(1)0}=0,\,\,\,\,  U^{(1)i}  \propto \frac{1}{a^2}.
\ee
In fact, for a pressureless dust
with the energy-momentum tensor  $T^{\mu\nu}=\rho U^\mu U^\nu$
in a  synchronous coordinate,
one can write $U^{\mu}=U^{(0)\mu}+U^{(1)\mu}$,
and show that $U^{(1)0}=0$ as a result of  $U^\mu U_\mu=-1$,
that $(a^2 U^{(1)i})'=0$ as a result of
the momentum conservation $T^{i\mu}\,_{;\mu}=0$  at the 1st order,
that is,  (\ref{U1Condition1}) is satisfied by a pressureless dust.
Thus, one can set the 1st-order velocity $\bar U^{(1)\mu}=0$
by  a 1st-order gauge transformation.
Similar calculations analysis can be also performed for the 2nd order,
leading to  a similar result,
i.e, the 2nd-order velocity $\bar U^{(2)\mu}$ can be brought zero
by  a 2nd-order gauge transformation.
Therefore, for a pressureless dust in a synchronous coordinate,
one can take the 4-velocity  $ U^\mu=(a^{-1},0,0,0)$.

Now  we  determine   the 2nd-order vector $\xi ^{(2)}$
of synchronous-to-synchronous  residual gauge transformations.
By the requirement
  $\bar g^{(2)}_{00}(x)=g^{(2)}_{00}(x)=0$,
the formula  (\ref{metricTrans2nd3}) gives
\bl\label{alpha2_1}
0 =0  -2 \mathcal L_{\xi^{(1)}} g^{(1)}_{00}
 +\mathcal L_{\xi^{(1)}}( \mathcal L_{\xi^{(1)}} g^{(0)}_{00} )
 -\mathcal L_{\xi^{(2)}} g^{(0)}_{00}.
\el
 As is checked, $ \mathcal{L}_{\xi^{(1)}} g^{(1)}_{00} =  0 $
 and  $\mathcal L_{\xi^{(1)}} \mathcal L_{\xi^{(1)}} g^{(0)}_{00}  =0$
 for the given $\xi^{(1)\mu}$,
so that the above reduces to $\mathcal L_{\xi^{(2)}} g^{(0)}_{00} =0$,
which    yields a solution
\be  \label{alpha2_3}
\alpha^{(2)}
=\frac{ A^{(2)}(\mathbf x)}{a(\tau)} \, ,
\ee
with $A^{(2)}({\bf x})$ being an arbitrary   function.
By the requirement    $\bar g^{(2)}_{0i}(x)=g^{(2)}_{0i}(x)=0$,
the formula   (\ref{metricTrans2nd3}) gives
\be \label{2nd0i}
0=0-2\mathcal L_{\xi^{(1)}} g^{(1)}_{0i}
+\mathcal L_{\xi^{(1)}}\l(\mathcal L_{\xi^{(1)}} g^{(0)}_{0i}\r)
-\mathcal L_{\xi^{(2)}} g^{(0)}_{0i}.
\ee
which leads to the equation
\bl\label{xi2dt}
   \frac{d  \xi^{(2)i} }{d\tau}
 & =   2\l( 2 \phi^{(1)}\delta_{ik} -  \chi^{(1)}_{ik} \r)
    \frac{A^{(1),\,k}}{a}
     +  2 a'  \frac{ A^{(1)}_{,\, i} A^{(1)} }{a^3 }    \nn  \\
&
   +  2 \frac{ A^{(1),\, k}}{a}   A^{(1)} _{,\,ik} \int^\tau \frac{d\tau'}{a(\tau')}
   +  2 \frac{A^{(1),\, k}}{a}   C^{||(1)} _{,\,ik}
   + \frac{ A^{(2)}_{,i}  }{a} .
\el
Its solution is
\bl\label{xi2ge}
\xi^{(2)}_i (\tau,{\bf x})
=&
4 A^{(1)}({\bf x})_{,i}\int^\tau \frac{\phi^{(1)}(\tau',{\bf x})}{a({\tau'})}d\tau'
- 2A^{(1)}({\bf x})^{,k}\int^\tau  \frac{\chi^{(1)}_{ki}(\tau',{\bf x})}{a(\tau')}d\tau'
\nn\\
&
-\frac{1}{a^2}A^{(1)}({\bf x})A^{(1)}({\bf x})_{,i}
+ 2 A^{(1)}({\bf x})^{,k}A^{(1)}({\bf x})_{,ki} \int^\tau    \frac{ d\tau' }{a(\tau')}
    \int^{\tau'} \frac{d\tau'' }{a(\tau'')}
\nn\\
&
+ 2 A^{(1)}({\bf x})^{,k}C^{||(1)}({\bf x})_{,ki}\int^{\tau} \frac{d\tau'}{a(\tau')}
+A^{(2)}({\bf x})_{,i}\int^\tau \frac{d\tau'}{a(\tau')}
+ C^{(2)}_i ({\bf x}),
\el
where $ C^{(2)}_i $  is an integration constant 3-vector,
and can be decomposed as
\be
 C^{(2)}_i   =     C^{||(2)}_{,\,i}   +C^{\perp(2)}_{\,i} .
\ee
$A^{(2)} ({\bf x}) $ and $C^{(2)}_i ({\bf x}) $ together
represent  4 degrees of freedom of 2nd-order gauge transformation
under consideration.
From Eq.(\ref{xi2ge}),
one has
\bl\label{beta2_1}
\beta^{(2)}=&
\nabla^{-2}\Big\{
[\nabla^2A^{(1)}({\bf x})]\int^\tau\frac{4\phi^{(1)}(\tau',{\bf x})}{a({\tau'})}d\tau'
+A^{(1)}({\bf x})_{,k}\int^\tau\frac{4\phi^{(1)}(\tau',{\bf x})^{,k}}{a({\tau'})}d\tau'
\nn\\
&
-A^{(1)}({\bf x})^{,\,ki}\int^\tau\frac{2\chi^{(1)}_{ki}(\tau',{\bf x})}{a(\tau')}d\tau'
-A^{(1)}({\bf x})^{,\,k}\int^\tau\frac{2\chi^{(1)}_{ki}(\tau',{\bf x})^{,\,i}}{a(\tau')}d\tau'
\nn\\
&
+ 2 A^{(1)}({\bf x})^{,\,k i}C^{||(1)}({\bf x})_{,\,k i}\int^\tau \frac{d\tau'}{a(\tau')}
+ 2 A^{(1)}({\bf x})^{,k}\nabla^2C^{||(1)}({\bf x})_{,k} \int^\tau\frac{d\tau'}{a(\tau')}\Big\}
\nn\\
&-\frac{1}{2a^2(\tau)}A^{(1)}({\bf x})A^{(1)}({\bf x})
+ A^{(1)}({\bf x})^{,k}A^{(1)}({\bf x})_{,k}
    \int^\tau  \frac{d\tau'}{a(\tau')}
    \int^{\tau'}\frac{d\tau'' }{a(\tau'')}
\nn\\
&
+A^{(2)}({\bf x})\int^\tau \frac{d\tau'}{a(\tau')}
+C^{||(2)} ({\bf x})
\,,
\el
\bl   \label{d2_2}
d^{(2)}_i=&\xi^{(2)}_i-\beta^{(2)}_{,i}
\nn\\
=&
\partial_i\nabla^{-2}\Big\{
-[\nabla^2A^{(1)}({\bf x})]\int^\tau \frac{4\phi^{(1)}(\tau',{\bf x})}{a({\tau'})}d\tau'
-A^{(1)}({\bf x})_{,k}\int^\tau \frac{4\phi^{(1)}(\tau',{\bf x})^{,k}}{a({\tau'})}d\tau'
\nn\\
&
+2A^{(1)}({\bf x})^{,\,kl}\int^\tau \frac{ \chi^{(1)}_{kl}(\tau',{\bf x})}{a(\tau')}d\tau'
+2A^{(1)}({\bf x})^{,\,k}\int^\tau \frac{\chi^{(1)}_{kl}(\tau',{\bf x})^{,\,l}}{a(\tau')}d\tau'
\nn\\
&
- 2 A^{(1)}({\bf x})^{,\,kl}C^{||(1)}({\bf x})_{,\,kl}\int^\tau \frac{d\tau'}{a(\tau')}
- 2 A^{(1)}({\bf x})^{,k} \nabla^2C^{||(1)}({\bf x})_{,k}  \int^\tau \frac{d\tau'}{a(\tau')}\Big\}
\nn\\
&
+ 4A^{(1)}({\bf x})_{,i}\int^\tau  \frac{\phi^{(1)}(\tau',{\bf x})}{a({\tau'})}d\tau'
-2A^{(1)}({\bf x})^{,k}\int^\tau  \frac{\chi^{(1)}_{ki}(\tau',{\bf x})}{a(\tau')}d\tau'
\nn\\
&+ 2 A^{(1)}({\bf x})^{,k}C^{||(1)}({\bf x})_{,ki}\int^\tau \frac{d\tau'}{a(\tau')}
+ C^{\perp(2)}_i ({\bf x}).
\el
The   results  (\ref{alpha2_3}), (\ref{beta2_1}),
and (\ref{d2_2})
are  valid for a general $a(\tau)$.

We now determine the transformation of 2nd-order metric perturbations.
Applying  the formula  (\ref{metricTrans2nd3})
to  the  (ij) components yields
\bl
\bar \phi^{(2)}  = &\phi^{(2)}
- \l[\l(4\frac{a'}{a}\phi^{(1)}+2\phi^{(1)'}\r)
    +\l(\frac{a''}{a}+\frac{a'\,^2}{a^2}\r)\alpha^{(1)}
    +\frac{a'}{a}\alpha^{(1)'}\r]\alpha^{(1)}   \nn\\
&  -\frac13\l(4\phi^{(1)}
    +\alpha^{(1)}\partial_0
    +\beta^{(1),k}\partial_k
    +4\frac{a'}{a}\alpha^{(1)} \r)  \nabla^2  \beta^{(1)}
    - \Big(2\phi^{(1)}_{,\,k}  + \frac{a'}{a}\alpha^{(1)}_{,\,k}\Big) \beta^{(1),k} \nn \\
&  -\frac23\l(-\chi^{(1)}_{kl}
   +\beta^{(1)}_{,kl}\r)\beta^{(1),\,kl}
   +\frac{a'}{a}\alpha^{(2)}  +\frac13\nabla^2\beta^{(2)}, \label{phi2transform0}
            \\
 \bar\chi^{(2)}_{ij}   = & \chi^{(2)}_{ij} +W_{ij} \,,  \label{chi2transF2}
\el
where
\bl\label{Wij1}
W_{ij}
 \equiv   &
-\frac{2}{a}\chi^{(1)'}_{ij}A^{(1)}
-4\frac{a'}{a^2}\chi^{(1)}_{ij}A^{(1)}
-2\chi^{(1)}_{ij,\,k}A^{(1),k} \int^{\tau} \frac{d\tau' }{a(\tau')}
-2 \chi^{(1)}_{ij,\,k}C^{||(1),\,k}
\nn\\
&
+8 \phi^{(1)}D_{ij}A^{(1)}  \int^{\tau}  \frac{d\tau' }{a(\tau')}
+8\phi^{(1)}D_{ij}C^{||(1)}
+\frac{2}{a^2}A^{(1)}D_{ij}A^{(1)}
\nn\\
&
-2 A^{(1),k}D_{ij}A^{(1)}_{,k} \l[\int^{\tau}  \frac{d\tau' }{a(\tau')}\r]^2
-2 A^{(1),k}D_{ij}C^{||(1)}_{,k} \int^{\tau}  \frac{d\tau' }{a(\tau')}
\nn\\
&
-2  C^{||(1),k}D_{ij}A^{(1)}_{,k}   \int^{\tau}  \frac{d\tau' }{a(\tau')}
-2C^{||(1),k}D_{ij}C^{||(1)}_{,k}
\nn\\
&
+\frac{8a'}{a^2} A^{(1)}D_{ij}A^{(1)}   \int^{\tau}  \frac{d\tau'}{a(\tau')}
+\frac{8a'}{a^2}A^{(1)}D_{ij}C^{||(1)}
\nn\\
&
-4
 \Big(
 \chi^{(1)}_{k(i}A^{(1),k}_{,j)}
-\frac13\chi^{(1)}_{kl}A^{(1),kl}\delta_{ij}
\Big)  \int^{\tau}  \frac{d\tau' }{a(\tau')}
-4\Big(  \chi^{(1)}_{k(i}C^{||(1),k}_{,j)}
-\frac13\chi^{(1)}_{kl}C^{||(1),kl}\delta_{ij}
\Big)
\nn\\
&
+2D_{ij}\Big(
A^{(1)}_{,k}A^{(1),k} \l[\int^{\tau}  \frac{d\tau' }{a(\tau')}\r]^2
+2 A^{(1)}_{,k}C^{||(1),k}  \int^{\tau}  \frac{d\tau' }{a(\tau')}
+C^{||(1)}_{,k}C^{||(1),k}
\Big)
\nn\\
&
 -2\l( D_{ij}\beta^{(2)} + d^{(2)}_{(i,j)}  \r) .
\el
Eq.(\ref {chi2transF2}) is to be decomposed into
  scalar,  vector,   tensor:
$\bar\chi^{ (2)}_{ij}
 =  \bar\chi^{\top(2)}_{ij}+\bar\chi^{\perp(2)}_{ij}
 +D_{ij}\bar\chi^{||(2)} $.
 By   calculations,
we get
\be\label{chi||2transF1}
 \bar\chi^{||(2)}
 =    \chi^{||(2)}
+\frac32    \nabla^{-2}\nabla^{-2}W_{kl}^{,kl} \,  ,
\ee
\be\label{chiVec2TransF1}
\bar\chi^{\perp(2)}_{ij}
=
\chi^{\perp(2)}_{ij}
+\nabla^{-2}\l( W^{, \, k}_{ki,j} + W^{,\, k}_{kj,i}\r )
-2\nabla^{-2}\nabla^{-2}W_{kl,ij}^{,kl}\, ,
\ee
\be\label{chiT2transF1}
\bar\chi^{\top(2)}_{ij}
=\chi^{\top(2)}_{ij}
+W_{ij}-\nabla^{-2}\l(W^{,k}_{ki,j}+W^{,k}_{kj,i}\r)
+\frac12\nabla^{-2}\nabla^{-2}W_{kl,ij}^{,kl}
+\frac{ \delta_{ij}}{2} \nabla^{-2}W_{kl}^{,kl} \,.
\ee
Substituting    (\ref{Wij1}) into
Eq.(\ref{chi||2transF1}),  Eq.(\ref{chiVec2TransF1}),   and Eq.(\ref{chiT2transF1}),
we obtain the residual gauge transformations
of the  scalar, vector and tensor  as the following
{\large
\bl\label{chi||2transF2}
 \bar\chi^{||(2)}
=&
 \chi^{||(2)}
+ \frac{1}{a^2}   \big[
A^{(1)}A^{(1)}
+2\nabla^{-2}\big(A^{(1)}\nabla^2A^{(1)}\big)
+3\nabla^{-2}\nabla^{-2}\big(
A^{(1),kl}A^{(1)}_{,\,kl}
-\nabla^2A^{(1)}\nabla^2A^{(1)}\big)
\big]
\nn\\
&
+\frac{4a'}{a^2}  \l[\int^\tau  \frac{d\tau' }{a(\tau')}\r]
  \big[
2\nabla^{-2}\big(A^{(1)}\nabla^2A^{(1)}\big)
+3\nabla^{-2}\nabla^{-2}\big(
A^{(1),kl}A^{(1)}_{,\,kl}
-\nabla^2A^{(1)}\nabla^2A^{(1)}\big)
\big]
\nn\\
&
+  \frac{a'}{a^2}   \big[
8\nabla^{-2}\big(A^{(1)}\nabla^2C^{||(1)}\big)
+6\nabla^{-2}\nabla^{-2}\big(
-\chi^{(1),kl}_{kl}A^{(1)}
-\chi^{(1)}_{kl}A^{(1),kl}
-2\chi^{(1),k}_{kl}A^{(1),l}
\nn\\
&
+2A^{(1),kl}C^{||(1)}_{,\,kl}
-2\nabla^2A^{(1)}\nabla^2C^{||(1)}\big)
\big]
-  \frac{3}{a}   \nabla^{-2} \nabla^{-2}\big[
\chi^{(1)',kl}_{kl}A^{(1)}
+\chi^{(1)'}_{kl}A^{(1),kl}
+2\chi^{(1)',k}_{kl}A^{(1),l} \big]
\nn \\
&
-2A^{(1),k}A^{(1)}_{,k}   \int^\tau  d\tau'  \frac{1}{a(\tau')}
    \int^{\tau'}\frac{d\tau'' }{a(\tau'')}
\nn \\
&
+\l[\int^\tau  \frac{d\tau' }{a(\tau')}\r]^2   \big[
2A^{(1)}_{,k}A^{(1),k}
-2\nabla^{-2}\big(A^{(1),k}\nabla^2A^{(1)}_{,k}\big)
+3\nabla^{-2}\nabla^{-2}\big(
\nabla^2A^{(1),k}\nabla^2A^{(1)}_{,k}
-A^{(1),klm}A^{(1)}_{,\,klm}\big)
\big]
\nn\\
&
+ \l[\int^\tau \frac{d\tau' }{a(\tau')} \r] \Big[
2A^{(1),k}C^{||(1)}_{,k}
+2\nabla^{-2}\big(4\phi^{(1)}\nabla^2A^{(1)}
+\chi^{(1)}_{kl}A^{(1),kl}
-2A^{(1),k}\nabla^2C^{||(1)}_{,k}
\big)
\nn\\
&
+3\nabla^{-2}\nabla^{-2}\big(
-\chi^{(1),klm}_{kl}A^{(1)}_{,m}
-3\chi^{(1)}_{kl,\,m}A^{(1),klm}
-4\chi^{(1),\,k}_{kl,\,m}A^{(1),\,lm}
-2\chi^{(1),\,k}_{kl}\nabla^2A^{(1),\,l}
\nn\\
&
-2\chi^{(1)}_{kl}\nabla^2A^{(1),\,kl}
+4\phi^{(1),kl}A^{(1)}_{,\,kl}
-4\nabla^2\phi^{(1)}\nabla^2A^{(1)}
+2\nabla^2A^{(1),k}\nabla^2C^{||(1)}_{,k}
-2A^{(1),klm}C^{||(1)}_{,\,klm}
\big)
\Big]
\nn\\
&
+   \Big[2C^{||(1)}_{,k}C^{||(1),k}
+\nabla^{-2}\big(8\phi^{(1)}\nabla^2C^{||(1)}
+2\chi^{(1)}_{kl}C^{||(1),kl}
-2 C^{||(1),k}\nabla^2C^{||(1)}_{,k}\big)
\nn\\
&
+3\nabla^{-2}\nabla^{-2}\big(
4\phi^{(1),kl}C^{||(1)}_{,\,kl}
-4\nabla^2\phi^{(1)}\nabla^2C^{||(1)}
-\chi^{(1),klm}_{kl}C^{||(1)}_{,m}
-3\chi^{(1)}_{kl,m}C^{||(1),klm}
\nn\\
&
-4\chi^{(1),k}_{kl,m}C^{||(1),lm}
-2\chi^{(1),\,k}_{kl}\nabla^2C^{||(1),\,l}
-2\chi^{(1)}_{kl}\nabla^2C^{||(1),\,kl}
+\nabla^2C^{||(1),k}\nabla^2C^{||(1)}_{,k}
-C^{||(1),klm}C^{||(1)}_{,\,klm}
\big)
\Big]  \nn \\
&
-4 \nabla^{-2}\Big[
\nabla^2A^{(1)} \int^\tau\frac{2\phi^{(1)}(\tau',{\bf x})}{a({\tau}')}d\tau'
+A^{(1)}_{,k} \int^\tau\frac{2\phi^{(1)}(\tau',{\bf x})^{,k}}{a({\tau}')}d\tau'
- A^{(1),\,kl} \int^\tau \frac{\chi^{(1)}_{kl}(\tau',{\bf x})}{a(\tau')}d\tau'
\nn\\
&
- A^{(1),\,k}  \int^\tau \frac{\chi^{(1)}_{km}(\tau',{\bf x})^{,\,m}}{a(\tau')}d\tau' \Big]
-2A^{(2)}\int^\tau \frac{d\tau' }{a(\tau')}
-2C^{||(2)}
   \, ,
\el
}
{\large
\bl\label{chiPerp2TransF}
\bar\chi^{\perp(2)}_{ij}
= &\chi^{\perp(2)}_{ij}
+ \frac{1}{a^2} \Big[
-2\partial_i\nabla^{-2}\big( A^{(1)}_{,j}\nabla^2A^{(1)} \big)
+\partial_i\partial_j\nabla^{-2}\big(
A^{(1),k}A^{(1)}_{,\,k}\big)
\nn\\
&
-2\partial_i\partial_j\nabla^{-2}\nabla^{-2}\big(
A^{(1),kl}A^{(1)}_{,\,kl}
-\nabla^2A^{(1)}\nabla^2A^{(1)}
\big)
\Big]
+\frac{4a'}{a^2} \l[\int^\tau \frac{d\tau' }{a(\tau')}\r] \Big[
-2\partial_i\nabla^{-2}\big( A^{(1)}_{,j}\nabla^2A^{(1)} \big)
\nn\\
&
+\partial_i\partial_j\nabla^{-2}\big(
A^{(1),k}A^{(1)}_{,\,k}
\big)
-2\partial_i\partial_j\nabla^{-2}\nabla^{-2}\big(
A^{(1),kl}A^{(1)}_{,\,kl}
-\nabla^2A^{(1)}\nabla^2A^{(1)}
\big)
\Big]
\nn\\
&
+\frac{2a'}{a^2} \Big[
2\partial_i\nabla^{-2}\big( 2A^{(1),k}C^{||(1)}_{,\,kj}
-2A^{(1)}_{,j}\nabla^2C^{||(1)} -\chi^{(1),k}_{kj}A^{(1)}
-\chi^{(1)}_{kj}A^{(1),k} \big)
\nn\\
&
+2\partial_i\partial_j\nabla^{-2}\nabla^{-2}\big(
2\nabla^2A^{(1)}\nabla^2C^{||(1)}
-2A^{(1),kl}C^{||(1)}_{,\,kl}
+\chi^{(1),kl}_{kl}A^{(1)}
+\chi^{(1)}_{kl}A^{(1),kl}
+2\chi^{(1),k}_{kl}A^{(1),l}
\big)
\Big] \nn \\
&
- \frac{1}{a} \Big[
2\partial_i\nabla^{-2}\big( \chi^{(1)',k}_{kj}A^{(1)} +\chi^{(1)'}_{kj}A^{(1),k} \big)
-2\partial_i\partial_j\nabla^{-2}\nabla^{-2}\big(
\chi^{(1)',kl}_{kl}A^{(1)}
+\chi^{(1)'}_{kl}A^{(1),kl}
+2\chi^{(1)',k}_{kl}A^{(1),l}
\big)
\Big] \nn \\
&
+ \l[\int^\tau \frac{d\tau' }{a(\tau')}\r]^2\big[
2\partial_i\nabla^{-2}\big( A^{(1)}_{,kj}\nabla^2A^{(1),k} \big)
-\partial_i\partial_j\nabla^{-2}\big(
A^{(1),km}A^{(1)}_{,\,km}
\big)
\nn\\
&
+2\partial_i\partial_j\nabla^{-2}\nabla^{-2}\big(
A^{(1),klm}A^{(1)}_{,\,klm}
-\nabla^2A^{(1),k}\nabla^2A^{(1)}_{,k}
\big)
\big]  \nn \\
&
- \l[\int^\tau \frac{d\tau' }{a(\tau')} \r]\Big[
2\partial_i\nabla^{-2}\big(
4\phi^{(1)}_{,j}\nabla^2A^{(1)}
-4\phi^{(1),k}A^{(1)}_{,\,kj}
+\chi^{(1),k}_{kj,m}A^{(1),m}
+2\chi^{(1)}_{kj,m}A^{(1),km}
\nn\\
&
+\chi^{(1),\,l}_{kl}A^{(1),k}_{,j}
+\chi^{(1)}_{kl}A^{(1),kl}_{,j}
+\chi^{(1)}_{kj}\nabla^2A^{(1),\,k}
+2A^{(1),km}C^{||(1)}_{,kmj}
-2A^{(1)}_{,kj}\nabla^2C^{||(1),k}
\big)
\nn\\
&
-2\partial_i\partial_j\nabla^{-2}\nabla^{-2}\big(
4\nabla^2\phi^{(1)}\nabla^2A^{(1)}
-4\phi^{(1),kl}A^{(1)}_{,\,kl}
+\chi^{(1),kl}_{kl,m}A^{(1),m}
+3\chi^{(1)}_{kl,m}A^{(1),klm}
+4\chi^{(1),k}_{kl,m}A^{(1),lm}
\nn\\
&
+2\chi^{(1),\,k}_{kl}\nabla^2A^{(1),\,l}
+2\chi^{(1)}_{kl}\nabla^2A^{(1),\,kl}
+2A^{(1),klm}C^{||(1)}_{,\,klm}
-2\nabla^2A^{(1),k}\nabla^2C^{||(1)}_{,k}
\big)
\Big]
\nn\\
&
+\Big[
-2\partial_i\nabla^{-2}\big(
4\phi^{(1)}_{,j}\nabla^2C^{||(1)}
-4\phi^{(1),k}C^{||(1)}_{,\,kj}
+\chi^{(1),k}_{kj,m}C^{||(1),m}
+2\chi^{(1)}_{kj,m}C^{||(1),km}
\nn\\
&
+\chi^{(1),\,l}_{kl}C^{||(1),k}_{,j}
+\chi^{(1)}_{kl}C^{||(1),kl}_{,j}
+\chi^{(1)}_{kj}\nabla^2C^{||(1),\,k}
+C^{||(1),km}C^{||(1)}_{,\,kmj}
-C^{||(1)}_{,kj}\nabla^2C^{||(1),k}
\big)
\nn\\
&
+2\partial_i\partial_j\nabla^{-2}\nabla^{-2}\big(
4\nabla^2\phi^{(1)}\nabla^2C^{||(1)}
-4\phi^{(1),kl}C^{||(1)}_{,\,kl}
+\chi^{(1),kl}_{kl,m}C^{||(1),m}
+4\chi^{(1),k}_{kl,m}C^{||(1),lm}
\nn\\
&
+3\chi^{(1)}_{kl,m}C^{||(1),klm}
+2\chi^{(1),\,k}_{kl}\nabla^2C^{||(1),\,l}
+2\chi^{(1)}_{kl}\nabla^2C^{||(1),\,kl}
+C^{||(1),klm}C^{||(1)}_{,\,klm}
-\nabla^2C^{||(1),k}\nabla^2C^{||(1)}_{,k}
\big)
\Big]
\nn\\
&
+\partial_i\Big[
A^{(1),k}\int^\tau \frac{2\chi^{(1)}_{kj}(\tau',{\bf x})}{a(\tau')}d\tau'
-A^{(1)}_{,j}\int^\tau \frac{4\phi^{(1)}(\tau',{\bf x})}{a({\tau'})}d\tau'
\Big]   \nn \\
&
+\partial_i\partial_j\nabla^{-2}\Big[
\nabla^2A^{(1)} \int^\tau \frac{4\phi^{(1)}(\tau',{\bf x})}{a({\tau'})}d\tau'
+A^{(1)}_{,k}\int^\tau \frac{4\phi^{(1)}(\tau',{\bf x})^{,k}}{a(\tau')}d\tau'
-A^{(1),\,kl}\int^\tau \frac{2\chi^{(1)}_{kl}(\tau' ,{\bf x})}{a(\tau' )}d\tau
\nn\\
&
-A^{(1),\,k}\int^\tau \frac{2\chi^{(1)}_{km}(\tau' ,{\bf x})^{,\, m}}{a(\tau' )}d\tau' \Big]
  -C^{\perp(2)}_{i,j}
+(i \leftrightarrow j ) \ ,
\el
}
showing  that transformation of $ \chi^{\perp(2)}_{ij}$
  depends on   $\xi^{(2)\mu}$ only through $C^{\perp(2)}_{(i,j)}$,
      {\large
\bl\label{chiT2transF2}
\bar\chi^{\top(2)}_{ij}
=& \chi^{\top(2)}_{ij}
+\l[\frac{1}{a^2}\r]\big[
\delta_{ij}\nabla^{-2}\big(
A^{(1),kl}A^{(1)}_{,\,kl}
-\nabla^2A^{(1)}\nabla^2A^{(1)}
\big)
+4\nabla^{-2}\big(
A^{(1)}_{,ij}\nabla^2A^{(1)}
-A^{(1),k}_{,i}A^{(1)}_{,\,kj}
\big)
\nn\\
&
+\partial_i\partial_j\nabla^{-2}\nabla^{-2}\big(
A^{(1),kl}A^{(1)}_{,\,kl}
-\nabla^2A^{(1)}\nabla^2A^{(1)}
\big)
\big]
+\frac{4a'}{a^2}\l[\int^\tau  \frac{d\tau' }{a(\tau')}\r]\big[
\delta_{ij}\nabla^{-2}\big(
A^{(1),kl}A^{(1)}_{,\,kl}
\nn\\
&
-\nabla^2A^{(1)}\nabla^2A^{(1)}
\big)
+4\nabla^{-2}\big(
A^{(1)}_{,ij}\nabla^2A^{(1)}
-A^{(1),k}_{,i}A^{(1)}_{,\,kj}
\big)
+\partial_i\partial_j\nabla^{-2}\nabla^{-2}\big(
A^{(1),kl}A^{(1)}_{,\,kl}
\nn\\
&
-\nabla^2A^{(1)}\nabla^2A^{(1)}
\big)
\big]
+\l[\frac{2a'}{a^2}\r]\big[
-\delta_{ij}\nabla^{-2}\big(
\chi^{(1),kl}_{kl}A^{(1)}
+\chi^{(1)}_{kl}A^{(1),kl}
+2\chi^{(1),k}_{kl}A^{(1),l}
\nn\\
&
+2\nabla^2A^{(1)}\nabla^2C^{||(1)}
-2A^{(1),kl}C^{||(1)}_{,\,kl}
\big)
-2\chi^{(1)}_{ij}A^{(1)}
+2\partial_i\nabla^{-2}\big(
\chi^{(1),k}_{kj}A^{(1)}
+\chi^{(1)}_{kj}A^{(1),k}
\big)
\nn\\
&
+2\partial_j\nabla^{-2}\big(
\chi^{(1),k}_{ki}A^{(1)}
+\chi^{(1)}_{ki}A^{(1),k}
\big)
+4\nabla^{-2}\big(
A^{(1)}_{,ij}\nabla^2C^{||(1)}
+C^{||(1)}_{,ij}\nabla^2A^{(1)}
-A^{(1),k}_{,i}C^{||(1)}_{,\,kj}
\nn\\
&
-A^{(1),k}_{,j}C^{||(1)}_{,\,ki}
\big)
-\partial_i\partial_j\nabla^{-2}\nabla^{-2}\big(
\chi^{(1),kl}_{kl}A^{(1)}
+\chi^{(1)}_{kl}A^{(1),kl}
+2\chi^{(1),k}_{kl}A^{(1),\,l}
+2A^{(1),kl}C^{||(1)}_{,\,kl}
\nn\\
&
-2\nabla^2A^{(1)}\nabla^2C^{||(1)}
\big)
\big]
-\l[\frac{1}{a}\r]\big[
\delta_{ij}\nabla^{-2}\big(\chi^{(1)',kl}_{kl}A^{(1)}
+\chi^{(1)'}_{kl}A^{(1),kl}
+2\chi^{(1)',k}_{kl}A^{(1),l}
\big)
\nn\\
&
+2\chi^{(1)'}_{ij}A^{(1)}
-2\partial_i\nabla^{-2}\big(
\chi^{(1)',k}_{kj}A^{(1)}
+\chi^{(1)'}_{kj}A^{(1),k}
\big)
-2\partial_j\nabla^{-2}\big(
\chi^{(1)',k}_{ki}A^{(1)}
+\chi^{(1)'}_{ki}A^{(1),k}
\big)
\nn\\
&
+\partial_i\partial_j\nabla^{-2}\nabla^{-2}\big(
\chi^{(1)',kl}_{kl}A^{(1)}
+\chi^{(1)'}_{kl}A^{(1),kl}
+2\chi^{(1)',k}_{kl}A^{(1),l}
\big) \big]
-\l[\int^\tau  \frac{d\tau' }{a(\tau')}\r]^2\big[
\delta_{ij}\nabla^{-2}\big(
A^{(1),klm}A^{(1)}_{,\,klm}
\nn\\
&
-\nabla^2A^{(1),k}\nabla^2A^{(1)}_{,k}
\big)
+2\nabla^{-2}\big(-2A^{(1),kl}_{,i}A^{(1)}_{,\,klj}
+2A^{(1)}_{,kij}\nabla^2A^{(1),k}
\big)
+\partial_i\partial_j\nabla^{-2}\nabla^{-2}\big(
A^{(1),klm}A^{(1)}_{,\,klm}
\nn\\
&
-\nabla^2A^{(1),k}\nabla^2A^{(1)}_{,k}
\big)
\big]
+\l[\int^\tau  \frac{d\tau' }{a(\tau')}\r]\big[
\delta_{ij}\nabla^{-2}\big(
4\phi^{(1),kl}A^{(1)}_{,\,kl}
-4\nabla^2\phi^{(1)}\nabla^2A^{(1)}
-\chi^{(1),kl}_{kl,m}A^{(1),m}
\nn\\
&
-4\chi^{(1),k}_{kl,m}A^{(1),lm}
+\chi^{(1)}_{kl,m}A^{(1),klm}
-2\chi^{(1),\,k}_{kl}\nabla^2A^{(1),\,l}
+2 A^{(1),kl}\nabla^2\chi^{(1)}_{kl}
-2A^{(1),klm}C^{||(1)}_{,\,klm}
\nn\\
&
+2\nabla^2A^{(1),k}\nabla^2C^{||(1)}_{,k}
\big)
-2\chi^{(1)}_{ij,k}A^{(1),k}
-2\chi^{(1)}_{ki}A^{(1),k}_{,j}
-2\chi^{(1)}_{kj}A^{(1),k}_{,i}
-4\nabla^{-2}\big(
2\phi^{(1),k}_{,i}A^{(1)}_{,\,kj}
+2\phi^{(1),k}_{,j}A^{(1)}_{,\,ki}
\nn\\
&
-2\phi^{(1)}_{,ij}\nabla^2A^{(1)}
-2A^{(1)}_{,ij}\nabla^2\phi^{(1)}
-A^{(1),kl}_{,i}C^{||(1)}_{,\,klj}
-A^{(1),kl}_{,j}C^{||(1)}_{,\,kli}
+A^{(1)}_{,kij}\nabla^2C^{||(1),k}
+C^{||(1)}_{,kij}\nabla^2A^{(1),k}
\big)
    \nn\\
    &
+2\partial_i\nabla^{-2}\big(
\chi^{(1),k}_{kj,\,l}A^{(1),l}
+2\chi^{(1)}_{kj,\,l}A^{(1),\,kl}
+\chi^{(1),\,l}_{kl}A^{(1),k}_{,j}
+\chi^{(1)}_{kl}A^{(1),kl}_{,j}
+\chi^{(1)}_{kj}\nabla^2A^{(1),\,k}
\big)
\nn\\
&
+2\partial_j\nabla^{-2}\big(
\chi^{(1),k}_{ki,\,l}A^{(1),l}
+2\chi^{(1)}_{ki,\,l}A^{(1),kl}
+\chi^{(1),\,l}_{kl}A^{(1),k}_{,i}
+\chi^{(1)}_{kl}A^{(1),kl}_{,i}
+\chi^{(1)}_{ki}\nabla^2A^{(1),\,k}
\big)
\nn\\
&
+\partial_i\partial_j\nabla^{-2}\nabla^{-2}\big(
4\phi^{(1),kl}A^{(1)}_{,\,kl}
-4\nabla^2\phi^{(1)}\nabla^2A^{(1)}
-\chi^{(1),kl}_{kl,m}A^{(1),m}
-4\chi^{(1),k}_{kl,m}A^{(1),lm}
-7\chi^{(1)}_{kl,m}A^{(1),klm}
\nn\\
&
-2\chi^{(1),\,k}_{kl}\nabla^2A^{(1),\,l}
-4\chi^{(1)}_{kl}\nabla^2A^{(1),\,kl}
-2 A^{(1),kl}\nabla^2\chi^{(1)}_{kl}
-2A^{(1),klm}C^{||(1)}_{,\,klm}
+2\nabla^2A^{(1),k}\nabla^2C^{||(1)}_{,k}
\big)
\big]
\nn\\
&
+\Big[
\delta_{ij}\nabla^{-2}\big(
4\phi^{(1),kl}C^{||(1)}_{,\,kl}
-4\nabla^2\phi^{(1)}\nabla^2C^{||(1)}
-\chi^{(1),kl}_{kl,m}C^{||(1),m}
-4\chi^{(1),\,k}_{kl,\,m}C^{||(1),\,lm}
+\chi^{(1)}_{kl,m}C^{||(1),klm}
\nn\\
&
-2\chi^{(1),\,k}_{kl}\nabla^2C^{||(1),\,l}
+2C^{||(1),kl}\nabla^2\chi^{(1)}_{kl}
+\nabla^2C^{||(1),k}\nabla^2C^{||(1)}_{,k}
-C^{||(1),klm}C^{||(1)}_{,\,klm}
\big)
-2 \chi^{(1)}_{ij,k}C^{||(1),k}
\nn\\
&
-2\chi^{(1)}_{ki}C^{||(1),k}_{,j}
-2\chi^{(1)}_{kj}C^{||(1),k}_{,i}
-4\nabla^{-2}\big(2\phi^{(1),k}_{,i}C^{||(1)}_{,\,kj}
+2\phi^{(1),k}_{,j}C^{||(1)}_{,\,ki}
-2\phi^{(1)}_{,ij}\nabla^2C^{||(1)}
-2C^{||(1)}_{,ij}\nabla^2\phi^{(1)}
\nn\\
&
-C^{||(1),kl}_{,i}C^{||(1)}_{,\,klj}
+C^{||(1)}_{,kij}\nabla^2C^{||(1),k}
\big)
+2 \partial_i\nabla^{-2}\big(
\chi^{(1),k}_{kj,\,l}C^{||(1),l}
+2\chi^{(1)}_{kj,\,l}C^{||(1),kl}
+\chi^{(1),\,l}_{kl}C^{||(1),k}_{,j}
\nn\\
&
+\chi^{(1)}_{kl}C^{||(1),kl}_{,j}
+\chi^{(1)}_{kj}\nabla^2C^{||(1),\,k}
\big)
+2 \partial_j\nabla^{-2}\big(
\chi^{(1),k}_{ki,\,l}C^{||(1),l}
+2\chi^{(1)}_{ki,\,l}C^{||(1),kl}
+\chi^{(1),\,l}_{kl}C^{||(1),k}_{,i}
\nn\\
&
+\chi^{(1)}_{kl}C^{||(1),kl}_{,i}
+\chi^{(1)}_{ki}\nabla^2C^{||(1),\,k}
\big)
-\partial_i\partial_j\nabla^{-2}\nabla^{-2}\big(
4\nabla^2\phi^{(1)}\nabla^2C^{||(1)}
-4\phi^{(1),kl}C^{||(1)}_{,\,kl}
+\chi^{(1),kl}_{kl,m}C^{||(1),m}
\nn\\
&
+4\chi^{(1),\,k}_{kl,\,m}C^{||(1),\,lm}
+7\chi^{(1)}_{kl,m}C^{||(1),klm}
+2\chi^{(1),\,k}_{kl}\nabla^2C^{||(1),\,l}
+4\chi^{(1)}_{kl}\nabla^2C^{||(1),kl}
+2 C^{||(1),kl}\nabla^2\chi^{(1)}_{kl}
\nn\\
&
+C^{||(1),klm}C^{||(1)}_{,\,klm}
-\nabla^2C^{||(1),k}\nabla^2C^{||(1)}_{,k}
\big)
\Big]  \, .
\el
}
Eq.(\ref{chiT2transF2}) tells  that transformation of $ \chi^{\top(2)}_{ij}$
  involves $\xi^{(1)\mu}$ only,   not  $\xi^{(2)\mu}$.
The above formulas  are valid for a general scale factor $a(\tau)$.
Note that,
in (\ref{chi||2transF2}), (\ref{chiPerp2TransF}), and (\ref{chiT2transF2}),
$\chi^{ (1)}_{ij } $ contains  the tensor  $\chi^{\top(1)}_{ij } $,
which belongs to  the   scalar-tensor coupling
and will not be  considered  in this paper.

\section{ Gauge Transformations from Synchronous to Poisson}

The  perturbed metric  in the Poisson gauge up to 2nd order
is generally written as
\cite{MaBertschinger1995,Matarrese98}
\be
g_{00}=-a^2\Big[
1+2\psi^{(1)}_{P}
+ \psi^{(2)}_{P}\Big],
\ee
\be
g_{0i}=a^2\Big[
w^{(1)}_{P\,i}
+\frac{1}{2}w^{(2)}_{P\,i}\Big],
\ee
\be
g_{ij}=a^2\Big[
\delta_{ij}
-2\Big(\phi^{(1)}_{P}+\frac{1}{2}\phi^{(2)}_{P}\Big)\delta_{ij}
+\chi^{\top(1)}_{P\,ij}
+\frac{1}{2}\chi^{\top(2)}_{P\,ij}\Big].
\ee
where the vector (shift) is transverse
\be \label{poissonCon}
\partial^i w^{(A)}_{P\,i}=0,  \, \,\,\, A=1,2
\ee
and   the tensor is
\be  \label{poissonConTensor}
\chi^{\top(A)i}_{P\,i}=0,
~~~~~~
\partial^i\chi^{\top(A)}_{P\,ij}=0.
\ee

Consider  transformations of the  metric perturbations
from a synchronous   to a Poisson coordinate.
Given the 1st-order  solutions
$\phi^{(1)},D_{ij}\chi^{||(1)},\chi^{\top(1)}_{ij}$
 in synchronous gauge  without vector mode,
one gets,  by   the  formula   (\ref{metricTrans1st}),
the 1st-order  perturbation in Poisson gauge
             as the following \cite{Matarrese98}
\be\label{psi1P}
\psi^{(1)}_P=-\alpha^{(1)'}-\frac{a'}{a}\alpha^{(1)},
\ee
\be\label{wi1P}
w^{(1)}_{P\,i}=\alpha^{(1)}_{,\,i}
-\beta^{(1)'}_{,\,i}
-d^{(1)'}_i,
\ee
\be\label{phi1P}
\phi^{(1)}_{P}=\phi^{(1)}
+\frac{1}{3}\nabla^2\beta^{(1)}
+\frac{a'}{a}\alpha^{(1)},
\ee
\be\label{chi1P}
\chi^{\top(1)}_{P\,ij}=
D_{ij}\chi^{||(1)}
+\chi^{\top(1)}_{ij}
-2D_{ij}\beta^{(1)}
-d^{(1)}_{i,j}
-d^{(1)}_{j,\,i}.
\ee
By the conditions  (\ref{poissonCon}) and (\ref{poissonConTensor})
and the solution
$D_{ij}\chi^{||(1)}$ of (\ref{Dchi1sol}),
one gets  the 1st-order  vector field of transformation
\be\label{alpha1bar}
\alpha^{(1)}=
-\frac{\tau}{3}\varphi
+\frac{9}{\tau^4}\nabla^{-2} X \, ,
~~~~~~
\beta^{(1)}=
-\frac{\tau^2}{6}\varphi
-\frac{3}{\tau^3}\nabla^{-2} X \, ,
~~~~~~
d^{(1)}_i=0 .
\ee
Substituting (\ref{alpha1bar}) into  (\ref{psi1P})-(\ref{chi1P}),
one obtains the 1st-order  perturbations in Poisson gauge \cite{Matarrese98}
\be
\psi^{(1)}_P = \phi^{(1)}_P  =  \varphi
+\frac{18}{\tau^5}\nabla^{-2} X \, ,
~~~~~ w^{(1)}_{P\,i}=0,
~~~~~ \chi^{\top(1)}_{P\,ij}=  \chi^{\top(1)}_{\,ij}   ,
\ee
which tells that
the two scalars  are   equal and contain decaying modes,
the vector  is   absent,
and the tensor  is equal to that in synchronous gauge.

Next   the 2nd-order transformation.
By the  formula   (\ref{metricTrans2nd3}),
keeping  only the scalar-scalar coupling,
using  the   solutions $\phi^{(1)}$ of   (\ref{phi1sol}),
 $\beta^{(1)}$ of (\ref{alpha1bar}),
 $\chi^{||(2)}_{S}$ of  (\ref{chi||2S}), $\chi^{\perp(2)}_{Sij}$  of  (\ref{chi2Sperp4}),
after  lengthy calculations,
one obtains
\bl\label{alpha2bar}
\alpha^{(2)}=&
\frac{\tau }{10} \nabla^{-2}F
-\frac{2\tau}{9}\Big[
5\varphi \varphi
-6\nabla^{-2} \l(\varphi_{,\,k} \varphi^{,\,k}\r)
+9\nabla^{-2} \nabla^{-2} \l( \varphi_{,\,kl}  \varphi^{,\,kl}
 -\nabla^2 \varphi \nabla^2 \varphi\r)
\Big]
\nonumber \\
&-\frac{\tau^3}{21}\Big[
\nabla^{-2}\l(\nabla^2\varphi \nabla^2\varphi
        - \varphi^{,\,kl}\varphi_{,\,kl}\r)
\Big]
\nn\\
&
+\frac{162}{\tau^9}
\Big[
-3\nabla^{-2} X\nabla^{-2} X
+\nabla^{-2}(
5\nabla^{-2} X_{,k}\nabla^{-2} X^{,k})
\nn\\
&
+\nabla^{-2}\nabla^{-2}
(
6\nabla^2X\nabla^{-2} X
+6 X_{,k}\nabla^{-2} X^{,k}
)
\Big]
    \nn\\
    &
+\frac{3}{2\tau^7}
\Big[
2 \nabla^{-2} X_{,k}\nabla^{-2} X^{,\,k}
+\nabla^{-2}(
9 X^2
-9\nabla^{-2}X^{,kl}\nabla^{-2}X_{,kl}
)
\Big]
    \nn\\
    &
+\frac{3}{\tau^4}
\Big[
3\nabla^{-2} Z
+\nabla^{-2}\nabla^{-2}
(
5\nabla^{-2} X\nabla^2\nabla^2\varphi
-5\varphi \nabla^2X
    \nn\\
    &
-6 X\nabla^2\varphi
+6\varphi_{,k} X^{,k}
-4\nabla^2\varphi^{,k}\nabla^{-2} X_{,k}
+8\varphi^{,kl}\nabla^{-2} X_{,kl}
)
\Big]
    \nn\\
    &
+\frac{1}{2\tau^2}
\Big[
5\varphi^{,\,k}\nabla^{-2} X_{,\,k}
+\nabla^{-2}(
-2 X\nabla^2\varphi
-8\varphi^{,\,kl}\nabla^{-2} X_{,\,kl}
-10\nabla^2\varphi^{,\,k}\nabla^{-2} X_{,\,k}
)
\Big]
,
\el
\bl\label{beta2bar}
\beta^{(2)}
=&
\frac{1}{2}\nabla^{-2}\nabla^{-2} A
-\frac{\tau^2}{6}\Big[
- \frac{3}{10}\nabla^{-2}F
+7\varphi \varphi
-4\nabla^{-2} \l(\varphi_{,\,k} \varphi^{,\,k}\r)
 \nn \\
& +6\nabla^{-2} \nabla^{-2} \l( \varphi_{,\,kl}  \varphi^{,\,kl}
 -\nabla^2 \varphi \nabla^2 \varphi\r)
\Big]
\nonumber \\
&-\frac{\tau^4}{504}\Big[
7\varphi^{,\,k}\varphi_{,\,k}
+6\nabla^{-2}\l(\nabla^2\varphi \nabla^2\varphi
        - \varphi^{,\,kl}\varphi_{,\,kl}\r)
\Big]
\nn\\
&
+\frac{81}{4\tau^8}
\Big[
\nabla^{-2}X\nabla^{-2} X
+\nabla^{-2}(
-5\nabla^{-2} X_{,k}\nabla^{-2} X^{,k}
)
\nn\\
&
+\nabla^{-2}\nabla^{-2}
(
-6\nabla^2X\nabla^{-2} X
-6 X_{,k}\nabla^{-2} X^{,k}
)
\Big]
\nn\\
&
+\frac{9}{4\tau^6}
\Big[
\nabla^{-2}(
- 4 X_{,k}\nabla^{-2} X^{,\,k}
- X^2
)
\nn\\
&
+\nabla^{-2}\nabla^{-2}
(
-6X_{,k l}\nabla^{-2} X^{,\,kl}
-6\nabla^{-2}X^{,klm}\nabla^{-2}X_{,klm}
)
\Big]
\nn\\
&
+\frac{1}{\tau^3}
\Big[
\nabla^{-2}(
3\nabla^{-2} X\nabla^2\varphi
-3 Z
)
\nn\\
&
+\nabla^{-2}\nabla^{-2}
(
-17\varphi\nabla^2X
-11 X\nabla^2\varphi
-64\varphi_{,k} X^{,k}
-36\varphi^{,kl}\nabla^{-2} X_{,kl}
)
\Big]
\nn\\
&
+\frac{1}{2\tau}
\Big[
-\varphi^{,\,k}\nabla^{-2} X_{,k}
+\nabla^{-2}(
2 X\nabla^2\varphi
-2\varphi_{,\,kl}\nabla^{-2} X^{,\,kl}
)
\Big]
,
\el
\bl\label{d2bar2}
d^{(2)}_{i}
=&
\nabla^{-2}G_i
-\frac{4\tau^2}{3}\nabla^{-2} \Big[
-\varphi_{,\,i}\nabla^2\varphi
+\varphi^{,\,k}\varphi_{,\,ki}
+\partial_i\nabla^{-2}(
\nabla^2\varphi\nabla^2\varphi
-\varphi_{,\,kl}\varphi^{,\,kl}
)
\Big]
\nn\\
&
+\frac{162}{\tau^8}
\nabla^{-2}
\Big[
- X_{,i}\nabla^{-2} X
+\partial_i\nabla^{-2}
\big(
\nabla^2X\nabla^{-2} X
+X_{,k}\nabla^{-2} X^{,k}
\big)
\Big]
\nn\\
&
+\frac{2}{\tau^3}
\nabla^{-2}
\Big[
7\varphi_{,i} X
+15\varphi_{,ki}\nabla^{-2} X^{,k}
-15\varphi^{ ,\,k }\nabla^{-2}X_{,ik}
-3 \nabla^{-2} X_{,i}\nabla^2\varphi
\nn\\
&
+\partial_i\nabla^{-2}
\big(
-4 X\nabla^2\varphi
+8 X_{,k}\varphi^{,k}
-12\nabla^2\varphi_{,k}\nabla^{-2} X^{,k}
\big)
\Big]
\nn\\
&
+\frac{1}{\tau}
\nabla^{-2}
\Big[
5\nabla^2\varphi^{,k}\nabla^{-2}X_{,ik}
-5 \varphi_{,ik}X^{,k}
+5\varphi^{,kl}\nabla^{-2}X_{,kli}
-5\varphi_{,\,kli}\nabla^{-2} X^{,\,kl}
\Big]
,
\el
where $F$, $Z$, $A$, $G_i$ are given by  (\ref{D}), (\ref{Zfixed}), (\ref{A}), and (\ref{Gi}),
all depending on the initial values at $\tau_0$,
and the 2nd metric perturbations in Poisson gauge are
\bl\label{psi2P2}
\psi^{(2)}_{P}=&-\frac{3}{10} \nabla^{-2}F
+\Big[
\frac{16}{3}\varphi \varphi
-4\nabla^{-2} \l(\varphi_{,\,k} \varphi^{,\,k}\r)
+6\nabla^{-2} \nabla^{-2} \l( \varphi_{,\,kl}  \varphi^{,\,kl}
 -\nabla^2 \varphi \nabla^2 \varphi\r)
\Big]
\nonumber \\
&
+\tau^2\Big[\frac{1}{6}\varphi^{,\,k}\varphi_{,\,k}
+\frac{5}{21}\nabla^{-2}\l(\nabla^2\varphi \nabla^2\varphi
        - \varphi^{,\,kl}\varphi_{,\,kl}\r)
\Big]
\nn\\
&
+\frac{162}{\tau^{10}}
\Big[
-12\nabla^{-2} X\nabla^{-2} X
+\nabla^{-2}(
35\nabla^{-2} X_{,k}\nabla^{-2} X^{,k}
)
\nn\\
&
+\nabla^{-2}\nabla^{-2}
(
42\nabla^2X\nabla^{-2} X
+42 X_{,k}\nabla^{-2} X^{,k}
)
\Big]
\nn\\
&
+\frac{3}{2\tau^8}
\Big[
46\nabla^{-2} X_{,\,k}\nabla^{-2} X^{,\,k}
+\nabla^{-2}(
45 X^2
-45\nabla^{-2}X^{,kl}\nabla^{-2}X_{,kl}
)
\Big]
\nn\\
&
+\frac{6}{\tau^5}
\Big[
3\nabla^{-2} Z
+7\varphi\nabla^{-2} X
+\nabla^{-2}\nabla^{-2}
(
5\nabla^{-2} X\nabla^2\nabla^2\varphi
-5\varphi \nabla^2X
-6 X\nabla^2\varphi
\nn\\
&
+6\varphi_{,k} X^{,k}
-4\nabla^2\varphi^{,k}\nabla^{-2} X_{,k}
+8\varphi^{,kl}\nabla^{-2} X_{,kl}
)
\Big]
+\frac{6}{\tau^3}\varphi^{,\,k}\nabla^{-2} X_{,\,k}
\ ,
\el
\bl\label{phi2P2}
\phi^{(2)}_{P}=&\Big[\frac{3 }{5} \phi^{(2)}_{S0}
+\frac{1}{10}\nabla^2\chi^{\parallel(2) }_{S0}
+ \frac{\tau_0^2}{6}\varphi_{,\,k}\varphi^{,\,k}
+ \frac{\tau_0^2}{6}\nabla^{-2}(
 \nabla^2\varphi \nabla^2\varphi
-\varphi^{,\,kl}\varphi_{,\,kl}   )
\nn\\
&
-\frac{\tau_0^4}{120} \nabla^{-2}(
 \nabla^2\varphi^{,\,k}\nabla^2\varphi_{,\,k}
                  - \varphi^{,\,klm} \varphi_{,\,klm} )
\Big]
\nn\\
&
+\Big[\frac{4}{3}  \varphi\varphi
+\frac{8}{3}\nabla^{-2} \l(\varphi_{,\,k} \varphi^{,\,k}\r)
+4\nabla^{-2} \nabla^{-2} \l( \nabla^2 \varphi \nabla^2 \varphi
 -\varphi_{,\,kl}  \varphi^{,\,kl}\r)
\Big]
\nonumber \\
&
+\tau^2\Big[\frac{1}{6}\varphi_{,\,k}\varphi^{,\,k}
+\frac{5}{21}\nabla^{-2} \l(\nabla^2 \varphi \nabla^2 \varphi
 -\varphi_{,\,kl}  \varphi^{,\,kl}\r)
 \Big]
\nn\\
&
+\frac{324}{\tau^{10}}
\Big[
-\nabla^{-2}(
5 X\nabla^{-2} X
)
+\nabla^{-2}\nabla^{-2}
(
6\nabla^2X\nabla^{-2} X
+6 X_{,k}\nabla^{-2} X^{,k}
)
\Big]
\nn\\
&
+\frac{15}{2\tau^8}
\Big[
8\nabla^{-2} X_{,\,k}\nabla^{-2} X^{,\,k}
+\nabla^{-2}(
9 X^2
-9\nabla^{-2}X^{,kl}\nabla^{-2}X_{,kl}
)
\Big]
\nn\\
&
+\frac{6}{\tau^5}
\Big[
3\nabla^{-2} Z
-17\varphi\nabla^{-2} X
+\nabla^{-2}\nabla^{-2}
(
5\nabla^{-2} X\nabla^2\nabla^2\varphi
-5\varphi \nabla^2X
\nn\\
&
-6 X\nabla^2\varphi
+6\varphi_{,k} X^{,k}
-4\nabla^2\varphi^{,k}\nabla^{-2} X_{,k}
+8\varphi^{,kl}\nabla^{-2} X_{,kl}
)
\Big]
\nn\\
&
+\frac{6}{\tau^3}\varphi^{,\,k}\nabla^{-2} X_{,\,k}
+\Big[
\frac{27}{10\tau_0^6}\nabla^{-2}(
\nabla^{-2}X^{,klm}\nabla^{-2}X_{,klm}
- X^{,k}X_{,k}
)
\nn\\
&
+\frac{3}{\tau_0^3}\nabla^{-2}(
 X\nabla^2\varphi
+\varphi^{ ,\,kl }\nabla^{-2}X_{,kl}
+2 X^{,k}\varphi_{,k}
)
\nn\\
&
+\frac{3}{10\tau_0} \nabla^{-2}
(
\varphi_{,klm}\nabla^{-2}X^{,klm}
-X_{,k}\nabla^2\varphi^{,k}
)
\Big]
,
\el
\bl\label{wi2P2}
w^{(2)}_{P\,i}    =&
\frac{8}{3}\tau\nabla^{-2} \Big[
-\varphi_{,\,i}\nabla^2\varphi
+\varphi^{,\,k}\varphi_{,\,ki}
+\partial_i\nabla^{-2}(
\nabla^2\varphi\nabla^2\varphi
-\varphi_{,\,kl}\varphi^{,\,kl}
)
\Big]
\nn\\
&
+\frac{1296}{\tau^9}
\nabla^{-2}
\Big[
- X_{,i}\nabla^{-2} X
+\partial_i\nabla^{-2}
\big(
\nabla^2X\nabla^{-2} X
+X_{,k}\nabla^{-2} X^{,k}
\big)
\Big]
\nn\\
&
+\frac{6}{\tau^4}
\nabla^{-2}
\Big[
-8\varphi_{,\,i} X
+12 \nabla^{-2} X_{,i}\nabla^2\varphi
\nn\\
&
+\partial_i\nabla^{-2}
\big(
-4 X\nabla^2\varphi
+8 X_{,k}\varphi^{,k}
-12\nabla^2\varphi_{,k}\nabla^{-2} X^{,k}
\big)
\Big]
,
\el
\bl\label{chi2P2}
\chi^{\top(2)}_{P\,ij}=&
\frac{20}{3} \delta_{ij}\nabla^{-2}\nabla^{-2}(
\varphi_{,\,kl} \varphi^{,\,kl}
- \nabla^2 \varphi \nabla^2 \varphi)
+\frac{80}{3}\nabla^{-2}\nabla^{-2}  (\varphi_{,\,i j} \nabla^2\varphi
-\varphi^{,\,k}_{,\,i}\varphi_{,\,kj})
\nn\\
&
+\frac{20}{3} \nabla^{-2}\nabla^{-2}\nabla^{-2}\partial_i\partial_j(
\varphi_{,\,kl} \varphi^{,\,kl}
- \nabla^2 \varphi\nabla^2\varphi   )
\nn\\
&
+\frac{1}{\tau^8}
\Big[
\nabla^{-2}\Big(
81X^2
-81\nabla^{-2} X_{,kl}\nabla^{-2} X^{,kl}
\Big)\delta_{ij}
-162\nabla^{-2} X_{,ij}\nabla^{-2} X
\nn\\
&
+\nabla^{-2}\partial_i\Big(
162X_{,j}\nabla^{-2} X
\Big)
+\nabla^{-2}\partial_j\Big(
162X_{,i}\nabla^{-2} X
\Big)
\nn\\
&
+\nabla^{-2}\partial_i\partial_j
\Big(
81\nabla^{-2} X_{,k}\nabla^{-2} X^{,k}
-81X\nabla^{-2} X
\Big)
\nn\\
&
+\nabla^{-2}\nabla^{-2}\partial_i\partial_j
\Big(
81\nabla^{-2} X_{,kl}\nabla^{-2} X^{,kl}
-81\nabla^2X\nabla^{-2} X
\Big)
\Big]
\nn\\
&
+\frac{9}{2\tau^6}
\Big[
\Big(
 X^2
-\nabla^{-2}X_{,kl}\nabla^{-2}X^{,kl}
\Big)\delta_{ij}
-4X\nabla^{-2}X_{,ij}
+4\nabla^{-2}X^{,k}_{,i}\nabla^{-2}X_{,kj}
\nn\\
&
+\nabla^{-2}\partial_i\partial_j
\,\Big(
X^2
-\nabla^{-2} X_{,kl}\nabla^{-2} X^{,\,kl}
\Big)
\Big]
\nn\\
&
+\frac{9}{8\tau^4}
\Big[
\delta_{ij}
\nabla^2\Big(
X^2
-\nabla^{-2}X^{,kl}\nabla^{-2}X_{,kl}
\Big)
+\partial_i\partial_j
\,\Big(
 X^2
-\nabla^{-2}X^{,kl}\nabla^{-2}X_{,kl}
\Big)
\nn\\
&
+\nabla^2 \Big(
4\nabla^{-2}X^{,k}_{,i}\nabla^{-2}X_{,kj}
-4 X\nabla^{-2}X_{,ij}
\Big)
\Big]
\nn\\
&
+\frac{1}{(2\pi)^{3/2}}
   \int d^3k   e^{i \,\bf{k}\cdot\bf{x}} \,
     \Bigg[
      \Big(
\frac{k\cos(k\tau)\bar B_{5\,ij}}{8\tau^3}
-\frac{k^3\cos(k\tau)\bar B_{4\,ij}}{144\tau^3}
+\frac{k^2\sin(k\tau)\bar B_{5\,ij}}{8\tau^2}
\nn\\
&
-\frac{k^4\sin(k\tau)\bar B_{4\,ij}}{144\tau^2}
\Big)\l(\int_{0}^{k\tau}\frac{\sin s}{s}ds\r)
+\Big(
\frac{k\sin(k\tau)\bar B_{5\,ij}}{8\tau^3}
-\frac{k^3\sin(k\tau)\bar B_{4\,ij}}{144\tau^3}
\nn\\
&
-\frac{k^2\cos(k\tau)\bar B_{5\,ij}}{8\tau^2}
+\frac{k^4\cos(k\tau)\bar B_{4\,ij}}{144\tau^2}
\Big)\l(\int_{k\tau}^{+\infty}\frac{\cos s}{s}ds\r)
+ \sum_{s={+,\times}} {\mathop \epsilon
    \limits^s}_{ij}(k) ~ {\mathop h\limits^s}_k(\tau)
    \Bigg]
\ .
\el
The results   in (\ref{psi2P2})-(\ref{chi2P2})
contain decaying modes,
and extend  (6.8) of Ref.~\cite{Matarrese98}
to general initial conditions at time $\tau_0 $.
The results tell that
the scalars $ \propto\tau^0, \tau^{2}$, $\tau^{-10}, \tau^{-8}$,
$\tau^{-5}, \tau^{-3}$,
 the vector $ \propto\tau^1$, $\tau^{-9}, \tau^{-4}$,
 and the  tensor $\propto\tau^{-8}, \tau^{-6}$, $\tau^{-4}, \tau^{0}$
   in Poisson gauge.


\begin{thebibliography}{34}

\bibitem{Lifshitz1946} E. M. Lifshitz,  Zh. Eksp. Theor. Fiz \textbf{16},   587 (1946).
	

\bibitem{LifshitzKhalatnikov1963} E. M. Lifshitz and I. M. Khalatnikov,
          Adv. Phys. {\bf 12} 185 (1963).

	

\bibitem{PressVishniac1980} W. H. Press and E. T. Vishniac,
 Astrophys. J. \textbf{239},   1 (1980).
	

\bibitem{Bardeen1980} J.  M.  Bardeen,     Phys.  Rev.  D \textbf{22},     1882 (1980).


\bibitem{KodamSasaki1984} H.  Kodama and  M.  Sasaki,
Prog. Theor. Phys. Suppl. \textbf{78},   1 (1984).
	

\bibitem{Grishchuk1994} L. P. Grishchuk,    Phys. Rev. D \textbf{50}, 7154, (1994)

\bibitem{Peebles1980} P. J. E. Peebles, \emph{The Large-Scale Structure of the Universe},
         (Princeton University Press, Princeton, New Jersey, 1980).


\bibitem{BaskoPolnarev1984} M.M. Basko and A.G. Polnarev,
                   Mon. Not. R. Astron. Soc. \textbf{191},   207 (1980);
                   Sov. Astron. \textbf{24},   268 (1980).


\bibitem{Polnarev1985} A.G. Polnarev, Sov. Astron. {\bf29}, 607  (1985).


\bibitem{MaBertschinger1995} C. P. Ma and E. Bertschinger,
Astrophys. J.  {\bf 455}, 7 (1995).


      E. Bertschinger,   Arxiv: astro-ph/9503125.


\bibitem{ZaldarriagaHarari1995} M. Zaldarriaga and D. D.Harari,
                   Phys. Rev. D {\bf 52}, 3276 (1995).

\bibitem{Kosowsky1996} A. Kosowsky, Ann. Phys.{\bf 246},  49 (1996).


\bibitem{ZaldarriagaSeljak1997} M. Zaldarriaga and U. Seljak,
                  Phys. Rev. D {\bf55},  1830 (1997).

\bibitem{Kamionkowski1997} M. Kamionkowski, and A.  Kosowsky, A. Stebbins,
                  Phys. Rev. D {\bf55}, 7368 (1997).

\bibitem{KeatingTimbie1998}  B. Keating, P.  Timbie, A. Polnarev, and J. Steinberger,
Astrophys. J. {\bf 495} 580 (1998).
	

\bibitem{ZhaoZhang2006} W. Zhao and Y. Zhang,
               Phys. Rev. D {\bf74},  083006 (2006);

 T.Y. Xia and Y. Zhang,  Phys. Rev. D {\bf78},  123005 (2008);

 T.Y. Xia and Y. Zhang,  Phys. Rev. D {\bf 79}, 083002  (2009);


   Z. Cai and Y. Zhang,
    Classical Quandum Gravity {\bf 29}  (2012)  105009.


\bibitem{Baskaran} D. Baskaran,  L. P. Grishchuk,  and A. G. Polnarev,
                      Phys. Rev. D {\bf 74}, 083008 (2006).


\bibitem{Polnarevmiller2008} A. G. Polnarev, N. J. Miller, and B. G. Keating,
           Mon. Not. R. Astron. Soc., {\bf386}, 1053 (2008);



\bibitem{Grishchuk}L. P. Grishchuk, Sov. Phys. JETP {\bf40}, 409 (1975);

         L. P. Grishchuk ,  Classical Quandum Gravity {\bf14} 1445 (1997);
            Lect. Notes Phys. {\bf 562} 167, (2001).


\bibitem{FordParker1977GW}  L. H. Ford and L. Parker, Phys. Rev. D {\bf16}, 1601 (1977).


\bibitem{Starobinsky}A. A. Starobinsky, JETP Lett. {\bf 30}, 682(1979).


\bibitem{Rubakov}   V. A. Rubakov, M. V. Sazhin  and  A. V. Veryaskin,
           Phys. Lett. {\bf115B}, 189 (1982 ).

\bibitem {Fabbri}  R. Fabbri and M.D. Pollock, Phys. Lett. {\bf125B},  445 (1983).

\bibitem {AbbottWise1984}  L.F. Abbott and M.B. Wise,  Nucl. Phys. {\bf B224}, 541 (1984).



\bibitem{Allen1988}B. Allen, Phys. Rev. D {\bf37}, 2078 (1988);

            B. Allen and S. Koranda, Phys. Rev. D {\bf 50} 3713  (1994).


\bibitem{Giovannini}     M. Giovannini,
            Phys. Rev. D {\bf 60}, 123511 (1999);
             PMC Phys. {\bf A4}, 1 (2010).


\bibitem{Tashiro}
         H. Tashiro,  T. Chiba and  M. Sasaki, Classical Quandum Gravity {\bf 21} 1761 (2004).

\bibitem{zhangyang05} Y. Zhang {\it et al}.,    Classical Quandum Gravity {\bf22}, 1383 (2005);

                  Y. Zhang {\it et al}.,   Classical Quandum Gravity  {\bf23}, 3783 (2006);

                   W. Zhao and Y. Zhang,   Phys. Rev. D \textbf{74},   043503   (2006);

             M.L. Tong and Y. Zhang, Phys. Rev. D {\bf80}, 084022 (2009);

                      Y. Zhang, M.L. Tong, and  Z.W. Fu,
             Phys. Rev. D   {\bf81}, 101501(R)  (2010);


            D.Q. Su and  Y. Zhang, Phys. Rev. D {\bf 85}, 104012  (2012);

            D.G. Wang, Y. Zhang and J.W. Chen,  Phys. Rev. D \textbf{94},   044033 (2016).
	



\bibitem{Morais2014} J. Morais, M. Bouhmadi-Lopez, and A. B. Henriques,
           Phys. Rev. D \textbf{89}, 023513 (2014).

\bibitem{PyneCarroll1996} T. Pyne and S.M. Carroll, Phys. Rev. D    {\bf 53}, 2920   (1996).


\bibitem{MollerachHarariMatarrese2004} S. Mollerach, D. Harari, and S. Matarrese,
      Phys. Rev. D {\bf 69}, 063002 (2004).



\bibitem{Bartolo2010}
V. Acquaviva, N.  Bartolo, S.  Matarrese, and A.  Riotto,
                    Nucl. Phys. \textbf{B667},   119 (2003);
	


 N. Bartolo, S.  Matarrese,   and   A.  Riotto,
                      Phys. Rev. D {\bf 69},    043503   (2004).

 N. Bartolo  , S. Matarrese    and A. Riotto,  J. Cosmol. Astropart. Phys. 01 (2004) 003;


 N. Bartolo  , S. Matarrese    and A. Riotto,   J. Cosmol. Astropart. Phys.  10 (2005) 010.



\bibitem{AnandaClarksonWands2007} K. N. Ananda, C. Clarkson and D. Wands,
               Phys.Rev.D {\bf 75},  123518 (2007).


\bibitem{Baumann2007} D. Baumann, P. Steinhardt, K. Takahashi and K. Ichiki,  Phys. Rev. D \textbf{76},   084019 (2007).
	

\bibitem{GW150914} B. P. Abbott {\it et al}.,   Phys. Rev. Lett. \textbf{116},  061102 (2016)

                  B. P. Abbott {\it et al}.,   Phys. Rev. Lett. \textbf{116}, 241103 (2016).

\bibitem{aLIGO O1} B. P. Abbott {\it et al}., Phys. Rev. Lett. \textbf{116},   131102 (2016).
	



\bibitem{WMAP9Bennett}  G. Hinshaw, D. Larson, E. Komatsu, {\it et al}.,  Astrophys. J. Suppl. Ser. \textbf{208},   19 (2013).
	


\bibitem{Planck2014} P. A. Ade, {\it et al}.,  Astron. Astrophys. \textbf{571},   A22 (2014).
	

    P. A. Ade,  {\it et al}.,  Astron. Astrophys. \textbf{594},   A13 (2016).
	

\bibitem{Tomita1967}   K. Tomita,  Prog. Theor. Phys. \textbf{37},   831 (1967).
	

\bibitem{Tomita1971}   K. Tomita,  Prog. Theor. Phys. \textbf{45},   1747 (1971);
	
 K. Tomita,  Prog. Theor. Phys. \textbf{47},   416 (1972).
	

\bibitem{MatarresePantanoSa'ez1994} S. Matarrese, O. Pantano and D. Saez,
             Phys. Rev. Lett. {\bf 72}, 320 (1994);
              Mon. Not. R. Astron. Soc. {\bf 271}, 513      (1994).

              S.  Matarrese and D. Terranova,
              Mon. Not.  R. Astron.  Soc. {\bf 283}, 400 (1996).

\bibitem{Russ1996} H. Russ, M. Morita, M. Kasai, and G. Borner,
              Phys. Rev. D {\bf 53}  6881 (1996).

\bibitem{Salopek}  D. S. Salopek, J. M. Stewart, and K. M. Croudace,
             Mon. Not. R. Astron. Soc. {\bf 271} , 1005 (1994).





\bibitem{MalikWands2004} K. A.  Malik  and  D.   Wands,
     Classical Quandum Gravity \textbf{21}, L65 (2004).



\bibitem{NohHwang2004}  H. Noh and J.-c. Hwang, Phys. Rev. D {\bf 69}, 104011   (2004);

    H. Noh and J.-c. Hwang, Classical Quandum Gravity \textbf{22} (2005) 3181;

    J.-c. Hwang, and  H. Noh, Phys. Rev. D {\bf73}, 044021 (2006);


  J.-c. Hwang, and  H. Noh,  Phys. Rev. D {\bf 76},  103527 (2007).



\bibitem{HwangNoh2012}  J.-c. Hwang,   H. Noh, and J.-O. Gong,  Astrophys. J. \textbf{752},   50 (2012).
	


\bibitem{HwangNoh2005}  J.-c. Hwang,  and H. Noh, Phys. Rev. D \textbf{72} 044011   (2005).


 \bibitem{Nakamura2003}   K. Nakamura,  Prog. Theor. Phys. \textbf{110},   723  (2003);
	
 Prog.   Theor. Phys. \textbf{113},   481  (2005);
	

           Phys. Rev. D {\bf 74},  101301  (2006);

           Phys. Rev. D {\bf 80}, 124021 (2009).


\bibitem{Domenech&Sasaki2017} G. Dom\`{e}nech and M. Sasaki,  arXiv:1709.09804.

\bibitem{Bruni97} M. Bruni, S. Matarrese, S. Mollerach and S. Sonego,
            Classical Quandum Gravity \textbf{14}, 2585 (1997).



\bibitem{Matarrese98} S.  Matarrese, S.  Mollerach,  and M.  Bruni,
                  Phys.  Rev.  D {\bf 58}, 043504 (1998).




\bibitem{Lu2008} T.  H.-C. Lu, K.  Ananda, C.  Clarkson,
                          Phys. Rev. D {\bf 77} 043523, (2008)

    T.  H.-C. Lu, K.  Ananda, C.  Clarkson, and R.  Maartens,
    J. Cosmol. Astropart. Phys. 02 (2009) 023.

\bibitem{Brilenkov&Eingorm2017} R.Brilenkov and M. Eingorm, Astrophys. J. {\bf 845}, 153 (2017)


\bibitem{Weinberg2004} S.  Weinberg,    Phys.  Rev.   D  {\bf69},  023503    (2004).



\bibitem{MiaoZhang2007} H. X.  Miao   and  Y.  Zhang,
             Phys.   Rev.  D   {\bf75},    104009 (2007).

\bibitem{WangZhang2008} S. Wang, Y. Zhang, T. Y. Xia, and H. X. Miao,
         Phys. Rev. D {\bf 77}, 104016 (2008).


\bibitem{Abramo1997} L. R. W. Abramo, R. H. Brandenberger and V. F. Mukhanov,
    Phys. Rev. D \textbf{56},   3248 (1997).
	

\bibitem{Gleiser1996} R. J. Gleiser, C. O. Nicasio, R. H. Price and J. Pullin,  Class. Quantum Grav. \textbf{13},   L117 (1996).
	


 \bibitem{Weinberg1972}     S. Weinberg, \emph{Gravitation and Cosmology: Principles and Applications of the General Theory of Relativity} (John Wiley and Sons, New York, 1972).
	

\bibitem{ZhangQinWang2017}
Y. Zhang {\it et al}., following article, Phys. Rev. D {\bf 96}, 103523
(2017).



\end{thebibliography}
\end{document}